\pdfoutput=1
\documentclass[prb,reprint, epsfig,amssymb,bibliography,amsmath,footinbib]{revtex4-2}

\usepackage{blkarray}
\usepackage[dvipsnames]{xcolor}

\usepackage{amssymb,amsmath,bm,epsfig,graphicx,subfigure,hyperref}

\usepackage[utf8]{inputenc}
\usepackage[vietnamese, english]{babel}
\usepackage{chngcntr}
\usepackage{xcolor}
\usepackage{braket}
\usepackage{physics}
\usepackage{sidecap}
\usepackage{lipsum}
\usepackage{amsfonts}
\usepackage{amsmath}
\usepackage{amssymb}
\usepackage{tikz}
\usetikzlibrary{shapes}
\usepackage{xcolor}
\usepackage{wasysym}
\usepackage{array}
\usepackage{bbold}
\usepackage{tabularx}
\usepackage{mathdots}
\usepackage[normalem]{ulem}

\begin{document}
\title{\color{Blue}{Valley Valves at Domain Walls in Symmetry-Broken Rhombohedral Graphene}}
\author{\foreignlanguage{vietnamese}{Võ Tiến Phong}$^{1,2}$}
\email{vophong@magnet.fsu.edu}
\author{Francisco Lobo$^{3}$}
\author{Elsa Prada$^{3}$}
\author{Pablo San-Jose$^{3}$}
\author{Francisco Guinea$^{4,5}$}
\email{paco.guinea@imdea.edu}
\author{Eugene Mele$^{6}$}
\email{mele@physics.upenn.edu}
\affiliation{$^{1}$Department of Physics, Florida State University, Tallahassee, FL, 32306, U.S.A.}
\affiliation{$^{2}$National High Magnetic Field Laboratory, Tallahassee, FL, 32310, U.S.A.}
\affiliation{$^{3}$Instituto de Ciencia de Materiales de Madrid (ICMM), CSIC, 28049 Madrid, Spain}
\affiliation{$^{4}$IMDEA Nanoscience, C/ Faraday 9, 28049 Madrid, Spain}
\affiliation{$^{5}$Donostia International Physics Center, Paseo Manuel de Lardiz\'{a}bal 4, 20018 San Sebastián, Spain}
\affiliation{$^{6}$Department of Physics and Astronomy, University of Pennsylvania, Philadelphia, PA 19104, U.S.A.}
\date{\today}

\begin{abstract}
    Rhombohedral multilayer graphene polarized by a moderate perpendicular displacement field hosts a time-reversal-symmetry-breaking valley-and-spin-polarized metallic phase that may condense into a chiral superconductor. Recent magnetic imaging and transport measurements in this unconventional system suggest the presence of domain walls both in the metallic and superconducting phases. In this work, we show that valley domain walls are impenetrable barriers to transport in the metallic regime. Transmission through such a domain wall must therefore be mediated by intervalley interactions. We derive the symmetry-allowed terms and show via microscopic numerical simulations that they enable the transmission of electrons across the domain wall. In the superconducting phase, we find that intervalley mixing is crucial for supporting an appreciable supercurrent through a SNS' Josephson junction that connects opposite-chirality superconducting regions. Taken together, our work elucidates the nature of domain walls in these experimentally relevant multilayer systems and emphasizes the critical role of intervalley hybridization in governing their transport properties.
\end{abstract}

\maketitle

\textcolor{Blue}{\textit{Introduction}}: Rhombohedral graphite is  a three-dimensional periodic arrangement of graphene layers which, although less energetically favorable than the more common Bernal graphite, is now routinely available in the laboratory. The semi-metallic behavior of a single graphene plane, combined with the special $ABC$ stacking of the planes, endows rhombohedral graphite with  nontrivial electronic properties. Its quasi-one-dimensional helical energy isosurfaces render it very close to a nodal semi-metal \cite{armitage_weyl_2018, Rudi2024Interfaces, Slizovskiy2019,Weht2026Topological}, and, as a result, it features flat surface bands at the top and bottom of a finite stack. The quenched bandwidths of these bands and their spectral isolation from the bulk states lead to the prediction of many correlated interaction-driven effects, including superconductivity \cite{heikkila_dimensional_2011,heikkila_flat_2011,kopnin_high-temperature_2011,Jiang2026Ideal}. Beyond the surface states, the bulk properties also imply that moderate magnetic fields can induce gaps in the three-dimensional Landau levels, leading to the three-dimensional quantum Hall effect \cite{arovas_stacking_2008}. 

Recent experiments have confirmed that multilayer rhombohedral graphene stacks indeed show flat electronic bands at the top and bottom surfaces \cite{pierucci2015evidence,Henck2018Flat,hagymasi_observation_2022, zhang2024correlated, Seifert2024Increasing}. Electronic interactions within these flat bands lead to a variety of experimentally-observed symmetry-broken phases, which are currently intensively studied \cite{lee_multicomponent_2016,shi_electronic_2020,zhou_half-_2021,zhou_superconductivity_2021,hagymasi_observation_2022,arp_intervalley_2024,han_large_2024,liu_spontaneous_2024,han_correlated_2024,winterer_ferroelectric_2024,zheng_switchable_2025,yang_impact_2025,xie_tunable_2025,kumar_superconductivity_2025,kumar_superconductivity_2025,holleis_fluctuating_2025,deng_superconductivity_2025,yang_magnetic_2025,qin_stripe_2025, Han2025Chiral, morissette_evidence_2025,li_fractional_2025,qin_stripe_2025,deng_superconductivity_2025,nguyen_hierarchy_2025,seo_family_2025,qin_extreme_2026,han_evidence_2026,kalantre_fermiology_2026,dutta_reconfigurable_2026,butler_13_2026}. One of the most striking observations recently made is the existence of superconductivity emerging from a fully valley-and-spin-polarized quarter metal \cite{dutta_reconfigurable_2026}. This superconducting phase is particularly remarkable because the full polarization of the parent metallic state implies that Cooper pairs cannot be formed with states related by time-reversal symmetry. When this is combined with the existence of a finite trigonal warping, pairs formed near the Fermi surface do not possess a logarithmically divergent pairing susceptibility as would occur in a conventional metal. This exotic form of superconductivity has inspired a number of theoretical works, which discuss both the overall phase diagram of rhombohedral graphene and specific simplified models for superconductivity \cite{ghazaryan_unconventional_2021, muten_exchange_2021, cea_superconductivity_2022, chou_acoustic-phonon-mediated_2022, chatterjee_inter-valley_2022, park_topological_2023, qin_functional_2023, davydova_nonreciprocal_2024, dong_theory_2024, koh_correlated_2024, kumar_unconventional_2024, dong_stability_2024, wang_chiral_2025, shi_doping_2025, huang_fractional_2025, Geier2026Chiral, wolf_magnetism_2024, Qin2026Chiral, parra-martinez_band_2025, han_exact_2025, lee_self-consistent_2025, li_berry_2025, maymann2025pairing, divic_anyon_2025,nosov_anyon_2025, Shi_2025, pichler_microscopic_2025, shavit_quantum_2025, yang_topological_2025,sedov_quantum_2025}.

\begin{figure}
    \centering
    \includegraphics[width=1\linewidth]{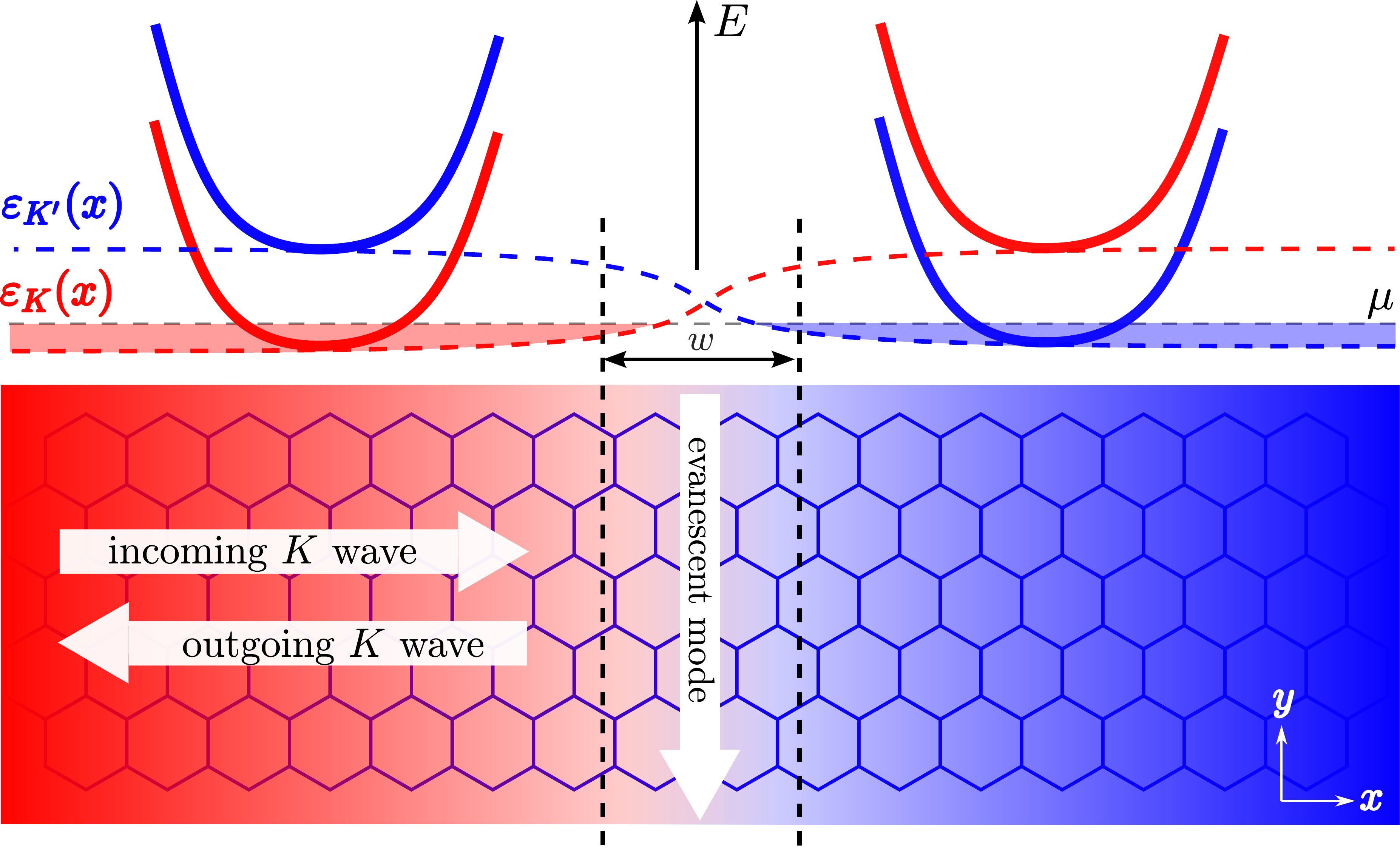}
    \caption{\textbf{Schematic illustration of a valley domain wall along the armchair direction.} Energy bands on the two sides of a domain wall illustrating the spatial variation of valley polarization $\varepsilon_\tau(x)$ along the $x$ direction, with $\mu$ being the chemical potential. The valley flavor is exchanged across the domain wall: $K$ for $x<0$ and $K'$ for $x > 0.$ For a step-function wall, an incoming $K$ wave is reflected off of an evanescent $K$ mode, which carries current along the domain wall, into an outgoing $K$ wave. Perfect reflection also occurs when the domain wall has finite width $w$ as along as there is no intervalley coherence in the wall.}
    \label{fig:fig1}
\end{figure}

Recently, experiments have further shown that this chiral superconducting phase exhibits hysteretic behavior when trained  by an applied  magnetic field \cite{yang_magnetic_2025,dutta_reconfigurable_2026}, suggesting the existence of some form of real-space domain structure \cite{dutta_reconfigurable_2026}. Motivated by these observations, we develop in this work a theory of the  transport properties by considering a domain wall that separates domains of valley-contrasting quarter metals.  The domain wall is assumed to be straight, which we expect  describes reasonably well a region of a realistic wall of length $l$ such that $k_F^{-1} \lesssim l \lesssim R$, where $k_F$ is the Fermi wavevector and $R$ is the local radius of the wall. Contrary to the naive expectation that aligning the momenta of valley contrasting states allows transmission across a valley domain wall, we find that an armchair domain wall of any width $w \lesssim l$ is completely opaque to electron transmission in the metallic phase. In order to mediate transport across such a domain wall, intervalley coupling near the domain wall is necessary. To this end, we classify symmetry-allowed intervalley-mixing terms and derive their tight-binding representations in possible $\sqrt{3} \times \sqrt{3}$ reconstructions of the graphene lattice. We then show numerically that these valley-coupled states indeed enable robust transmission across the domain wall.

In the superconducting phase, symmetry dictates that the pairing order parameter has odd angular momentum due to its full valley-spin polarization. We choose a $p_+$ order parameter in valley $K$ and a $p_-$ order parameter in valley $K'$. These order parameters have been justified by phenomenological arguments \cite{davydova_nonreciprocal_2024,Qin2026Chiral,Yoon2026Quarter} and microscopic calculations \cite{parra-martinez_band_2025}. We compute the DC supercurrent across an SNS' Josephson junction that joins a $p_+$ superconductor on one side to a $p_-$ superconductor on the other side through a metallic region. We again find that intervalley mixing in the metallic region is needed to sustain an appreciable supercurrent across a finite-width junction. In both phases, our results illustrate the crucial role of intervalley interactions: they act as a \textit{valve} to control transport across domain walls that separate valley-contrasting regions.

\textcolor{Blue}{\textit{Opacity of an Abrupt Domain Wall}}: We first show that an abrupt domain wall that separates two regions of opposite valley polarizations is \textit{perfectly} opaque. This simplified problem allows for an analytic solution that illustrates the main features of the physics. For a domain wall that runs along a zigzag direction, this conclusion follows immediately from momentum conservation since  the two valleys reside in two momentum-separated energy minima, rendering the domain wall completely opaque. For a domain wall that runs along or near the armchair direction, the situation is less apparent. The projected energy spectrum in this case maps both valleys near the same crystal momenta (identical for the armchair structure) and thus, mixing of the valleys is kinematically allowed. Contrary to this naive expectation, we now show analytically that for an abrupt wall, no such transmission is possible. This is similar to the blocking of transmission between Weyl semimetals with reversed velocities \cite{Takahashi2011Gapless}. Because the zigzag and armchair directions represent the two distinct limiting termination geometries in graphene, analytic continuity between both null results implies that, in general, ballistic transmission across an abrupt domain wall separating the two different valley-polarized states does not occur.

For illustration, we consider first a monolayer model with the following spin-polarized low-energy effective Hamiltonian
\begin{equation}
\begin{split}
    \hat{\mathcal{H}}_{\tau_z}(x,k_y) &= \hbar v_0 \left(-i\partial_x,k_y \right) \cdot \left( \tau_z \sigma_x,\sigma_y \right) \\
    &+  \left
    (\varepsilon_K(x)\tau_+ + \varepsilon_{K'}(x)\tau_- \right)\sigma_z,    
\end{split}
\end{equation}
where $\sigma$ and $\tau$ matrices act on sublattice and valley spaces, respectively, and $\tau_\pm = \left(\tau_0\pm \tau_z\right)/2.$ For now, we set $\hbar v_0 = 1.$ The valley-dependent sublattice asymmetry is taken to be  
\begin{equation}
    \varepsilon_K(x) = \left\{
\begin{array}{ll}
      \varepsilon_- & x\leq 0 \\
      \varepsilon_+ & x > 0
\end{array} 
\right. \quad \text{and} \quad \varepsilon_{K'}(x) = \left\{
\begin{array}{ll}
      \varepsilon_+ & x\leq 0 \\
      \varepsilon_- & x > 0
\end{array} 
\right. ,
\end{equation}
where $0 < \varepsilon_- \leq \varepsilon_+,$ as shown in Fig. \ref{fig:fig1}. That is, on the left-hand side, we have propagating states from the $K$ valley with energies $E = \sqrt{\varepsilon_-^2+k_x^2+k_y^2}$ {\it and} evanescent states from the $K'$ valley with energies $E = \sqrt{\varepsilon_+^2-\kappa_x^2+k_y^2}.$ The situation is reversed on the right-hand side. The scattering state for an incoming $K$ wave on the left has the following contribution to the wavefunction from the $K'$ valley:
\begin{equation}
\begin{split}
    \begin{pmatrix}
        \psi_{A,K'}  \\
        \psi_{B,K'}  \\
    \end{pmatrix}_{x<0} &=  \mathfrak{r}_{K'}e^{+\kappa_xx} \begin{pmatrix}
         \varepsilon_+ + E  \\
         i (\kappa_x+k_y) 
    \end{pmatrix}  ,   \\
    \begin{pmatrix}
        \psi_{A,K'}  \\
        \psi_{B,K'} \\
    \end{pmatrix}_{x>0} &= \mathfrak{t}_{K'}e^{+ik_xx} \begin{pmatrix}
         \varepsilon_- + E  \\
          -k_x+ik_y
    \end{pmatrix}  ,
\end{split}
\end{equation}
where $\mathfrak{r}$ and $\mathfrak{t}$ are reflection and transmission amplitudes, respectively. To enforce current conservation, we demand that the wavefunction be continuous at $x = 0.$ From this, it is obvious that $\mathfrak{r}_{K'} = \mathfrak{t}_{K'} = 0$ as long as $\varepsilon_+ \neq \varepsilon_-.$ Consequently, the $K'$ valley is a spectator in the dynamics. This immediately implies that  the incoming $K$ wave must be completely reflected into an outgoing $K$ wave because there are \textit{no} propagating modes on the other side through which electrons can be transmitted. Indeed, we have \cite{Takahashi2011Gapless}
 \begin{equation}
\begin{split}
    \begin{pmatrix}
        \psi_{A,K} \\
        \psi_{B,K} \\
    \end{pmatrix}_{x<0} &=  e^{ik_xx} \begin{pmatrix}
         \varepsilon_- + E \\
         k_x+ik_y
    \end{pmatrix} + \mathfrak{r}_K e^{-ik_xx} \begin{pmatrix}
         \varepsilon_- + E \\
         ik_y-k_x
    \end{pmatrix}  ,   \\
    \begin{pmatrix}
        \psi_{A,K} \\
        \psi_{B,K}
    \end{pmatrix}_{x>0}  &= \mathfrak{t}_K e^{-\kappa_xx} \begin{pmatrix}
         \varepsilon_+ + E \\
         i (\kappa_x+k_y) 
    \end{pmatrix}  ,
\end{split}
\end{equation}
where the coefficients are given by
\begin{equation}
    \begin{split}
        \mathfrak{r}_{K} &= \frac{2 (\varepsilon_+ + E) k_x}{
    E (i \kappa_x + k_x) + \varepsilon_+ (k_x - i k_y) + i \varepsilon_- (\kappa_x + k_y)}-1, \\
  \mathfrak{t}_{K} &= -\frac{2 i (\varepsilon_- + E) k_x}{
    E (\kappa_x - i k_x) -  \varepsilon_+ (ik_x +  k_y) + \varepsilon_- (\kappa_x + k_y)}.
    \end{split}
\end{equation}
We confirm that $|\mathfrak{r}_K|^2=1$ when the energy is within the quarter-metal phase, demonstrating perfect reflection to the same valley. Fascinatingly, the evanescent modes from the $K$ valley {\it do} carry current parallel to the domain wall, even for $k_y = 0,$ given by $J_y \propto (\varepsilon_++E) \mathfrak{t}_K^\dagger \mathfrak{t}_K,$ reminiscent of a current carried by boundary states at the edge of a topological insulator.

The preceding result can be generalized to $N$-layer rhombohedral graphene using a minimal two-band model for the Hamiltonian that incorporates the $k^N$ dispersion (without including trigonal warping):
\begin{equation}
\begin{split}
    \hat{\mathcal{H}}_{\tau_z}(x,k_y) &= \begin{pmatrix}
        0 & (-i\tau_z \partial_x-ik_y)^N \\
        (-i\tau_z \partial_x+ik_y)^N & 0
    \end{pmatrix} \\
    &+  \left
    (\varepsilon_K(x)\tau_+ + \varepsilon_{K'}(x)\tau_- \right)\sigma_z.
\end{split}
\end{equation}
In this case, there are many more evanescent modes at a given energy. If we write $E - \varepsilon_\pm = \mathcal{E}e^{i\varrho},$ where $\varrho \in \lbrace0,\pi\rbrace$ depending on the sign of $E - \varepsilon_\pm,$ then the wavevectors are
\begin{equation}
    k_{\pm n} = \pm \sqrt{\mathcal{E}^{1/N} \exp \left( \frac{i\varrho}{N} + \frac{2\pi ni}{N} \right) - k_y^2},
\end{equation}
where $n = 0,1,...,N-1$. For $\sqrt{\varepsilon_-^2+k_y^{2N}} < E < \sqrt{\varepsilon_+^2+k_y^{2N}},$ we have two propagating modes (one incoming and one outgoing) and $N-1$ normalizable evanescent modes on the left-hand side and $N$ normalizable evanescent modes on the right-hand side, all residing in the $K$ valley. The existence of evanescent modes on \textit{both} sides of the domain wall differs from the monolayer situation where there is only one evanescent mode on the opposite side of the propagating region. This situation also differs from the one in topological insulators where boundary states only exist in a spectral gap; these evanescent states reside deep in the bulk spectrum of the propagating domain. For the $K'$ valley, the situation in reversed with the propagating modes residing on the right-hand side. To ensure continuity of current, we demand continuity of the wavefunction and its first $N-1$ derivatives at the boundary. The explicit calculations are shown in Ref. \cite{SI}. Essentially, we can again match the boundary condition for an incoming $K$ wave using wavefunctions entirely in the $K$ valley in the absence of intervalley mixing. Therefore, in the $N$-layer system, we still find that an abrupt domain wall is perfectly opaque.

\begin{table}[h]
\centering
\footnotesize
\renewcommand{\arraystretch}{1.25}
\begin{tabular*}{\columnwidth}{@{\extracolsep{\fill}} ll l}
\hline\hline
 & $\mathcal{T}$ even & $\mathcal{T}$ odd \\
\hline
\multicolumn{3}{l}{$N \bmod 3 = 0$, effective lattice is triangular} \\[-2pt]
Intravalley
& $\sigma_z,\ \sigma_x,\ \tau_z \sigma_y$
& $\sigma_y,\ \tau_z,\ \tau_z \sigma_z,\ \tau_z  \sigma_x$ \\
Intervalley
& $\tau_x,\ \tau_y,\ \tau_{x}  \sigma_{x},\ \tau_{x}  \sigma_{z},\ \tau_{y}  \sigma_{x},\ \tau_{y}  \sigma_{z} $
& $\tau_{x}  \sigma_y, \ \tau_{y}  \sigma_y$ \\[4pt]
\multicolumn{3}{l}{$N \bmod 3 = \pm1$, effective lattice is honeycomb} \\[-2pt]
Intravalley
& $\sigma_z$
& $\tau_z,\ \tau_z  \sigma_z$ \\
Intervalley
& $\tau_{x}  \sigma_+,\ \tau_{y} \sigma_+$
& none \\
\hline\hline
\end{tabular*}
\caption{
\textbf{Symmetry classification of valley--orbital operators for rhombohedral $N$‑layer graphene at the $\overline{\Gamma}$ point in the reconstructed Kekulé lattice.}
$\mathcal{T}$ denotes spinless time‑reversal symmetry. $\tau$ acts on valley while $\sigma$ acts on $A^{(1)}$-$B^{(N)}$ sublattice degrees of freedom. $\sigma_\pm=(\sigma_0\pm\sigma_z)/2.$
}
\label{tab:valley_orbital_operators - main}
\end{table}

\begin{figure}
    \centering
    \includegraphics[width=1\linewidth]{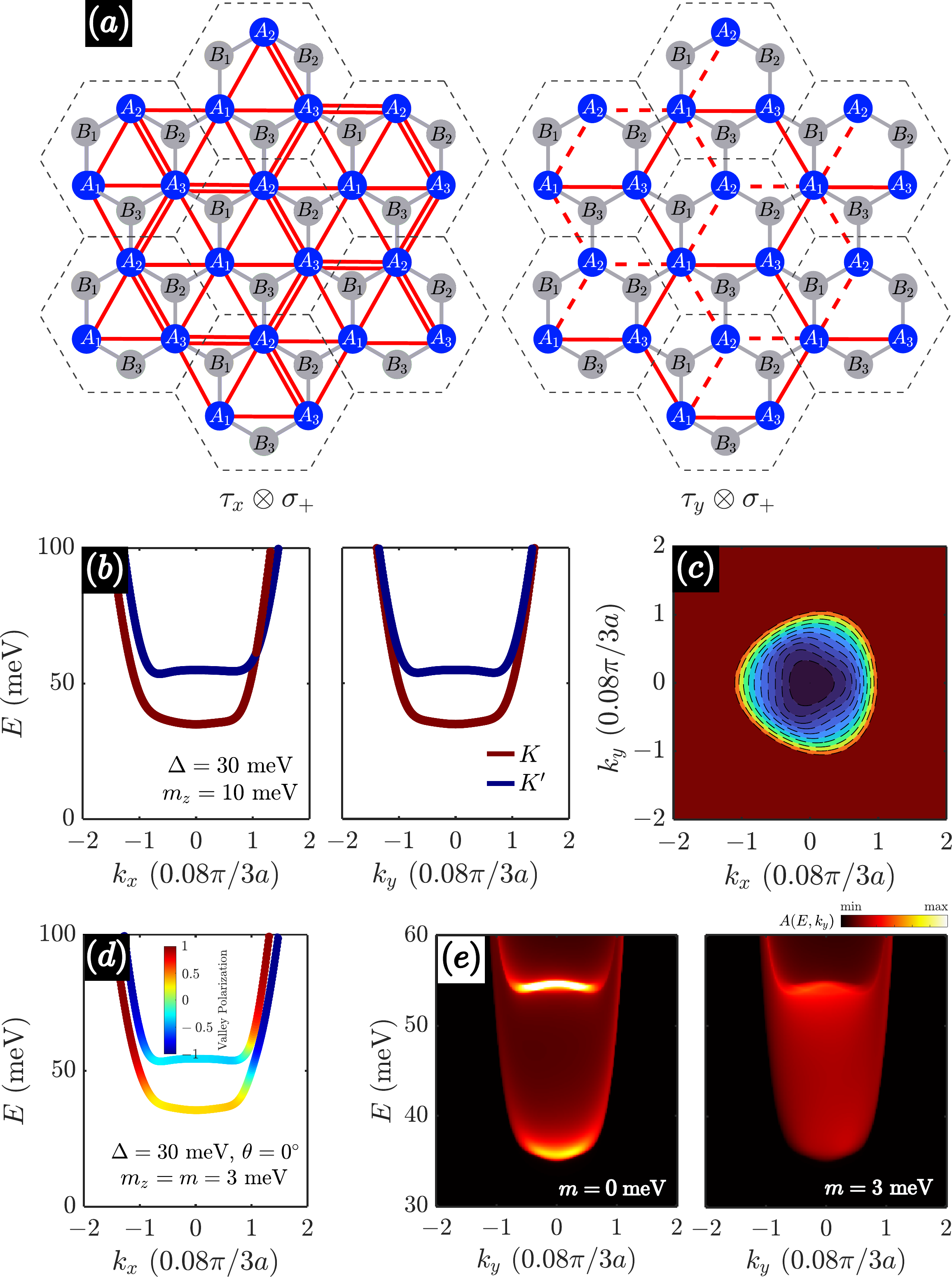}
    \caption{\textbf{Intervalley scattering on the active sublattice.} (a) Schematic hopping model realizing the intervalley mixing interactions $\tau_x\sigma_+$ and $\tau_y\sigma_+$ with bond strengths distinguished by single, double, and dashed red lines connecting atoms. (b) Bulk band structure of tetralayer rhombohedral graphene along the $k_x$ (left) and $k_y$ (right) directions with $\Delta = 30$ meV and $m_z = 10$ meV. The color represents valley polarization. (c) Energy map for the conduction band ($K$ valley) in (b) showing the isoenergy contours at carrier densities $n = 0,0.1,0.2,...,1.0\times10^{12}$ cm$^{-2}$. (d) Band structure with intervalley coupling $\tau_x \sigma_+$ for $m_z = m=3$ meV. (e) Spectral function $A(E,k_y)$ of a domain wall without (left) and with (right) intervalley coupling $\tau_x\sigma_+$. Here, unless stated otherwise, $m_z = 10$ meV, and the domain wall is $10$ nm wide.} 
    \label{fig:fig2}
\end{figure}

\textcolor{Blue}{\textit{Transmission through a Finite-Width Domain Wall}}: We now validate the above results with a microscopic model of $N$-layer rhombohedral graphene using the numerical Green's function approach to scattering. We use a standard tight-binding Hamiltonian for $N$-layer rhombohedral graphene that includes 5 hopping parameters between carbon atoms: $\lbrace \gamma_0,\gamma_1,\gamma_2,\gamma_3, \gamma_4 \rbrace$. A displacement field is implemented as a progressive layer-dependent potential energy with energy difference between adjacent layers denoted by $\Delta.$ This field opens a gap at charge neutrality with magnitude approximately $\Delta (N-1).$ By convention, the sign of $\Delta$ is chosen so that electrons in the conduction band are localized on the $A$ sublattice of layer 1. Valley polarization is implemented by a Haldane-like next-nearest neighbor hopping only on the $A^{(1)}$ sublattice \cite{haldane_model_1988}\footnote{There are other ways to implement valley imbalance, as shown in Ref. \cite{SI}.}
\begin{equation}
    \delta\hat{\mathcal{H}}_{\tau_z}(\mathbf{k}) = \frac{2m_z}{3\sqrt{3}} \sum_{j=1}^3 \sin \left[\mathbf{k} \cdot \mathbf{a}_j \right]\sigma_+,
\end{equation}
where $\mathbf{a}_{j+1} = a\left(\cos\left[2\pi j/3 \right],\sin\left[2\pi j/3 \right] \right)$ and $a$ is the lattice constant. To model a domain wall of width $w$ that switches the valley flavor, we let the mass be a function of position $m_z(x<-w/2) = m_z,$ $m_z(x>+w/2) = -m_z,$ and $m_z(|x|\leq w/2) = -2m_zx/w $. To ensure hermiticity, we take $x$ to be the mid-bond position. We use typical parameters for ferromagnetic domain walls, $ w \lesssim 10 \text{ }\mathrm{nm}$ \cite{Catalan2012Domain, Kumar2022Domain}, in our simulations. Throughout this work, we shall assume that the energy scale of valley polarization is much smaller than the displacement-field-induced trivial gap. We focus on $N = 4$ in the main text, with $\Delta = 30$ meV and $m_z = 10$ meV. 

To calculate transmission, we use the Green's function formalism for quantum transport \cite{Datta1995,Ryndyk2016}. The left and right leads are semi-infinite quarter metals with different flavors of valley polarization joined by the domain wall. The entire system is periodic in the $y$ direction; thus, $k_y$ is a good quantum number. The transmission function is calculated from the Caroli-Fisher-Lee formula \cite{caroli1971direct,Fisher1981Relation,Guinea1983Effective,sancho1985highly} $T(E,k_y) = \tr \left[ \Gamma_L \mathcal{G}_D \Gamma_R \mathcal{G}_D^\dagger \right],$ where $\Gamma_{L/R}$ is the level-width functions of the left and right domains computed from their respective self energies, and $\mathcal{G}_D$ is the Green's function of the domain wall, which accounts for the self energies of the leads. We have verified numerically that the transmission is always zero, as long as $E$ resides in a spectral region that populates opposite valley flavors across the domain wall, even if the domain wall has finite width. This is true for waves at normal ($k_y =0$) as well as oblique ($k_y \neq 0$) incidence, demonstrating numerically that a domain wall between two valley-contrasting quarter metals is perfectly opaque, in agreement with the previous analytic calculation for an abrupt wall.

\begin{figure}[h]
    \centering
    \includegraphics[width=1\linewidth]{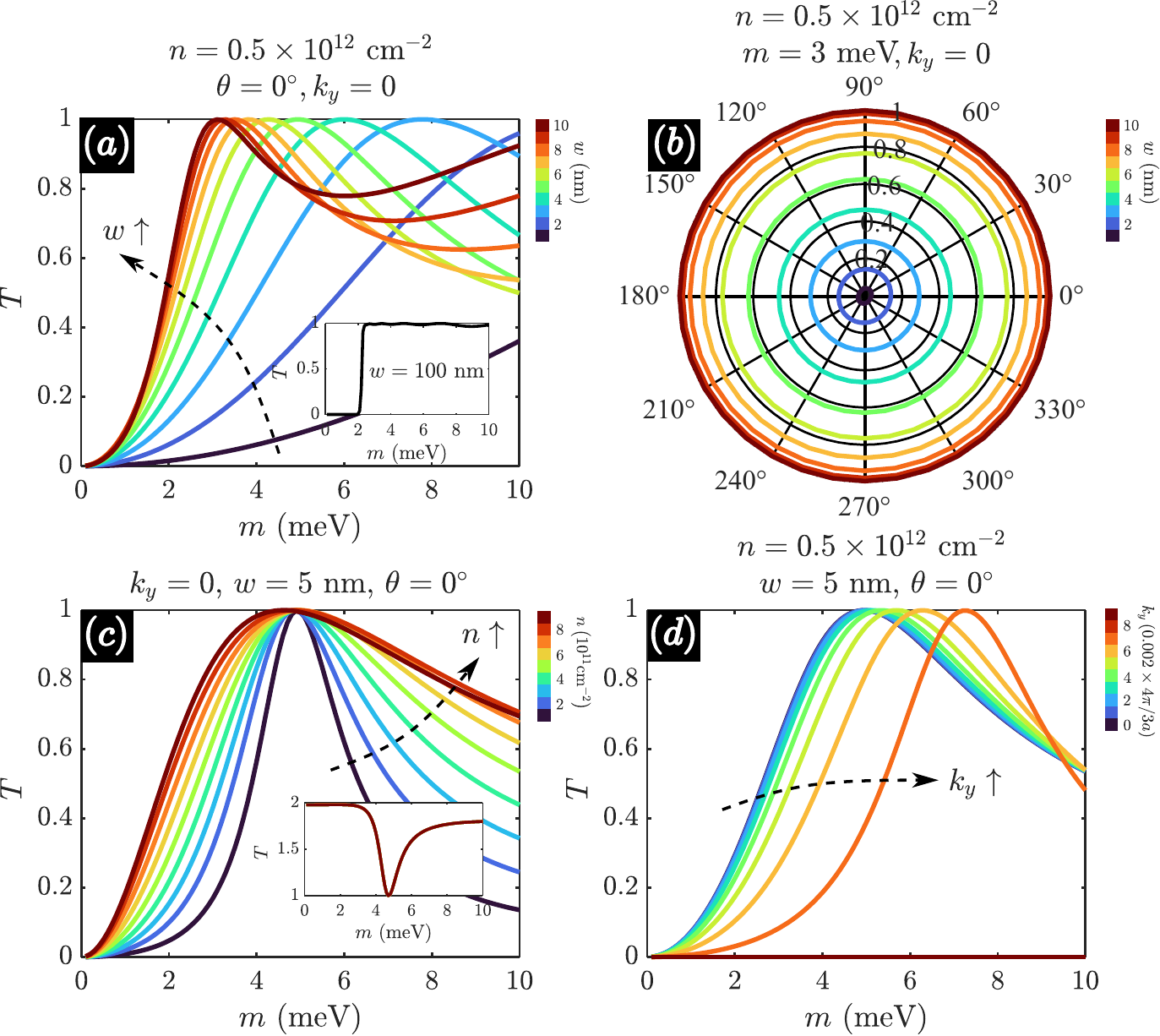}
    \caption{\textbf{Transmission across a domain wall with intervalley coupling.} (a) Transmission as a function of the strength of intervalley coupling $m$ at $\theta = 0^\circ, k_y = 0$, and $n = 5\times10^{11}$ cm$^{-2}$. The inset shows the transmission of a wide domain wall $w =100$ nm. (b) Dependence of the transmission on the angle $\theta$ of the intervalley interaction at $m = 3$ meV, $k_y = 0$, and  $n = 5\times10^{11}$ cm$^{-2}$. Different colored curves in (a,b) correspond to different domain wall widths $w.$ (c) Transmission at different filling densities in the quarter-metal phase for $k_y = 0,$ $w = 5$ nm, and $\theta = 0^\circ$. The inset shows transmission at a density \textit{outside} the quarter-metal phase where both valleys are occupied on both sides of the domain wall. (d) Transmission at oblique incidence, $k_y \neq 0,$ for $\theta = 0^\circ, w = 5$ nm, and $n = 5\times10^{11}$ cm$^{-2}$.  }
    \label{fig:fig3}
\end{figure}

\textcolor{Blue}{\textit{Intervalley-Coupled Domain Walls}}: The preceding considerations have established  that electron transmission across an intrinsic domain wall is forbidden. Thus, there has to be some kind of intervalley mixing or \emph{coherence} at the domain wall in order for electrons to cross from one side of it to the other. Microscopically, this mixing may arise from atomic-like impurities or disorder, the non-equilibrium effects of injected currents \cite{dutta_reconfigurable_2026}, or from the same interactions that cause symmetry breaking in the bulk, now inside a domain wall. Here, we model its effect phenomenologically, using symmetry-based arguments. In order to hybridize the two valleys, we consider Kekul\'{e}-type perturbations that reconstruct the honeycomb unit cell into a $\sqrt3 \times \sqrt{3}$ supercell. In the reconstructed Brillouin zone, both $K$ and $K'$ are mapped to $\bar{\Gamma},$ the zone center. We seek perturbations that locally respect $C_{3z}$ symmetry on one of the three $A$ sites on layer 1 in the Kekul\'{e} basis. For $N\mod3=0,$ the effective lattice constructed from the $A^{(1)}$ and $B^{(N)}$ sublattices is triangular, while for $N\mod3 =\pm1,$ the effective lattice is honeycomb. All the symmetry-allowed terms that do not vanish at $\bar{\Gamma}$, including those that are intravalley, are listed in Table \ref{tab:valley_orbital_operators - main}. Of particular interest to us are the intervalley terms $\tau_x\sigma_+$ and $\tau_y\sigma_+$ that are generic because they are symmetry allowed for all values of $N$ \footnote{If the rotation center is chosen on the $B$ sublattice, the allowed terms would involve $\sigma_-$}.  These terms only act on the $A^{(1)}$ sublattice. We use the following momentum-space regularization to implement these valley-mixing terms on the Kekul\'{e} lattice:
\begin{equation}
\begin{split}
 \delta \hat{\mathcal{H}}_{\mathrm{intervalley}}(\mathbf{k}) &= m \left[h_x(\mathbf{k}) \cos \theta  + h_y(\mathbf{k}) \sin \theta\right],   \\
 h_x(\mathbf{k}) &= \frac{1}{2}\begin{pmatrix}
     0 & - g^{\dagger}(\mathbf{k}) & -  g(\mathbf{k}) \\
     -  g(\mathbf{k}) & 0 & 2g^\dagger(\mathbf{k}) \\
     -  g^\dagger(\mathbf{k}) & 2g(\mathbf{k}) & 0
 \end{pmatrix}, \\
 h_y(\mathbf{k}) &= \frac{\sqrt{3}}{2}\begin{pmatrix}
     0 &  g^{\dagger}(\mathbf{k}) & - \ g(\mathbf{k}) \\
      g(\mathbf{k}) & 0 & 0 \\
     -  g^\dagger(\mathbf{k}) & 0 & 0
 \end{pmatrix},
\end{split}
\end{equation}
where $g(\mathbf{k}) =  \sum_{i=1}^3 e^{i \mathbf{k} \cdot \mathbf{a}_i},$ $m$ is the magnitude of the perturbation, and $\theta$ rotates in $(\tau_x,\tau_y)$ space. The real-space representations of these  valley-mixing terms are shown in Fig. \ref{fig:fig2}(a).   Note that we introduce the intervalley coupling \textit{only} inside the domain wall. 

The band structures of the bulk valley-polarized quarter metals are shown in Fig. \ref{fig:fig2}(b,c). We focus on the regime where the Fermi surface has a simply-connected, trigonally-warped topology, as shown in Fig. \ref{fig:fig2}(c). Intervalley coupling hybridizes the bands from different valleys, as shown in Fig. \ref{fig:fig2}(d). In Fig. \ref{fig:fig2}(e), we simulate a 10 nm domain wall without and with a $\tau_x\sigma_+$ intervalley coupling. The density plots show the spectral function $A(\omega = E+i0^{+},k_y) = -\textrm{Im} \Tr \mathcal{G}_D(\omega,k_y)/\pi$ summed over the entire domain wall width. Without $\tau_x\sigma_+$, we observe evanescent states which are highly localized inside the domain wall near the band edges. With $\tau_x \sigma_+,$ these evanescent states appear to penetrate deeper into the left and right domains, resulting in a more delocalized spectral function in energy space.

Next, we  calculate the transmission through a domain wall with valley mixing. We first focus on the case of normal incidence where $k_y = 0$ and fixed chemical potential. For a $\tau_x\sigma_+$ interaction of varying strengths and domain wall widths, the results are shown in Fig. \ref{fig:fig3}(a). For a moderate width $w \lesssim10$ nm, the transmission monotically increases nonlinearly as a function of increasing interaction strength $m.$ For the smallest widths, this nonlinear behavior is exceptionally well captured by a quadratic dependence on $m.$ At some critical $m^*,$ which is width dependent, the transmission reaches unity, attributed to a resonant condition inside the domain wall for optimal transmission. For $m > m^*,$ the transmission oscillates. For large widths (e.g $w = 100$ nm), the transmission displays an abrupt switching behavior. It remains practically zero below a threshold value of $m$, quickly jumping to near unity above it. The latter regime can be understood as an adiabatic evolution since the valley-polarization changes slowly over a large-width domain wall.

At a fixed $m = 3$ meV, still at normal incidence, we scan $\theta$ for various widths as shown in Fig. \ref{fig:fig3}(b). The transmission only shows a weak dependence on $\theta.$ In Fig. \ref{fig:fig3}(c), we sweep the chemical potential at a fixed domain wall width, always staying within the regime where both the left and right domains are quarter metals. The monotonic behavior at small $m$ remains the same for all densities. In the inset of Fig. \ref{fig:fig3}(c), we show the transmission for a large chemical potential where the domains host both valleys; we observe  that the transmission can saturate at $2$ instead of $1$ because of the presence of scattering modes from both valleys. Interestingly, intervalley coupling in this case introduces scattering, which can lower the transmission, essentially having the opposite effect to that in the quarter-metal phase. Lastly, the transmissions of oblique waves, $k_y \neq 0$, are shown in Fig. \ref{fig:fig3}(d). These results taken together demonstrate the importance of intervalley hybridization inside the domain wall in mediating electron transport from a quarter metal on one side of the domain wall to its dual on the other side.

\begin{figure}
    \centering
    \includegraphics[width=1\linewidth]{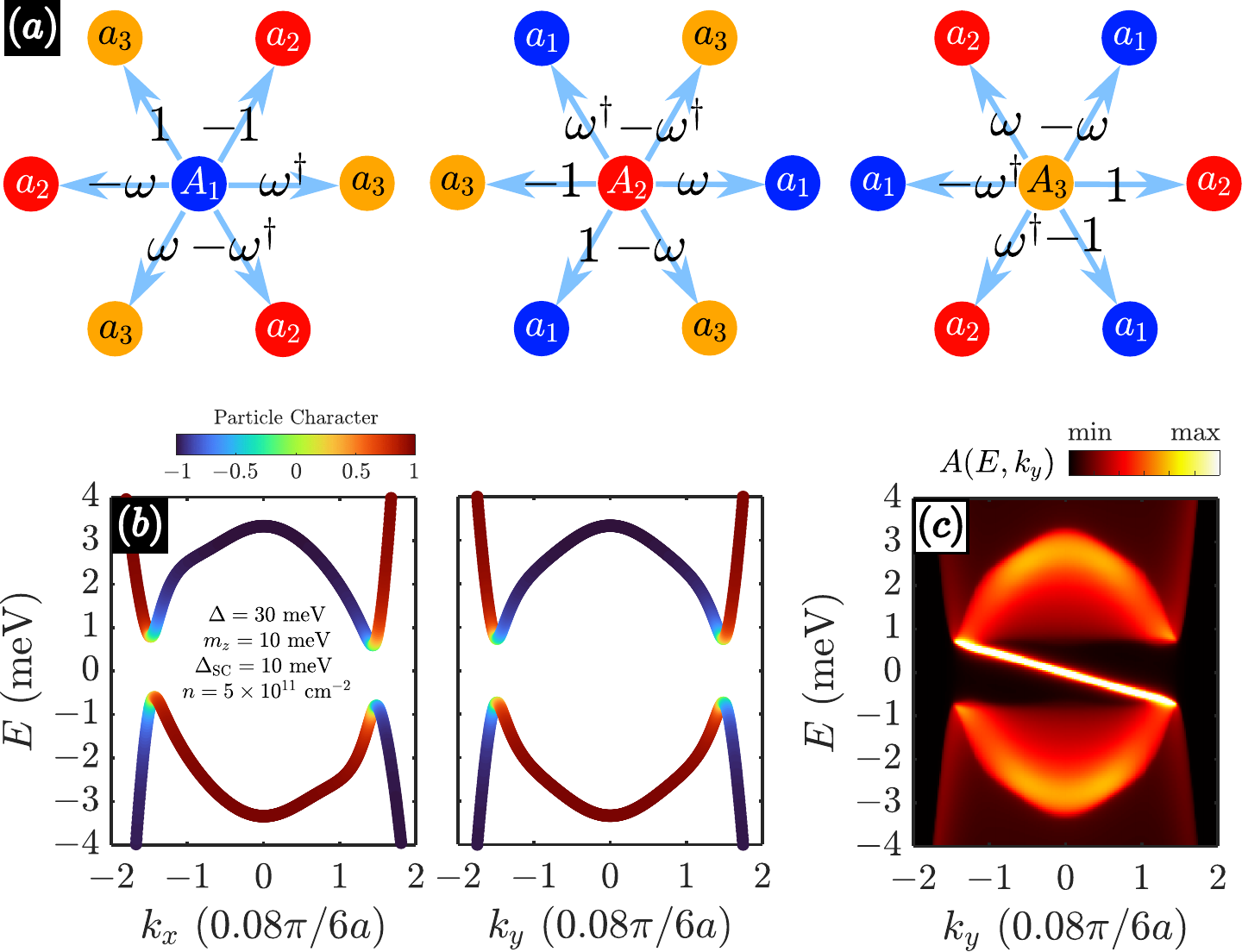}
    \caption{\textbf{Finite-momentum chiral $p$-wave superconductivity.} (a) Real-space representation of the hopping processes between the electron ($A_i$) and hole ($a_i$) sectors in Eq. \eqref{eq: BdG Hamiltonian}. (b) Bulk band structure along the $k_x$ and $k_y$ directions color-coded by particle character ($+1$ for electrons and $-1$ for holes). (c) Spectral function for a semi-infinite plane with an armchair termination showing the presence of a chiral Majorana branch traversing the bulk gap due to nontrivial Chern topology.  }
    \label{fig:fig4}
\end{figure}
\textcolor{Blue}{\textit{Models of Chiral Superconductors}}: Next, we address some unique aspects of transport in the exotic superconducting state. Because pairing occurs from the quarter-metal normal state condensed in one of the valleys, the pair momentum must necessarily be non-zero, $2\mathbf{K}_\pm \neq 0$, leading to a pair density wave on the scale of the reconstructed Kekul\'{e} lattice. Prior theoretical works have converged on $p_\pm$ wave as the likely candidate for the pair symmetry \cite{Yoon2026Quarter,Chou2025Intravalley,chen_intrinsic_2025,parra-martinez_band_2025,Qin2026Chiral,Geier2026Chiral,majorana_crystal_rhombohedral_2026}; it has also been suggested by studying analytic models that the pairing symmetry is valley-locked \cite{maymann2025pairing,Tavakol2026}. In this work, we do not solve for the pairing field self-consistently. Instead, we postulate its form based on symmetry. In the $K$ valley with $p_+$ pairing, the Bogoliubov-de Gennes (BdG) Hamiltonian takes the following functional form:
\begin{equation}
\label{eq: BdG Hamiltonian}
\begin{split}
    \hat{\mathcal{H}}_\mathrm{BdG}(\mathbf{k}) &= \begin{pmatrix}
        \hat{\mathcal{H}}(\mathbf{k}) -\mu & \hat{\Delta}_\mathrm{SC}(\mathbf{k}) \\
        \hat{\Delta}^\dagger_\mathrm{SC}(\mathbf{k}) & \mu-\hat{\mathcal{H}}^*(-\mathbf{k}) 
    \end{pmatrix},    \\
    \hat{\Delta}_\mathrm{SC}(\mathbf{k}) &= \begin{pmatrix}
        0 & -\omega f(-\mathbf{k}) & \omega^\dagger f(\mathbf{k}) \\
        \omega f(\mathbf{k}) & 0 & - f(-\mathbf{k}) \\
        - \omega^\dagger f(\mathbf{-k}) & f(\mathbf{k}) & 0 
    \end{pmatrix},
\end{split}
\end{equation}
where $f(\mathbf{k}) = \frac{\Delta_\mathrm{SC}}{2i} \left[e^{i\mathbf{k} \cdot \mathbf{a}_1} + \omega e^{i\mathbf{k} \cdot \mathbf{a}_2} + \omega^\dagger e^{i\mathbf{k} \cdot \mathbf{a}_3} \right]$, $\Delta_\mathrm{SC}$ is the pairing amplitude, $\omega = e^{2\pi i/3}$, and $\mu$ is the chemical potential. The pairing field is only implemented on the active $A^{(1)}$ sublattice, as shown in Fig. \ref{fig:fig4}(a). It satisfies antisymmetry $\hat{\Delta}_{\mathrm{SC},\sigma\sigma'}(\mathbf{k}) = -\hat{\Delta}_{\mathrm{SC},\sigma'\sigma}(-\mathbf{k})$ and, once projected onto the $K$ valley, takes the form $ \hat{\Delta}_\mathrm{SC}(\mathbf{k}) = \Delta_\mathrm{SC} \left[ \sin(\mathbf{k}\cdot \mathbf{a}_1)+  \omega\sin(\mathbf{k}\cdot \mathbf{a}_2) +  \omega^\dagger\sin(\mathbf{k}\cdot \mathbf{a}_3)\right] \approx 3\Delta_\mathrm{SC}a(k_x+ik_y)/2,$ showing that it has the correct phase winding for a $p_+$ order parameter. We choose $\Delta_\mathrm{SC}$ so that the gap opening is about $1$ meV since the best critical temperature measured in experiments is only about 300 mK \cite{Han2025Chiral}. A $p_\pm$-wave superconductor generically has unit Chern number if it is fully gapped. Using numerical integration of the non-Abelian Berry curvature, we have confirmed that this is indeed the case for our models. Representative bulk and armchair-edge-projected band structures are shown in Figs. \ref{fig:fig4}(b,c), respectively. A single Majorana branch traverses the gap at the boundary of the semi-infinite plane due to the unit Chern number of the superconducting state.


\textcolor{Blue}{\textit{Supercurrent through a DC Josephson Junction}}: Using the superconductivity model above, we assess the effect of intervalley coupling on the DC supercurrent across a Josephson junction. The supercurrent is computed using the standard formula derived from the phase dependence of the free energy assuming an abrupt change in the phase profile across the junction (i.e. there is no gradual spatial phase variation). It is given by
\begin{equation}
\begin{split}
    I(\phi,k_y) &= \frac{e}{\hbar}  \int_\mathbb{R} \frac{d\omega}{2\pi} n_F(\omega) \mathrm{Re}\Tr \left[ \mathcal{G}(\omega,\phi,k_y ) \left[\gamma_z,\Sigma_R(\omega,\phi,k_y) \right]\right] , \\
    J(\phi) &= \int_{\mathrm{BZ}} \frac{dk_y}{2\pi} I(\phi,k_y),
\end{split}
\end{equation}
where $\phi$ is the phase difference across the junction, $\gamma_z$ is a Pauli matrix that acts on the Nambu basis, $\omega \mapsto \lim_{\eta \rightarrow 0^+} \left(E+i\eta\right),$  $\mathcal{G}$ is the Green's function computed in the BdG basis, $\Sigma_R$ is the self-energy from the right lead, $n_F(\omega)$ is the fermion occupation function, and $T$ is the temperature. In our geometry, the left side of the junction hosts a $p_+$ superconductor in the $K$ valley whose pairing phase is fixed, and the right side of the junction hosts a $p_-$ superconductor in the $K'$ valley whose pairing phase is spatially constant but is tunable. The domain wall at the junction is modeled as a normal-metal region of width $w$, possibly including intervalley mixing. 
In all of our supercurrent calculations, we set the domain wall to be $w = 2$ nm, temperature to be $T = 0$ K, and use the same parameters for the normal state from the previous section. Unless otherwise noted, we set $\eta = 10^{-4}$ meV in our numerical calculations. 

\begin{figure}
    \centering
    \includegraphics[width=1\linewidth]{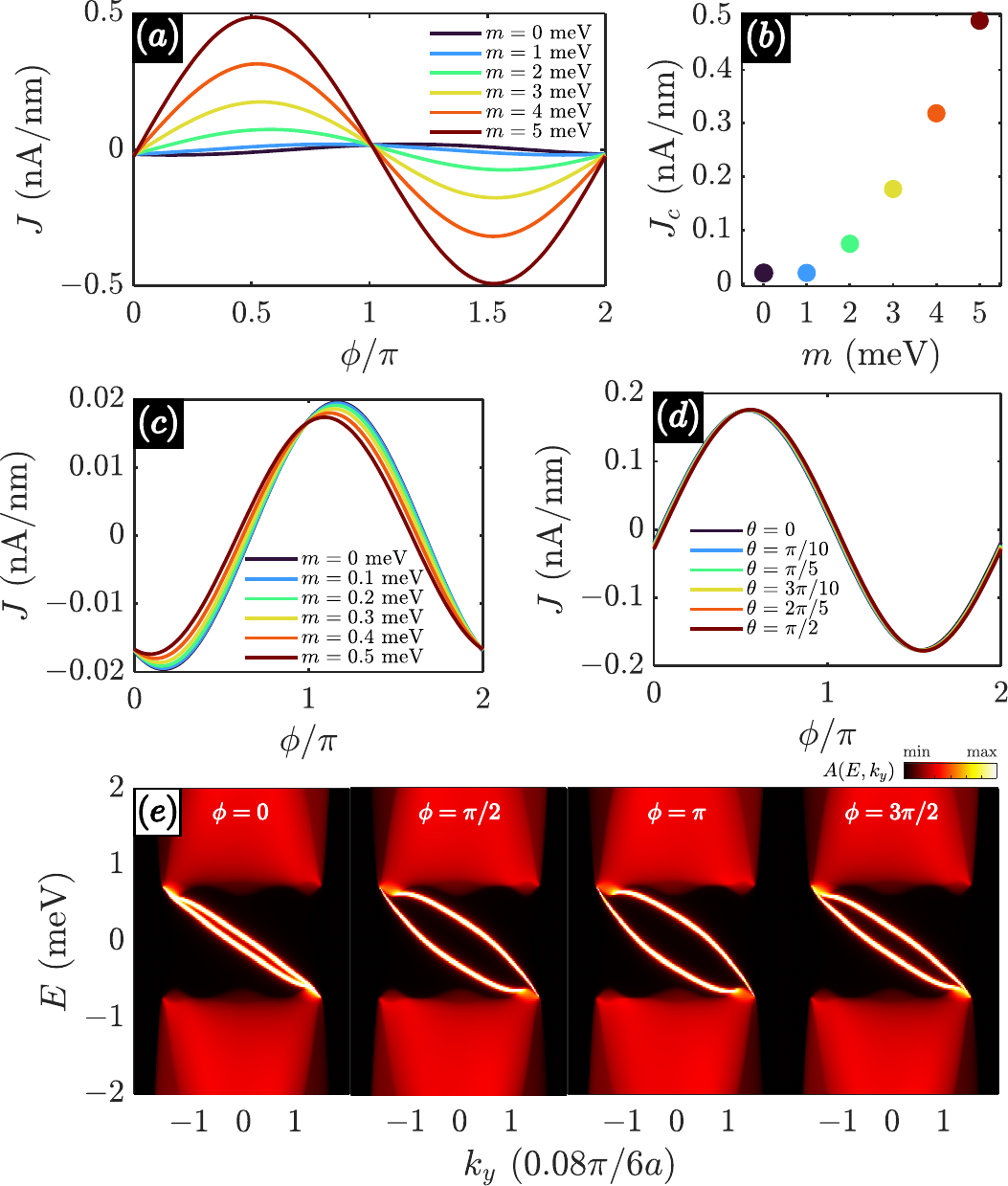}
    \caption{\textbf{Current-phase relation for supercurrent across an SNS' Josephson junction.} (a) CPR for different values of intervalley coherence $m \in [0,1,2,3,4,5]$ meV with $\theta = 0^\circ.$ (b) Plot of the supercurrent density amplitude $J_c$ as a function of $m$ at $\theta = 0^\circ.$ (c) Same as (a) but for a smaller range of $m \in [0,0.1,0.2,0.3,0.4,0.5]$ meV. (d) CPR at fixed $m = 3$ meV for different values of $\theta.$ (e) BdG spectral function for different values of $\phi$ with $m=5$ meV and $\theta = 0$. For all plots, $N = 4,$ $m_z = 10$ meV, $n=5\times10^{11}$ cm$^{-2},$ $w = 2$ nm, $\Delta = 30$ meV, and $\Delta_\mathrm{SC} = 10$ meV.}
    \label{fig:fig5}
\end{figure}

The results for the supercurrent simulations are shown in Fig. \ref{fig:fig5}. Turning on intervalley mixing via $\tau_x$ only (i.e. $\theta = 0^\circ$), we calculate the current-phase relation (CPR) for various magnitudes of the intervalley  coupling, shown in Fig. \ref{fig:fig5}(a). In all cases, the CPR follows the standard DC Josephson oscillation $J(\phi) \approx J_c \sin \left(\phi+\phi_0\right),$ where $J_c$ is the current density amplitude and $\phi_0$ is an angular offset. For $m \approx 0$ meV, the current density is suppressed, on the order of 0.02 nA/nm, as shown in Fig. \ref{fig:fig5}(c). However, it is important to note that even when $m$ identically vanishes, the current density does \textit{not} numerically vanish, even though it is small \footnote{ We have checked that numerically this is true even when $\eta = 10^{-9}$ meV}. This \textit{weakly} violates the Ambegaokar-Baratoff relation as one would otherwise have expected the supercurrent to identically vanish if the normal-state transmission is zero \cite{AmbegaokarBaratoff1963}. As we increase $m,$ thereby enhancing the transmission of the domain wall, the supercurrent density is appreciably amplified, as shown in  Fig. \ref{fig:fig5}(b). This enhancement is in qualitative agreement with the Ambegaokar-Baratoff relation.

Because the change in Chern numbers is two across the junction, $|\Delta \mathcal{C}| = 2$, there are two Majorana branches traversing the BdG gap with a group velocity of equal sign; they are localized at the NS boundaries of the domain wall. The oscillation of the CPR relation tracks the oscillation of these two Majorana branches, as shown in Fig. \ref{fig:fig5}(e). This suggests that the the Andreev states of the BdG spectrum make the dominant contribution to the supercurrent density. Furthermore, most of the contributions to the supercurrent come from states within a narrow window of $k_y$ around the $\bar{\Gamma}$ point.

Finally, we assess the CPR as a function of $\theta,$ the angle that rotates between $\tau_x$ and $\tau_y,$ for a fixed $m = 3$ meV. As shown in Fig. \ref{fig:fig5}(d), all of these curves demonstrate similar CPR behaviors; they are barely discernible from each other on the scale we plot. This is further confirmation that the supercurrent is directly influenced by the transmission transparency of the domain wall in the normal state, and since that transparency does not depend significantly on $\theta,$ the supercurrent does not either. In addition to results the reported here, which are computed using $w = 2$ nm, we have also checked that when the domain width is made much wider, the supercurrent density in the absence of intervalley mixing diminishes significantly, which is consistent with the experimental evidence from Ref. \cite{dutta_reconfigurable_2026}. The dependence of the supercurrent on domain width will be reported in the future \cite{forthcoming_publication}. Taken together, these results demonstrate the importance of intervalley coherence in transporting supercurrent across a Josephson junction that connects two valley-contrasting, opposite-chirality $p$-wave superconductors.

\textcolor{Blue}{\textit{Discussion and Conclusion}}: In this work, we have considered electron transport through domain walls in rhombohedral multilayer graphene at moderate perpendicular displacement fields, which can host quarter-metal and chiral superconducting phases due to strong electronic interactions. Here, we focus solely on the valley degree of freedom by assuming that spin is uniformly polarized throughout a two-dimensional sample, even across domains that alternate valleys. Spin flip transmission across a domain wall would require some additional mechanism to precess the spin, the exploration of which is an interesting subject for future studies. On the other hand, conservation of the valley pseudospin is not an \textit{exact} symmetry; consequently, one might naively expect that special domain wall orientations, such as armchair walls, would naturally mix them. It is therefore striking that in all cases valley domain walls are predicted to be transport blockades: they are  impenetrable boundaries in the absence of some form of intervalley mixing.  Recent experiments \cite{dutta_reconfigurable_2026} find  transport signatures between valley polarized domains in the normal state, strongly suggesting the presence of some form of intervalley mixing. It is possible that such mixing in the experiment is mediated by a dilute density of impurity scatterers near or at the ends of the domain walls  whose aggregate effect is qualitatively captured by our theory. Intriguingly, the metallic and superconducting phases are found to be hysteretic with a {\textit{different}} dependence on training by a weak magnetic field. These observations provide an important window for valley-valve behavior that must be mediated by intervalley coupling with a fundamentally different structure in the normal and superconducting states. The resolution of the microscopic origin of this behavior now poses an important problem for exposing the physics of these novel states.

We thank Cyprian Lewandowski, Thomas Scaffidi, Sandeep Joy, Eli Zeldov, Mathias Scheurer, and Andrea Young for fruitful conversations. V.T.P was supported in part by  Cyprian Lewandowski's start-up funds from Florida State University and the National High Magnetic Field Laboratory and in part by Cyprian Lewandowki's NSF CAREER grant No. DMR-2543710. The National High Magnetic Field Laboratory is supported by the National Science Foundation through NSF/DMR-2128556 and the State of Florida. F.L., E.P., and P.S-J. were supported by Grants PID2021-125343NB-I00 and PRE2022-104373, funded by MICIU/AEI/10.13039/501100011033, ``ERDF A way of making Europe'' and ``ESF+'', and the CSIC’s Quantum Technologies Platform (QTEP). 
F.G. acknowledges funding from NOVMOMAT project PID2022-142162NB-I00 funded by MICIU/AEI/10.13039/501100011033 and by FEDER, UE, from the EU NextGenerationEU/PRTR-C17.I1, as well as from the IKUR Strategy under the collaboration agreement between Ikerbasque Foundation and DIPC on behalf of the Department of Education of the Basque Government. IMDEA Nanociencia, Donostia International Physics Center, and ICMM-CSIC acknowledge support from the Severo Ochoa Centres of Excellence program in R\&D through Grants CEX2020-001039-S, CEX2024-001491-S, and CEX2024-001445-S, respectively.
Work by E.M. is supported by the Department of Energy Grant DE-FG02-84ER45118. Numerical calculations are done using the High Performance Compute Cluster  of the Research Computing Center (RCC) at Florida State University.


\clearpage

\setcounter{equation}{0}
\setcounter{figure}{0}
\renewcommand{\theequation}{S\arabic{equation}}
\renewcommand{\thefigure}{S\arabic{figure}}
\renewcommand{\thetable}{S\Roman{table}}
\renewcommand{\bibnumfmt}[1]{[#1]}
\renewcommand{\citenumfont}[1]{#1}

\onecolumngrid
\begin{center}
\Large
    Supplementary Information
\end{center}

\tableofcontents

\section{Kekul\'{e} Reconstruction}

To describe intervalley-hybridized states in real space, we consider a $\sqrt{3}\times \sqrt{3}$ reconstruction of the honeycomb unit cell to form a Kekul\'{e} lattice. In reciprocal space, this reconstruction maps both the $K$ and $K'$ valleys in the original Brillouin zone to $\overline{\Gamma}$ in the folded Brillouin zone. In this setting, momentum-preserving intervalley perturbations (which would necessarily be momentum-non-preserving in the honeycomb lattice) can mix the overlapping states from different valleys and generate intervalley-mixed Bloch wavefunctions. As such, this Kekul\'{e} reconstruction is a natural and elegant approach to describe intervalley coupling at domain walls.

\subsection{Monolayer Toy Models}
\label{sec: Monolayer Toy Models}

\subsubsection{Valley-Orbital Basis States and Their Symmetry Representations}
\begin{figure}[h!]
    \centering
    \includegraphics[width=0.3\linewidth]{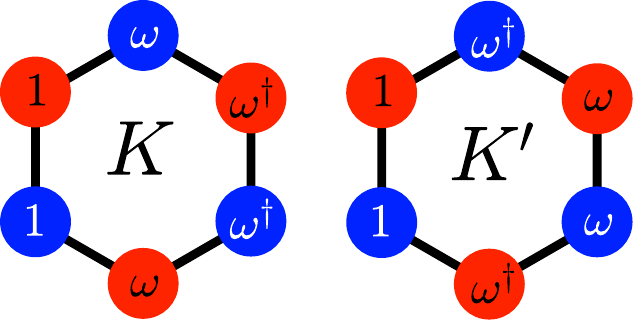}
    \caption{\textbf{Phase winding of valley basis states.} $\omega = \exp(2\pi i/3)$.}
    \label{fig:phase_winding}
\end{figure}

It is instructive to use simplified single-layer models to define the notion of intervalley interaction. We begin with the tight-binding model of graphene with two sublattices per unit cell. In the usual convention, basis states are written as
\begin{equation}
    \begin{split}
        \ket{\psi_A(\mathbf{k})} &= \frac{1}{\sqrt{N}} \sum_{\mathbf{r}} e^{i \mathbf{k} \cdot \mathbf{r}} \ket{\phi_A(\mathbf{r})}, \\
        \ket{\psi_B(\mathbf{k})} &= \frac{1}{\sqrt{N}} \sum_{\mathbf{r}} e^{i \mathbf{k} \cdot \left(\mathbf{r}+\boldsymbol{\delta_r}\right)} \ket{\phi_B(\mathbf{r})},
    \end{split}
\end{equation}
where $\mathbf{r} = n_1\mathbf{a}_1+n_2\mathbf{a}_2,$ where $\mathbf{a}_1 = a (1,0)$, $\mathbf{a}_2 = a(-1/2,\sqrt{3}/2)$, and $\mathbf{a}_3 = a(-1/2,-\sqrt{3}/2)$ are primitive lattice vectors, $a = 2.46 $ \AA ~ is the lattice constant, $n_1$ and $n_2$ are integers, and $\boldsymbol{\delta_r} = a/\sqrt{3}(0,1)$ is the basis vector. $\ket{\phi_A(\mathbf{r})}$ is an orbital located at $\mathbf{r}$ while $\ket{\phi_B(\mathbf{r})}$ is an orbital located at $\mathbf{r}+ \boldsymbol{\delta_r}.$ Exactly at the valleys $\mathbf{K} = 4\pi/3a (1,0)$ and $\mathbf{K}' = -4\pi/3a (1,0)$ the wavefunctions take the following form 
\begin{equation}
\label{eq: phases of K and K'}
    \begin{split}
        \ket{\psi_\sigma(\mathbf{K})} &= \frac{1}{\sqrt{N}} \sum_{n_1,n_2} e^{+ \frac{2\pi i}{3}(2n_1-n_2)} \ket{\phi_\sigma(n_1\mathbf{a}_1+n_2\mathbf{a}_2)} ,\\
        \ket{\psi_\sigma(\mathbf{K}')} &= \frac{1}{\sqrt{N}} \sum_{n_1,n_2} e^{- \frac{2\pi i}{3}(2n_1-n_2)} \ket{\phi_\sigma(n_1\mathbf{a}_1+n_2\mathbf{a}_2)} ,
    \end{split}
\end{equation}
where $\sigma = \lbrace A, B \rbrace$ labels sublattice. The phases at $K$ in Eq. \eqref{eq: phases of K and K'} are zero if $\mathrm{mod}(2n_1-n_2,3) = 0$ and are $\pm 2\pi/3 $ if $\mathrm{mod}(2n_1-n_2,3) = \pm 1.$ The phases at $K'$ are complex conjugates of those at $K.$ Around a hexagon, the phases on the $A$ ($B$) sublattice wind clockwise (counterclockwise) in the $K$ valley. For the $K'$ valley, the phases on the $A$ ($B$) sublattice wind counterclockwise (clockwise), as shown in Fig. \ref{fig:phase_winding}. In the space of these valley-orbital basis states, the Hamiltonian $\delta\hat{\mathcal{H}}$ (to be added to the pristine Hamiltonian $\hat{\mathcal{H}}_0$)  can be constructed from valley Pauli matrices $\boldsymbol{\tau}$ and orbital Pauli matrices $\boldsymbol{\sigma},$ giving 16 possible terms to the Hamiltonian (including the trivial $\tau_0\otimes \sigma_0$ term): 
$\delta\hat{\mathcal{H}} = \sum_{i,j \neq 0} \delta\mathcal{H}_{i,j}\tau_i\otimes\sigma_j.$ Even though the original Hamiltonian of graphene has $C_6$ symmetry, we allow this to be broken by additional hopping terms (since rhombohedral multilayers do not have this symmetry to begin with). Instead, we study $C_3$ rotation symmetry around various high-symmetry centers. To start, we examine $C_3(\varhexagon)$ symmetry, which is a threefold rotation about a hexagon center. In the valley-orbital basis, with ordering $\begin{pmatrix}
    KA&KB&K'A&K'B
\end{pmatrix}$, the $C_3(\varhexagon)$ operator takes the form
\begin{equation}
    \hat{C}_3(\varhexagon) =\exp \left( \frac{2\pi i}{3} \tau_3 \otimes \sigma_3 \right) = \begin{pmatrix}
        \omega & 0 & 0 & 0 \\
        0 & \omega^\dagger  & 0 & 0 \\
        0 & 0 & \omega^\dagger  & 0 \\
        0 & 0 & 0 & \omega
    \end{pmatrix},
\end{equation}
where $\omega = \exp(2\pi i /3)$ \footnote{$\omega$ here is not the same as frequency $\omega$ used later. It should be clear from context which $\omega$ is used as they do not overlap in usage.}. Applying this operator on all possible combinations of $\tau_i\otimes \sigma_j,$ we find
\begin{equation}
\begin{split}
    \hat{C}_3(\varhexagon)^\dagger \tau_0 \otimes \sigma_0\hat{C}_3(\varhexagon) &= \tau_0 \otimes \sigma_0, \quad
    \hat{C}_3(\varhexagon)^\dagger \tau_0 \otimes \sigma_3\hat{C}_3(\varhexagon) = \tau_0 \otimes \sigma_3, \quad
    \hat{C}_3(\varhexagon)^\dagger \tau_1 \otimes \sigma_1\hat{C}_3(\varhexagon) = \tau_1 \otimes \sigma_1, \\
    \hat{C}_3(\varhexagon)^\dagger \tau_1 \otimes \sigma_2\hat{C}_3(\varhexagon) &= \tau_1 \otimes \sigma_2, \quad
    \hat{C}_3(\varhexagon)^\dagger \tau_2 \otimes \sigma_1\hat{C}_3(\varhexagon) = \tau_2 \otimes \sigma_1, \quad
    \hat{C}_3(\varhexagon)^\dagger \tau_2 \otimes \sigma_2\hat{C}_3(\varhexagon) = \tau_2 \otimes \sigma_2, \\
    \hat{C}_3(\varhexagon)^\dagger \tau_3 \otimes \sigma_0\hat{C}_3(\varhexagon) &= \tau_3 \otimes \sigma_0, \quad
    \hat{C}_3(\varhexagon)^\dagger \tau_3 \otimes \sigma_3\hat{C}_3(\varhexagon) =\tau_3 \otimes \sigma_3, \\
    \hat{C}_3(\varhexagon)^\dagger \tau_0 \otimes \sigma_1\hat{C}_3(\varhexagon) &= -\frac{1}{2}\tau_0 \otimes \sigma_1 - \frac{\sqrt{3}}{2}\tau_3 \otimes \sigma_2, \quad
    \hat{C}_3(\varhexagon)^\dagger \tau_3 \otimes \sigma_2\hat{C}_3(\varhexagon) = +\frac{\sqrt{3}}{2}\tau_0 \otimes \sigma_1-\frac{1}{2}\tau_3 \otimes \sigma_2, \\
    \hat{C}_3(\varhexagon)^\dagger \tau_0 \otimes \sigma_2\hat{C}_3(\varhexagon) &= -\frac{1}{2}\tau_0 \otimes \sigma_2 + \frac{\sqrt{3}}{2}\tau_3 \otimes \sigma_1, \quad
    \hat{C}_3(\varhexagon)^\dagger \tau_3 \otimes \sigma_1\hat{C}_3(\varhexagon) = -\frac{\sqrt{3}}{2}\tau_0 \otimes \sigma_2-\frac{1}{2}\tau_3 \otimes \sigma_1, \\
    \hat{C}_3(\varhexagon)^\dagger \tau_1 \otimes \sigma_0\hat{C}_3(\varhexagon) &= -\frac{1}{2}\tau_1 \otimes \sigma_0 - \frac{\sqrt{3}}{2}\tau_2 \otimes \sigma_3, \quad
    \hat{C}_3(\varhexagon)^\dagger \tau_2 \otimes \sigma_3\hat{C}_3(\varhexagon) = + \frac{\sqrt{3}}{2}\tau_1 \otimes \sigma_0 -\frac{1}{2} \tau_2\otimes \sigma_3, \\
    \hat{C}_3(\varhexagon)^\dagger \tau_1 \otimes \sigma_3\hat{C}_3(\varhexagon) &= -\frac{1}{2}\tau_1 \otimes \sigma_3 - \frac{\sqrt{3}}{2}\tau_2 \otimes \sigma_0, \quad
    \hat{C}_3(\varhexagon)^\dagger \tau_2 \otimes \sigma_0\hat{C}_3(\varhexagon) = +\frac{\sqrt{3}}{2}\tau_1 \otimes \sigma_3 -\frac{1}{2}\tau_2 \otimes \sigma_0. \\
\end{split}
\end{equation}
It might be possible to form invariant linear combinations of matrices that are rotated into each other in the above equation. However, it turns out that they are all trivial combinations. For example, consider this
\begin{equation}
\begin{split}
     \hat{C}_3(\varhexagon)^\dagger \left[ a \tau_0\otimes \sigma_1+ b \tau_3 \otimes \sigma_2  \right]\hat{C}_3(\varhexagon) &= a\left(-\frac{1}{2}\tau_0 \otimes \sigma_1 - \frac{\sqrt{3}}{2}\tau_3 \otimes \sigma_2\right)+ b\left( +\frac{\sqrt{3}}{2}\tau_0 \otimes \sigma_1-\frac{1}{2}\tau_3 \otimes \sigma_2\right)   \\
     &= \left(- \frac{a}{2}+\frac{\sqrt{3}b}{2} \right)\tau_0\otimes \sigma_1 + \left( -\frac{\sqrt{3}a}{2}- \frac{b}{2} \right) \tau_3 \otimes \sigma_2.
\end{split}
\end{equation}
This leads to the following conditions on the complex numbers $a$ (not the lattice constant) and $b$ 
\begin{equation}
    \begin{split}
        a & = - \frac{a}{2}+\frac{\sqrt{3}b}{2} \quad \text{and} \quad b= -\frac{\sqrt{3}a}{2}- \frac{b}{2},
    \end{split}
\end{equation}
which only has the trivial solution $(a,b)=(0,0)$. Therefore, the only nontrivial $C_3(\varhexagon)$-invariant terms to the Hamiltonian are 
\begin{equation}
    \begin{split}
        \text{Intravalley:  }&  \sigma_3, \tau_3, \tau_3\otimes \sigma_3,\\
        \text{Intervalley:  }&  \tau_1 \otimes \sigma_1, \tau_1 \otimes \sigma_2, \tau_2 \otimes \sigma_1, \tau_2 \otimes \sigma_2.\\
    \end{split}
\end{equation}
We note that all the intravalley terms are known to have simple real-space representations: $\sigma_3$ is a Semenoff mass, $\tau_3$ is a valley Zeeman interaction (this is, actually, the term that we need to achieve a chiral phase by shifting the energy of the $K$ valley relative to the $K'$ valley), and $\tau_3 \otimes \sigma_3$ is the Haldane mass. It is also interesting to note that there are no intervalley terms that are diagonal in sublattice and also preserve $C_3(\varhexagon)$ symmetry, i.e., we do not have $C_3(\varhexagon)$-respecting $\tau_1$ and $\tau_2.$ We can also classify the operators according to \textit{spinless} time-reversal $\mathcal{T}$ symmetry (by assuming spin polarization, we are \textit{always} breaking \textit{physical, spinful} time-reversal symmetry), which is represented by 
\begin{equation}
    \hat{\mathcal{T}} = \tau_1 \mathcal{K},
\end{equation}
where $\mathcal{K}$ is the complex conjugation operator. Spinless time-reversal symmetry is important because in order to have a nonzero Hall conductivity, this symmetry needs to be broken in addition to spinful time-reversal symmetry. The classification result is shown in Table \ref{tab:valley-orbital operators}.

\begin{table}[]
\begin{center}
\begin{tabular}{c |c |c} 
 \hline
  $C_3(\varhexagon)$& $\mathcal{T}$ even & $\mathcal{T}$ odd\\
 \hline
 Intravalley & $\sigma_3$ & $\tau_3,\tau_3\otimes \sigma_3$ \\
 Intervalley & $\tau_1\otimes \sigma_1,\tau_2\otimes\sigma_1$ & $\tau_1\otimes \sigma_2,\tau_2\otimes \sigma_2$ \\
 \hline
 \hline
  $C_3(\vartriangle)$& $\mathcal{T}$ even & $\mathcal{T}$ odd\\
 \hline
 Intravalley & $\sigma_3$ & $\tau_3,\tau_3\otimes \sigma_3$ \\
 Intervalley & $\tau_1 \otimes (\sigma_0 +\sigma_3),\tau_2 \otimes (\sigma_0 + \sigma_3)$ & none \\
 \hline
 \hline
  $C_3(\triangledown)$& $\mathcal{T}$ even & $\mathcal{T}$ odd\\
 \hline
 Intravalley & $\sigma_3$ & $\tau_3,\tau_3\otimes \sigma_3$ \\
 Intervalley & $\tau_1 \otimes (\sigma_0 -\sigma_3),\tau_2 \otimes (\sigma_0 - \sigma_3)$ & none \\
 \hline
 \end{tabular}
\end{center}
    \caption{\textbf{Symmetry classification of operators in valley-orbital space.} $\mathcal{T}$ is spinless time-reversal symmetry. $C_3(\varhexagon),$ $C_3(\vartriangle),$ and $C_3(\triangledown)$ are threefold rotation symmetries about a hexagon center, an $A$ site, and a $B$ site, respectively.}
    \label{tab:valley-orbital operators}
\end{table}

We now repeat the above analysis for a three-fold rotation about a carbon site. If we choose the rotation center to be an $A$-site, denoted $C_3(\vartriangle)$, the symmetry operator in this case is
\begin{equation}
    \hat{C}_3(\vartriangle) = \begin{pmatrix}
        1 & 0 & 0 &0 \\
        0 & \omega & 0 & 0 \\
        0 & 0 & 1 & 0 \\
        0 & 0 & 0 & \omega^\dagger
    \end{pmatrix}.
\end{equation}
If we instead choose a $B$-site, denoted $C_3(\triangledown)$, the symmetry operator is 
\begin{equation}
    \hat{C}_3(\triangledown) = \begin{pmatrix}
        \omega^\dagger & 0 & 0 &0 \\
        0 & 1 & 0 & 0 \\
        0 & 0 & \omega & 0 \\
        0 & 0 & 0 & 1
    \end{pmatrix} =  \hat{M}_x\hat{C}_3(\vartriangle)\hat{M}_x,
\end{equation}
where $\hat{M}_x = \tau_1 \otimes \sigma_1.$ The action of these operators on the Pauli matrices is
\begin{equation}
    \begin{split}
        \hat{C}_3(\vartriangle)^\dagger \tau_0 \otimes \sigma_0\hat{C}_3(\vartriangle) &= \tau_0 \otimes \sigma_0, \quad
        \hat{C}_3(\vartriangle)^\dagger \tau_3 \otimes \sigma_0\hat{C}_3(\vartriangle) = \tau_3 \otimes \sigma_0,\\
        \hat{C}_3(\vartriangle)^\dagger \tau_0 \otimes \sigma_3\hat{C}_3(\vartriangle) &= \tau_0 \otimes \sigma_3,\quad
        \hat{C}_3(\vartriangle)^\dagger \tau_3 \otimes \sigma_3\hat{C}_3(\vartriangle) = \tau_3 \otimes \sigma_3,\\
        \hat{C}_3(\vartriangle)^\dagger \tau_0 \otimes \sigma_1\hat{C}_3(\vartriangle) &= - \frac{1}{2} \tau_0 \otimes \sigma_1 - \frac{\sqrt{3}}{2}\tau_3 \otimes \sigma_2,\quad
        \hat{C}_3(\vartriangle)^\dagger \tau_3 \otimes \sigma_2\hat{C}_3(\vartriangle) = +\frac{\sqrt{3}}{2}\tau_0 \otimes \sigma_1 - \frac{1}{2}\tau_3 \otimes \sigma_2,\\
        \hat{C}_3(\vartriangle)^\dagger \tau_0 \otimes \sigma_2\hat{C}_3(\vartriangle) &= - \frac{1}{2} \tau_0 \otimes \sigma_2 + \frac{\sqrt{3}}{2}\tau_3 \otimes \sigma_1,\quad
        \hat{C}_3(\vartriangle)^\dagger \tau_3 \otimes \sigma_1\hat{C}_3(\vartriangle) = -\frac{\sqrt{3}}{2}\tau_0 \otimes \sigma_2 - \frac{1}{2}\tau_3 \otimes \sigma_1,\\
        \hat{C}_3(\vartriangle)^\dagger \tau_1 \otimes \sigma_1\hat{C}_3(\vartriangle) &= -\frac{1}{2}\tau_1 \otimes \sigma_1 + \frac{\sqrt{3}}{2} \tau_2 \otimes \sigma_1,\quad
        \hat{C}_3(\vartriangle)^\dagger \tau_2 \otimes \sigma_1\hat{C}_3(\vartriangle) = -\frac{\sqrt{3}}{2}\tau_1 \otimes \sigma_1 - \frac{1}{2}\tau_2 \otimes \sigma_1,\\
        \hat{C}_3(\vartriangle)^\dagger \tau_1 \otimes \sigma_2\hat{C}_3(\vartriangle) &= -\frac{1}{2}\tau_1 \otimes \sigma_2 + \frac{\sqrt{3}}{2} \tau_2 \otimes \sigma_2,\quad
        \hat{C}_3(\vartriangle)^\dagger \tau_2 \otimes \sigma_2\hat{C}_3(\vartriangle) = -\frac{\sqrt{3}}{2} \tau_1 \otimes \sigma_2 - \frac{1}{2} \tau_2 \otimes \sigma_2,\\
        \hat{C}_3(\vartriangle)^\dagger \tau_1 \otimes \sigma_0\hat{C}_3(\vartriangle) &= \frac{1}{4}\tau_1 \otimes \sigma_0+\frac{3}{4} \tau_1 \otimes \sigma_3 - \frac{\sqrt{3}}{4} \tau_2 \otimes \sigma_0 + \frac{\sqrt{3}}{4}\tau_2 \otimes \sigma_3,\\
        \hat{C}_3(\vartriangle)^\dagger \tau_1 \otimes \sigma_3\hat{C}_3(\vartriangle) &= \frac{3}{4}\tau_1 \otimes \sigma_0 + \frac{1}{4}\tau_1 \otimes \sigma_3 + \frac{\sqrt{3}}{4} \tau_2 \otimes \sigma_0 - \frac{\sqrt{3}}{4} \tau_2 \otimes \sigma_3,\\
        \hat{C}_3(\vartriangle)^\dagger \tau_2 \otimes \sigma_0\hat{C}_3(\vartriangle) &= \frac{\sqrt{3}}{4}\tau_1 \otimes \sigma_0 - \frac{\sqrt{3}}{4}\tau_1 \otimes \sigma_3 + \frac{1}{4} \tau_2 \otimes \sigma_0 + \frac{3}{4} \tau_2 \otimes \sigma_3,\\
        \hat{C}_3(\vartriangle)^\dagger \tau_2 \otimes \sigma_3\hat{C}_3(\vartriangle) &= -\frac{\sqrt{3}}{4}\tau_1 \otimes \sigma_0 + \frac{\sqrt{3}}{4}\tau_1 \otimes \sigma_3 + \frac{3}{4} \tau_2 \otimes \sigma_0 + \frac{1}{4} \tau_2 \otimes \sigma_3.\\
    \end{split}
\end{equation}
By computing possible linear combinations of operators that are rotated into each other to search for invariant combinations, we find the following
\begin{equation}
    \begin{split}
        \text{Intravalley: } &\tau_3, \sigma_3, \tau_3\otimes \sigma_3,\\
        \text{Intervalley: } & \tau_1 \otimes \sigma_0 + \tau_1 \otimes \sigma_3,  \tau_2 \otimes \sigma_0 + \tau_2 \otimes \sigma_3.
    \end{split}
\end{equation}
In contrast to $C_3(\varhexagon)$-invariant couplings, we do not have $\tau_1\otimes \sigma_1,$ $\tau_1\otimes\sigma_2,$ $\tau_2\otimes\sigma_1,$ and $\tau_2\otimes\sigma_2$ in the present case, Instead, $\tau_1 \otimes \left(\sigma_0+ \sigma_3\right)$ and $\tau_2 \otimes \left(\sigma_0+ \sigma_3\right)$ are invariant under $C_3(\vartriangle)$ symmetry but not under $C_3(\varhexagon)$ symmetry. We note that $(\sigma_0+\sigma_3)/2 = \begin{pmatrix}
    1 & 0 \\
    0 & 0
\end{pmatrix}$ is the projector to the $A$ sublattice; so $\tau_1 \otimes \left(\sigma_0+ \sigma_3\right)$ and $\tau_2 \otimes \left(\sigma_0+ \sigma_3\right)$ are realizations of $\tau_1$ and $\tau_2$ if we consider only one sublattice, i.e. by considering a triangular lattice instead of a honeycomb lattice. To obtain the invariant terms under $\hat{C}_3(\triangledown),$ we simply apply the mirror operator $\hat{M}_x$ to the $\hat{C}_3(\vartriangle)$-invariant terms to find
\begin{equation}
    \begin{split}
        \text{Intravalley: } &\tau_3, \sigma_3, \tau_3\otimes \sigma_3,\\
        \text{Intervalley: } & \tau_1 \otimes \sigma_0 - \tau_1 \otimes \sigma_3,  \tau_2 \otimes \sigma_0 - \tau_2 \otimes \sigma_3.
    \end{split}
\end{equation}
In this case, we find $\tau_1$ and $\tau_2$ projected to the $B$ sublattice, as represented by the projector $(\sigma_0-\sigma_3)/2 = \begin{pmatrix}
    0 & 0 \\
    0 & 1
\end{pmatrix}.$

It is worth emphasizing that in the preceding analysis, we have only classified symmetry-allowed terms that are finite at exactly the $K$ and $K'$ points; there may be momentum-dependent terms that respect the symmetries enumerated above but which vanish as momentum approaches the zone corners. These momentum-dependent terms are not captured in our present analysis. Since we are interested in strong intervalley coupling that yields intervalley-mixed states, we suspect that the dominant terms are the ones which remain finite exactly at $K$ and $K',$ hence our choice to neglect momentum-dependent terms. That said, the other terms might be important as well, but we postpone their analysis to future works.

\subsubsection{Kekul\'{e} Basis States and $C_3(\varhexagon)$ Symmetry-Allowed Intervalley Terms}

\begin{figure}[h!]
    \centering
    \includegraphics[width=0.8\linewidth]{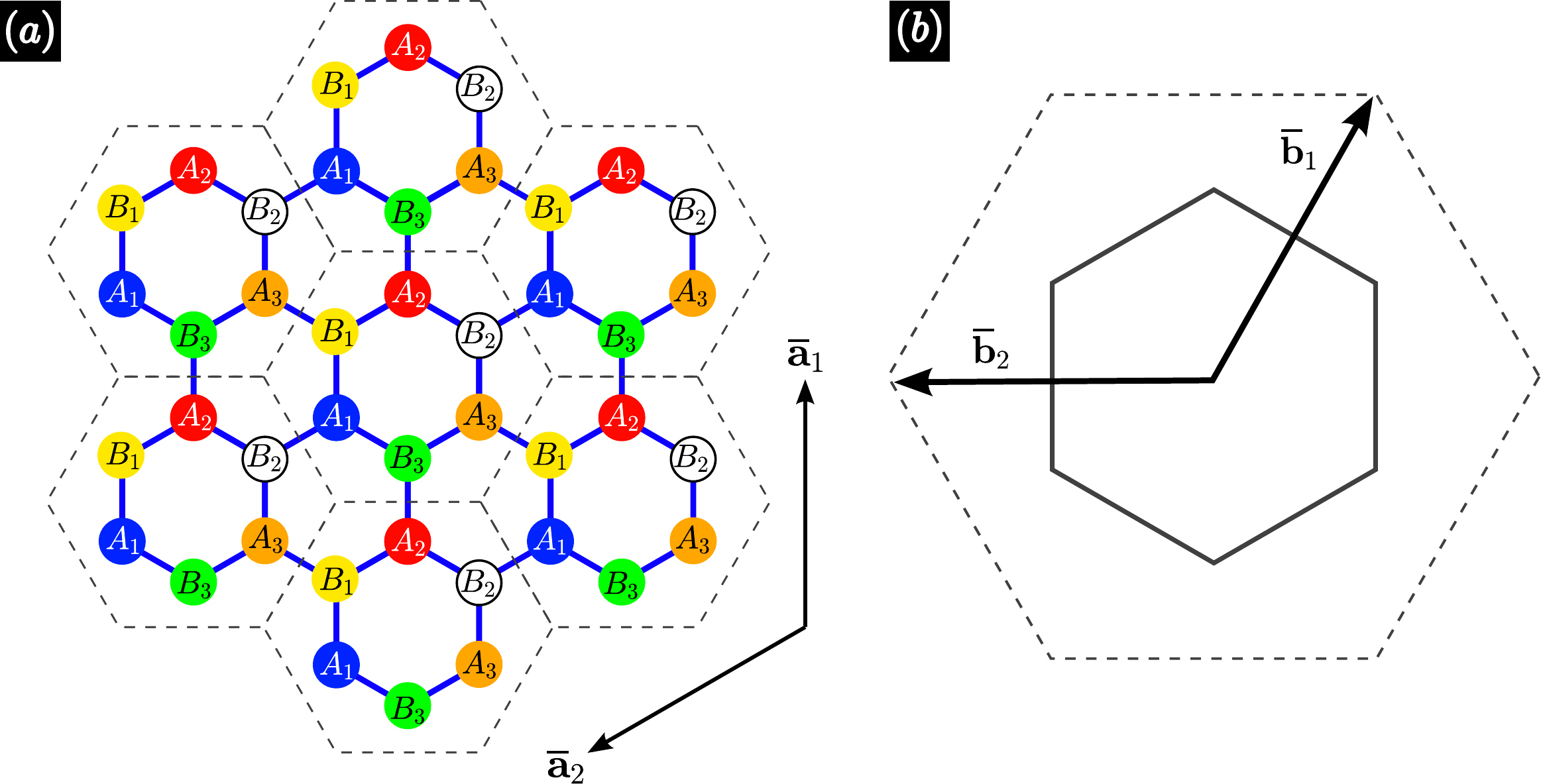}
    \caption{\textbf{$\sqrt{3}\times \sqrt{3}$ reconstructed unit cell.} (a) Real-space lattice with each unit cell, enclosed by dashed hexagons, consisting of six atoms. (b) Brillouin zone (solid hexagon) has been dilated by a factor of $1/3$ compared to the original Brillouin zone (dashed hexagon). }
    \label{fig:sqrt(3)sqrt(3)}
\end{figure}

\begin{figure}
    \centering
    \includegraphics[width=\linewidth]{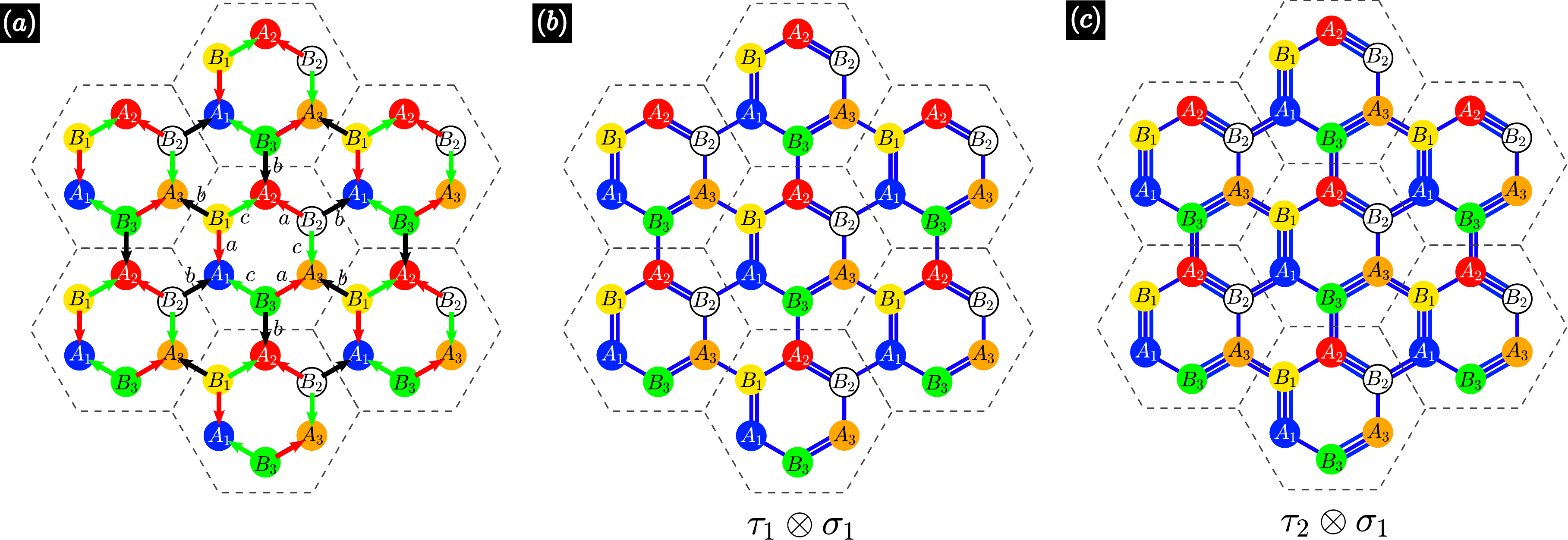}
    \caption{\textbf{Real-space representation of $C_3(\varhexagon)$ symmetric intervalley scattering on a honeycomb lattice.} (a) Directed hoppings are indicated by arrows, with notation specified in Eq. \eqref{eq: hoppings a b c}. (b) Example of a tight-binding model that realizes the $\tau_1\otimes \sigma_1$ interaction. Intracell bonds show benzene-like bond-strength modulation. (c) Example of a tight-binding model that realizes the $\tau_2\otimes \sigma_1$ interaction, which involves modulation of both intracell and intercell bonds.}
    \label{fig:monolayer_examples}
\end{figure}

Now that we have the complete classification of all symmetry-allowed terms, let us find explicit real-space representations for each of them. To do this, we implement an enlargement of the unit cell to include six orbitals as shown in Fig. \ref{fig:sqrt(3)sqrt(3)}(a). The lattice vectors for the enlarged unit cells are $\overline{\mathbf{a}}_1 = \sqrt{3}a\left( 0,1 \right) $ and  $\overline{\mathbf{a}}_2 = \sqrt{3}a \left( -\sqrt{3}/2,-1/2 \right)$. The lattice constant is now $\sqrt{3}a.$ The corresponding reciprocal lattice vectors are $\overline{\mathbf{b}}_1 = 4\pi/3a \left( -1/2, \sqrt{3}/2 \right)$ and $\overline{\mathbf{b}}_2 = 4\pi/3a\left(- 1, 0 \right).$ We notice that $\mathbf{K}_\tau = -\tau \overline{\mathbf{b}}_2,$ implying that both Dirac cones are zone-folded back to the $\overline{\Gamma}$ point in the new Brillouin zone, as advertised. Since both valleys are folded to the same point, we do not need to consider the valleys separately, which, of course, is the rationale for considering this augmented basis in the first place. In the augmented basis, $\begin{pmatrix}
    A_1& B_1 & A_2 & B_2 & A_3 & B_3
\end{pmatrix}$,  the unperturbed Hamiltonian is just
\begin{equation}
\label{eq: unperturbed monolayer Hamiltonian}
    \hat{\mathcal{H}}_0(\mathbf{k}) =  t_0\begin{pmatrix}
        0 &  e^{i \mathbf{k} \cdot \boldsymbol{\delta}_1} & 0 &   e^{i \mathbf{k} \cdot \boldsymbol{\delta}_2} & 0 &   e^{i \mathbf{k} \cdot \boldsymbol{\delta}_3} \\
          e^{-i \mathbf{k} \cdot \boldsymbol{\delta}_1} & 0 &   e^{-i \mathbf{k} \cdot \boldsymbol{\delta}_2} &0 &   e^{-i \mathbf{k} \cdot \boldsymbol{\delta}_3} & 0 \\
        0 &   e^{i \mathbf{k} \cdot \boldsymbol{\delta}_2} & 0 &   e^{i \mathbf{k} \cdot \boldsymbol{\delta}_3} & 0 &   e^{i \mathbf{k} \cdot \boldsymbol{\delta}_1} \\
          e^{-i \mathbf{k} \cdot \boldsymbol{\delta}_2} & 0 &   e^{-i \mathbf{k} \cdot \boldsymbol{\delta}_3} & 0 &   e^{-i \mathbf{k} \cdot \boldsymbol{\delta}_1} & 0 \\
        0 &   e^{i \mathbf{k} \cdot \boldsymbol{\delta}_3} & 0 &   e^{i \mathbf{k} \cdot \boldsymbol{\delta}_1} & 0 &   e^{i \mathbf{k} \cdot \boldsymbol{\delta}_2} \\
          e^{-i \mathbf{k} \cdot \boldsymbol{\delta}_3} & 0 &   e^{-i \mathbf{k} \cdot \boldsymbol{\delta}_1} & 0 &   e^{-i \mathbf{k} \cdot \boldsymbol{\delta}_2} & 0
    \end{pmatrix},
\end{equation}
where $\boldsymbol{\delta}_1 = (0,1)a/\sqrt{3},$ $\boldsymbol{\delta}_2 = (-\sqrt{3}/2,-1/2)a/\sqrt{3},$ and $\boldsymbol{\delta}_3 = (\sqrt{3}/2,-1/2)a/\sqrt{3}$ are the nearest-neighbor vectors. In this representation, the valley states are represented by (referring to Figs. \ref{fig:phase_winding} and \ref{fig:sqrt(3)sqrt(3)} for visual aid)
\begin{equation}
\label{eq: Kekula basis}
    \ket{\psi_A(\mathbf{K})} = \frac{1}{\sqrt{3}}\begin{pmatrix}
        1 \\ 0 \\ \omega \\ 0 \\ \omega^\dagger \\ 0 
    \end{pmatrix}, \quad     \ket{\psi_B(\mathbf{K})} = \frac{1}{\sqrt{3}} \begin{pmatrix}
        0 \\ 1 \\ 0 \\ \omega^\dagger \\ 0 \\ \omega
    \end{pmatrix} \quad \ket{\psi_A(\mathbf{K}')} = \frac{1}{\sqrt{3}}\begin{pmatrix}
        1 \\ 0 \\ \omega^\dagger \\ 0 \\ \omega \\ 0 
    \end{pmatrix}, \quad     \ket{\psi_B(\mathbf{K}')} = \frac{1}{\sqrt{3}} \begin{pmatrix}
        0 \\ 1 \\ 0 \\ \omega \\ 0 \\ \omega^\dagger
    \end{pmatrix}.
\end{equation}
We have verified explicitly that $\hat{\mathcal{H}}(\mathbf{0})\ket{\psi_\sigma(\mathbf{K}_\tau)} = 0,$ demonstrating the valley states are indeed mapped to $\overline{\Gamma}.$ Now, anticipating the $C_3(\varhexagon)$-symmetric intervalley terms are all intersublattice, we write the perturbation Hamiltonian as
\begin{equation}
    \delta\hat{\mathcal{H}}_{\tau_i\otimes\sigma_j}(\mathbf{k}=\mathbf{0}) = \begin{pmatrix}
        0 & f_1 & 0 & f_2 & 0 & f_3 \\
        f_1^\dagger & 0 & f_4 & 0 & f_5 & 0 \\
        0 & f_4^\dagger & 0 & f_6 & 0 & f_7 \\
        f_2^\dagger & 0 & f_6^\dagger & 0 & f_8 & 0 \\
        0 & f_5^\dagger & 0 & f_8^\dagger & 0  & f_9 \\
        f_3^\dagger & 0 & f_7^\dagger & 0 & f_9^\dagger & 0
    \end{pmatrix},
\end{equation}
where $f_i$ are complex numbers. In this basis, $C_3(\varhexagon)$ symmetry is represented by 
\begin{equation}
    \hat{C}_3(\varhexagon) = \begin{pmatrix}
        0 & 0 & 1  & 0 & 0 & 0 \\
        0 & 0 & 0  & 1 & 0 & 0  \\
        0 & 0 & 0  & 0 & 1 & 0 \\
        0 & 0 & 0  & 0 & 0 & 1 \\
        1 & 0 & 0  & 0 & 0 & 0 \\
        0 & 1 & 0  & 0 & 0 & 0
    \end{pmatrix}.
\end{equation}
Invariance under $C_3(\varhexagon)$ symmetry requires the following
\begin{equation}
    \begin{split}
    \label{eq: hoppings a b c}
        a= f_1 = f_6 =  f_9,\quad b = f_2= f_5^\dagger = f_7, \quad \text{and} \quad c = f_3 = f_4^\dagger = f_8^\dagger. \\
    \end{split}
\end{equation}
So perturbation Hamiltonian now simplifies significantly to 
\begin{equation}
\label{eq: tau_i sigma_j Hamiltonian}
    \delta\hat{\mathcal{H}}_{\tau_i\otimes\sigma_j}(\mathbf{k}=\mathbf{0}) = \begin{pmatrix}
        0 & a & 0 & b & 0 & c \\
        a^\dagger & 0 & c^\dagger & 0 & b^\dagger & 0 \\
        0 & c & 0 & a & 0 & b \\
        b^\dagger & 0 & a^\dagger & 0 & c^\dagger & 0 \\
        0 & b & 0 & c & 0  & a \\
        c^\dagger & 0 & b^\dagger & 0 & a^\dagger & 0
    \end{pmatrix}.
\end{equation}
Projecting this into the basis defined by Eq. \eqref{eq: Kekula basis}, we find that the Hamiltonian takes the form
\begin{equation}
    \delta\hat{\mathcal{H}}_{\tau_i\otimes\sigma_j}(\mathbf{k}=\mathbf{0}) = \begin{pmatrix}
        0 & 0 & 0 & a +  \omega b + \omega^\dagger c \\
        0 & 0 & a^\dagger + \omega b^\dagger + \omega^\dagger c^\dagger & 0 \\
        0 & a + \omega^\dagger b + \omega c & 0 & 0 \\
        a^\dagger + \omega^\dagger b^\dagger + \omega c^\dagger & 0 & 0 & 0
    \end{pmatrix}.
\end{equation}
Now, by choosing appropriate  complex numbers $a,b,c,$ we can realize any of the symmetry-allowed terms. If we define $a = a_r+ia_i,$ $b = b_r+ib_i,$ and $c = c_r+ic_i,$ we can write the Hamiltonian generically as 
\begin{equation}
\label{eq: hamil in Pauli matrices}
    \delta\hat{\mathcal{H}}_{\tau_i\otimes\sigma_j}(\mathbf{k}=\mathbf{0}) = \left[ a_r - \frac{b_r}{2} - \frac{c_r}{2}\right] \tau_1 \otimes \sigma_1 +   \left[ \frac{b_i}{2}+ \frac{c_i}{2} - a_i\right] \tau_1 \otimes \sigma_2 +  \left[ \frac{\sqrt{3}c_r-\sqrt{3}b_r}{2}\right] \tau_2 \otimes \sigma_1 +  \left[ \frac{\sqrt{3}b_i-\sqrt{3}c_i}{2}\right] \tau_2 \otimes \sigma_2.
\end{equation} 
Notice that the $\mathcal{T}$-even terms are associated with real hoppings while the $\mathcal{T}$-odd terms are associated with imaginary hoppings, as we would expect. Using this, we have the following special cases where each term in Eq. \eqref{eq: hamil in Pauli matrices} is isolated:
\begin{equation}
    \begin{split}
        \tau_1 \otimes \sigma_1 &: b = c = 0, a = a_r,\\
        \tau_1 \otimes \sigma_2 &: b = c = 0, a = a_i,\\
        \tau_2 \otimes \sigma_1 &: a_i=b_i=c_i = 0, 2a_r = b_r+c_r, b_r\neq c_r, \\
        \tau_2 \otimes \sigma_2 &: a_r=b_r=c_r = 0, 2a_i = b_i+c_i, b_i\neq c_i. \\
    \end{split}
\end{equation}
Terms involving $\tau_1 \otimes \sigma_1$ and $\tau_1 \otimes \sigma_2$ consist of modulation of the bonds around the hexagon of a single unit cell, as shown in Fig. \ref{fig:monolayer_examples}(b). For example, we can have alternating double and single bonds around the hexagon similar to the resonance bonding structures of benzene. On the other hand, terms involving $\tau_2 \otimes \sigma_1$ and $\tau_2 \otimes \sigma_2$ consist of modulation of both intracell and intercell bonds, as shown in Fig. \ref{fig:monolayer_examples}(c). In particular, for any triad of nearest-neighbor bonds, one must be the average of the other two. 

For illustration, let us consider one concrete example with the following Hamiltonian
\begin{equation}
    \hat{\mathcal{H}}(\mathbf{k}) =  \begin{pmatrix}
        0 &  (t_0+t_1)e^{i \mathbf{k} \cdot \boldsymbol{\delta}_1} & 0 &   t_0e^{i \mathbf{k} \cdot \boldsymbol{\delta}_2} & 0 &   t_0e^{i \mathbf{k} \cdot \boldsymbol{\delta}_3} \\
          (t_0+t_1)e^{-i \mathbf{k} \cdot \boldsymbol{\delta}_1} & 0 &   t_0e^{-i \mathbf{k} \cdot \boldsymbol{\delta}_2} &0 &   t_0e^{-i \mathbf{k} \cdot \boldsymbol{\delta}_3} & 0 \\
        0 &   t_0e^{i \mathbf{k} \cdot \boldsymbol{\delta}_2} & 0 &   (t_0+t_1)e^{i \mathbf{k} \cdot \boldsymbol{\delta}_3} & 0 &   t_0e^{i \mathbf{k} \cdot \boldsymbol{\delta}_1} \\
          t_0e^{-i \mathbf{k} \cdot \boldsymbol{\delta}_2} & 0 &   (t_0+t_1)e^{-i \mathbf{k} \cdot \boldsymbol{\delta}_3} & 0 &   t_0e^{-i \mathbf{k} \cdot \boldsymbol{\delta}_1} & 0 \\
        0 &   t_0e^{i \mathbf{k} \cdot \boldsymbol{\delta}_3} & 0 &   t_0e^{i \mathbf{k} \cdot \boldsymbol{\delta}_1} & 0 &   (t_0+t_1)e^{i \mathbf{k} \cdot \boldsymbol{\delta}_2} \\
          t_0e^{-i \mathbf{k} \cdot \boldsymbol{\delta}_3} & 0 &   t_0e^{-i \mathbf{k} \cdot \boldsymbol{\delta}_1} & 0 &   (t_0+t_1)e^{-i \mathbf{k} \cdot \boldsymbol{\delta}_2} & 0
    \end{pmatrix}.
\end{equation}
The eigenstates of this Hamiltonian at $\mathbf{k} = \mathbf{0}$ are 
\begin{align*}
        E_1 &= -3t_0-t_1, &\ket{\psi_1} &= \begin{pmatrix}
            -1 & 1 & -1&1&-1 & 1
        \end{pmatrix}/\sqrt{6}, \\
        E_2 &= - t_1, &\ket{\psi_2} &= \begin{pmatrix}
            1 & -1 & 0 & 0 & -1 & 1
        \end{pmatrix}/\sqrt{4}, \\
        E_3 &= - t_1, &\ket{\psi_3} &= \begin{pmatrix}
            1 & -1  & -1 & 1 & 0 & 0
        \end{pmatrix}/\sqrt{4}, \\
        E_4 &= + t_1, &\ket{\psi_4} &= \begin{pmatrix}
            -1 & -1  & 0 & 0 & 1 & 1
        \end{pmatrix}/\sqrt{4}, \\
        E_5 &= + t_1, &\ket{\psi_5} &= \begin{pmatrix}
            -1 & -1  & 1 & 1 & 0 & 0
        \end{pmatrix}/\sqrt{4}, \\
        E_6 &= +3t_0+t_1, &\ket{\psi_6} &= \begin{pmatrix}
            1 & 1 & 1 & 1 & 1 & 1
        \end{pmatrix}/\sqrt{6}.
\end{align*}
$\ket{\psi_1}$ and $\ket{\psi_6}$ are high-energy states that we shall neglect when only states near the Dirac cones matter to the physics (i.e. the entirety of this work). States near $E=0$ are doubly-degenerate and are hybridized from the valley-polarized states. Actually, this statement requires further inspection because it is possible that even though these states have no appearance of valley polarization, perhaps there are linear combinations of them which are valley-polarized. To prove that this is \textit{not} the case and that these degenerate states are indeed intervalley-hybridized (i.e. not valley-polarized) states, we calculate the valley character of the linear combination $\ket{\psi} = \alpha\ket{\psi_2}+\beta e^{i\vartheta} \ket{\psi_3},$ where $\alpha \geq 0, \beta \geq 0,\vartheta$ $(\alpha^2+\beta^2 = 1)$ are real numbers to be extremized for maximal valley polarization:
\begin{equation}
\begin{split}
    \abs{\bra{\psi}\ket{\psi_A(\mathbf{K})}}^2  + \abs{\bra{\psi}\ket{\psi_B(\mathbf{K})}}^2= \abs{\bra{\psi}\ket{\psi_A(\mathbf{K}')}}^2+\abs{\bra{\psi}\ket{\psi_B(\mathbf{K}')}}^2 &= \frac{1+ \alpha\beta \cos\vartheta }{2} \leq \frac{3}{4}, \\
\end{split}
\end{equation}
The maximal value of cosine is $\cos\vartheta = 1.$ For these values of $\vartheta,$ the numerator is $1+\alpha\sqrt{1-\alpha^2},$ which is maximal for $\alpha = \beta =  1/\sqrt{2}.$ The  overlaps with valley states are therefore bounded by $3/4,$ showing that these states are indeed intervalley-coherent states.

\subsubsection{$C_3(\vartriangle)$ Symmetry-Allowed Intervalley Terms}

In this section, we write down all the symmetry-allowed terms consistent with $C_3(\vartriangle)$ symmetry acting on the $A$ sublattice only. The most general such Hamiltonian is 
\begin{equation}
    \delta\hat{\mathcal{H}}_{\tau_i\otimes(\sigma_0+\sigma_3)}(\mathbf{k}=\mathbf{0}) = \begin{pmatrix}
        f_1 & 0 & a_r+ia_i & 0 & b_r+ib_i & 0 \\
        0 & 0 & 0 & 0 & 0 & 0 \\
        a_r-ia_i & 0 & f_2 & 0 & c_r+ic_i & 0 \\
        0 & 0 & 0 & 0 & 0 & 0 \\
        b_r-ib_i & 0 & c_r-ic_i & 0 & f_3 & 0 \\
        0 & 0 & 0 & 0 & 0 & 0 
    \end{pmatrix},
\end{equation}
where the diagonal elements are required to be real by Hermiticity while the off-diagonal elements can be complex in general. In the Kekul\'{e} basis, the symmetry operator for $C_3(\vartriangle)$ is 
\begin{equation}
    \hat{C}_3(\vartriangle) = \begin{pmatrix}
        1 & 0 & 0 & 0 & 0 & 0 \\
        0 & 0 & 0 & 0 & 0 & 1 \\
        0 & 0 & 1 & 0 & 0 & 0 \\
        0 & 1 & 0 & 0 & 0 & 0 \\
        0 & 0 & 0 & 0 & 1 & 0 \\
        0 & 0 & 0 & 1 & 0 & 0 \\
    \end{pmatrix}.
\end{equation}
We notice this operator matrix contains an invariant submatrix (a $3\times3$ identity submatrix). Therefore, it immediately follows that $\hat{C}_3(\vartriangle)^\dagger\delta\hat{\mathcal{H}}_{\tau_i\otimes(\sigma_0+\sigma_3)}(\mathbf{k}=\mathbf{0}) \hat{C}_3(\vartriangle) = \delta\hat{\mathcal{H}}_{\tau_i\otimes(\sigma_0+\sigma_3)}(\mathbf{k}=\mathbf{0}).$ Projecting this Hamiltonian to the valley-orbital basis, we find
\begin{equation}
\begin{split}
\delta\hat{\mathcal{H}}_{\tau_i\otimes(\sigma_0+\sigma_3)}(\mathbf{k}=\mathbf{0}) = &\left[ \frac{1}{6} \left(f_1+f_2+f_3-a_r-b_r-c_r \right) \tau_0  + \frac{1}{12} \left(2f_1-f_2-f_3-2a_r-2b_r+4c_r \right) \tau_1 \right. \\
&\left. \frac{1}{4\sqrt{3}} \left(f_3-f_2+2a_r-2b_r \right) \tau_2 + \frac{1}{2\sqrt{3}} \left( -a_i+b_i-c_i\right) \tau_3\right]\otimes \left[\sigma_0+\sigma_3\right].   
\end{split}
\end{equation}
The diagonal elements $(f_1,f_2,f_3)$ can originate from on-site potentials or from \textit{next}-nearest intra-sublattice hoppings. This means that we can obtain intervalley interaction even from purely local charge modulation on the reconstructed Kekul\'{e} lattice. For example, by setting $f_1=a=b=c=0,$ we have 
\begin{equation}
\begin{split}
\delta\hat{\mathcal{H}}_{\tau_i\otimes(\sigma_0+\sigma_3)}(\mathbf{k}=\mathbf{0}) = &\left[\frac{1}{6} \left(f_2 + f_3 \right) \tau_0-\frac{1}{12} \left(f_2+f_3 \right) \tau_1 + \frac{1}{4\sqrt{3}} \left(f_3-f_2 \right) \tau_2 \right]\otimes \left[\sigma_0+\sigma_3\right]  
\end{split}
\end{equation}
On the other hand, with nearest-neighbor intra-sublattice hoppings, we can have a $\tau_3$ coupling from the imaginary parts of $a,b,c$. While $\tau_1,\tau_2$ describe intervalley scattering, $\tau_3$ describes valley imbalance. This means that simultaneously varying the real and imaginary parts of $a,b,c,$ we can model both intervalley hybridization and valley polarization in the same model. For example, by setting $f_1=f_2=f_3 = 0,$ we have
\begin{equation}
\begin{split}
\delta\hat{\mathcal{H}}_{\tau_i\otimes(\sigma_0+\sigma_3)}(\mathbf{k}=\mathbf{0}) = &\frac{1}{6}\left[  \left(-a_r-b_r-c_r \right) \tau_0  +  \left(2c_r-a_r-b_r \right) \tau_1 + \sqrt{3} \left(a_r-b_r \right) \tau_2 + \sqrt{3} \left( -a_i+b_i-c_i\right) \tau_3\right]\otimes \left[\sigma_0+\sigma_3\right].   
\end{split}
\end{equation}
We set $a_r = -b_r-c_r$ to eliminate the $\tau_0$ term and also assume $a_i=b_i=0$ since we only need $c_i$ to control the $\tau_3$ term. By identifying $c = t_2e^{i\varphi}$ and $b_r = t_1,$ we have
\begin{equation}
\begin{split}
\delta\hat{\mathcal{H}}_{\tau_i\otimes(\sigma_0+\sigma_3)}(\mathbf{k}=\mathbf{0}) = &\left[  t_2\cos\varphi \tau_1  -  \frac{2t_1+t_2\cos\varphi}{\sqrt{3}} \tau_2 - \frac{t_2\sin\varphi}{\sqrt{3}}\tau_3\right]\otimes \frac{\sigma_0+\sigma_3}{2}.   
\end{split}
\end{equation}
By tuning $t_1,$ $t_2,$ and $\varphi,$ we can go from $\tau_1$ to $\tau_2$ to $\tau_3.$ The first two terms are even under $\mathcal{T}$ symmetry while the final term is odd under $\mathcal{T}$ symmetry. For $\varphi = 0, \pi,$ we recover $\mathcal{T}$ symmetry, which is sensible because these are the limits where all of the hoppings are real.

We end this section by mentioning that while terms involving $\tau_1$ and $\tau_2$ necessarily require at least a $\sqrt{3}\times \sqrt{3}$ reconstruction to coherently mix the valleys, terms involving $\tau_3$ are intravalley and do not require such a reconstruction, even though the example above does involve a Kekul\'{e} lattice. A famous example is the Haldane model. In the hexagonal Brillouin zone, it is given by 
\begin{equation}
    \delta \hat{\mathcal{H}}_{\tau_3}(\mathbf{k}) = \frac{2m_z}{3\sqrt{3}} \sum_{i=1}^3 \sin \left[\mathbf{k} \cdot \mathbf{a}_i \right] \sigma_3 \rightarrow \delta \hat{\mathcal{H}}_{\tau_3}(\mathbf{k} = \mathbf{0}) = - m_z \tau_3 \otimes \sigma_3.
\end{equation}
We can modify this Hamiltonian to have the desired sublattice structure by replacing $\sigma_3$ with the appropriate $\sigma$ matrices. We will use the Haldane model to simulate valley imbalance since it is simpler.

\subsubsection{General Transformation Between Valley-Orbital Basis and Kekul\'{e} Basis }

Here, we provide a prescription to transform between the two basis sets by classifying every possible perturbation without any symmetry constraint in the valley-orbital basis. We start with a few examples. Let us write down tight-binding Hamiltonians that realize $\tau_1\otimes \sigma_0,$ $\tau_1\otimes \sigma_3,$ $\tau_2 \otimes \sigma_0,$ and $\tau_2 \otimes \sigma_3$ and show explicitly that they break $C_3(\varhexagon)$ symmetry, consistent with our preceding analysis. Both of these terms must hop only within the $A$ or $B$ sublattice of the original unit cell (not the Kekul\'{e} unit cell).  The following Hamiltonian realizes the $\tau_1\otimes \sigma_0$ and $\tau_1\otimes \sigma_3$ interactions
\begin{equation}
\label{eq: tau1 Hamiltonian}
    \delta\hat{\mathcal{H}}_{\tau_1}(\mathbf{k}=\mathbf{0}) = \begin{pmatrix}
        0 & 0 & t_1 & 0 & t_1 & 0 \\
        0 & 0 & 0 & t_2 & 0 & t_2 \\
        t_1 & 0 & 0 & 0 & -2t_1 & 0 \\
        0 & t_2 & 0 & 0 & 0 & -2t_2 \\
        t_1 & 0 & -2t_1 & 0 & 0 & 0 \\
        0 & t_2 & 0 & -2t_2 & 0 & 0 \\
    \end{pmatrix} = -(t_1+t_2) \tau_1 \otimes \sigma_0 - (t_1-t_2) \tau_1 \otimes \sigma_3.
\end{equation}
On the other hand, the following Hamiltonian realizes the $\tau_2\otimes \sigma_0$ and $\tau_2\otimes \sigma_3$ interactions
\begin{equation}
\label{eq: tau2 Hamiltonian}
    \delta\hat{\mathcal{H}}_{\tau_2}(\mathbf{k}=\mathbf{0}) = \begin{pmatrix}
        0 & 0 & t_1 & 0 & -t_1 & 0 \\
        0 & 0 & 0 & t_2 & 0 & -t_2 \\
        t_1 & 0 & 0 & 0 & 0 & 0 \\
        0 & t_2 & 0 & 0 & 0 & 0 \\
        -t_1 & 0 & 0 & 0 & 0 & 0 \\
        0 & -t_2 & 0 & 0 & 0 & 0 \\
    \end{pmatrix} = \frac{1}{\sqrt{3}}(t_1-t_2) \tau_2 \otimes \sigma_0 + \frac{1}{\sqrt{3}} (t_1+t_2) \tau_2 \otimes \sigma_3.
\end{equation}
We have checked explicitly that these two Hamiltonians do not preserve $C_3(\varhexagon)$ symmetry by confirming that $\hat{C}_3(\varhexagon)^\dagger \delta \hat{{\mathcal{H}}}_{\tau_i} \hat{C}_3(\varhexagon) \neq \delta\hat{{\mathcal{H}}}_{\tau_i}.$ However, these can be made to respect $C_3(\vartriangle)$ symmetry by setting $t_2 = 0.$ Finally, we consider the most general perturbation of the form
\begin{equation}
\label{eq: hamiltonian in Kekule basis}
    \hat{\mathcal{H}}(\mathbf{k}=\mathbf{0}) = \sum_{\mu=0}^8 \sum_{\nu=0}^3 t_{\mu\nu} \lambda_\mu \otimes \sigma_\nu,
\end{equation}
where $\lambda_\mu$ are the Gell-Mann matrices that form the basis for the Lie algebra of $SU(3)$ and $t_{\mu\nu}$ are 36 real parameters the describe both hoppings and on-site energies. The nine Gell-Mann matrices are
\begin{equation}
\begin{split}
    \lambda_0 &=  \sqrt{\frac{2}{3}}\begin{pmatrix}
        1 & 0 & 0 \\
        0 & 1 & 0 \\
        0 & 0 & 1
    \end{pmatrix} \quad \lambda_1 = \begin{pmatrix}
        0 & 1 & 0 \\
        1 & 0 & 0 \\
        0 & 0 & 0
    \end{pmatrix} \quad \lambda_2 = \begin{pmatrix}
        0 & -i & 0 \\
        i & 0 & 0 \\
        0 & 0 & 0
    \end{pmatrix} \quad \lambda_3 = \begin{pmatrix}
        1 & 0 & 0 \\
        0 & -1 & 0 \\
        0 & 0 & 0
    \end{pmatrix} \quad \lambda_4 = \begin{pmatrix}
        0 & 0 & 1 \\
        0 & 0 & 0 \\
        1 & 0 & 0
    \end{pmatrix}    \\
    \lambda_5 &= \begin{pmatrix}
        0 & 0 & -i \\
        0 & 0 & 0 \\
        i & 0 & 0
    \end{pmatrix}   \quad \lambda_6 = \begin{pmatrix}
        0 & 0 & 0 \\
        0 & 0 & 1 \\
        0 & 1 & 0
    \end{pmatrix}   \quad \lambda_7 = \begin{pmatrix}
        0 & 0 & 0 \\
        0 & 0 & -i \\
        0 & i & 0
    \end{pmatrix}  \quad \lambda_8 = \frac{1}{\sqrt{3}} \begin{pmatrix}
        1 & 0 & 0 \\
        0 & 1 & 0 \\
        0 & 0 & -2
    \end{pmatrix}  .
\end{split}
\end{equation}
Like the Pauli matrices, these Gell-Mann matrices satisfy $\Tr{\lambda_\mu \lambda_\nu} = 2 \delta_{\mu\nu}.$ The previous examples can be written in terms of these Gell-Mann matrices, using the notation in Eqs. \eqref{eq: tau_i sigma_j Hamiltonian}, \eqref{eq: tau1 Hamiltonian}, and \eqref{eq: tau2 Hamiltonian},
\begin{equation}
    \begin{split}
        \delta\hat{\mathcal{H}}_{\tau_i\otimes \sigma_j}(\mathbf{k}=\mathbf{0}) &=  \sqrt{\frac{3}{2}} a_r \lambda_0 \otimes \sigma_1 - \sqrt{\frac{3}{2}}a_i \lambda_0 \otimes \sigma_2 \\
        &+ \frac{1}{2}(b_r+c_r)\left( \lambda_1+\lambda_4+\lambda_6 \right) \otimes \sigma_1 + \frac{1}{2}(b_r-c_r) \left( -\lambda_2+ \lambda_5-\lambda_7 \right)\otimes \sigma_2\\
        &+ \frac{1}{2} (b_i-c_i) \left(-\lambda_2+ \lambda_5-\lambda_7  \right) \otimes \sigma_1 + \frac{1}{2}(b_i+c_i) \left(-\lambda_1-\lambda_4-\lambda_6 \right)\otimes \sigma_2, \\
        \delta\hat{\mathcal{H}}_{\tau_1}(\mathbf{k}=\mathbf{0}) &= \frac{1}{2} (t_1+t_2) \left( \lambda_1  + \lambda_4  -2 \lambda_6  \right)\otimes \sigma_0+ \frac{1}{2}(t_1-t_2) \left( \lambda_1  + \lambda_4  -2 \lambda_6  \right) \otimes \sigma_3, \\
        \delta\hat{\mathcal{H}}_{\tau_2}(\mathbf{k}=\mathbf{0}) &= \frac{1}{2} (t_1+t_2) \left( \lambda_1  - \lambda_4  \right)\otimes \sigma_0+ \frac{1}{2}(t_1-t_2) \left( \lambda_1  - \lambda_4   \right) \otimes \sigma_3.
    \end{split}
\end{equation}
In general, we can project any $6\times6$ Hamiltonian in the Kekul\'{e} basis into the space of the valley-orbital states to obtain $ \hat{\mathcal{H}}(\mathbf{k}=\mathbf{0}) = \sum_{i=0}^3 \sum_{j=0}^3 \mathcal{H}_{ij} \tau_i \otimes \sigma_j,$ where the coefficients are given by 
\begin{equation}
\label{eq: hamiltonian in valley basis}
    \begin{split}
        \mathcal{H}_{00} &=  (\sqrt{6} t_{00}-t_{10}-t_{40}-t_{60})/3, \\
        \mathcal{H}_{01} &=  (-2 t_{11}+3 t_{31}-2 t_{41}+4 t_{61}+\sqrt{3} t_{81})/6, \\
        \mathcal{H}_{02} &= (-2 t_{12}+3 t_{32}-2 t_{42}+4 t_{62}+\sqrt{3} t_{82})/6, \\
        \mathcal{H}_{03} &= (\sqrt{6} t_{03}-t_{13}-t_{43}-t_{63})/3, \\
        \mathcal{H}_{10} &= (-2 t_{10}+3 t_{30}-2 t_{40}+4 t_{60}+\sqrt{3} t_{80})/6, \\
        \mathcal{H}_{11} &=  (\sqrt{6} t_{01}-t_{11}-t_{41}-t_{61})/3, \\
        \mathcal{H}_{12} &= (\sqrt{6} t_{02}-t_{12}-t_{42}-t_{62})/3, \\
        \mathcal{H}_{13} &= (-2 t_{13}+3 t_{33}-2 t_{43}+4 t_{63}+\sqrt{3} t_{83})/6, \\
        \mathcal{H}_{20} &= (2 \sqrt{3} t_{13}+\sqrt{3} t_{33}-2 \sqrt{3} t_{43}-3 t_{83})/6, \\
        \mathcal{H}_{21} &= (t_{22}-t_{52}+t_{72})/\sqrt{3}, \\
        \mathcal{H}_{22} &= (-t_{21}+t_{51}-t_{71})/\sqrt{3}, \\
        \mathcal{H}_{23} &= (2 \sqrt{3} t_{10}+\sqrt{3} t_{30}-2 \sqrt{3} t_{40}-3 t_{80})/6, \\
        \mathcal{H}_{30} &= (t_{23}-t_{53}+t_{73})/\sqrt{3},\\
        \mathcal{H}_{31} &= (-2 t_{12}-t_{32}+2 t_{42}+\sqrt{3} t_{82})/2 \sqrt{3}, \\
        \mathcal{H}_{32} &= (2 t_{11}+t_{31}-2 t_{41}-\sqrt{3} t_{81})/2 \sqrt{3}, \\
        \mathcal{H}_{33} &= (t_{20}-t_{50}+t_{70})/\sqrt{3}. \\        
    \end{split}
\end{equation}
Eq. \eqref{eq: hamiltonian in valley basis} allows us to write the Hamiltonian in the valley-orbital basis for any given Hamiltonian in the Kekul\'{e} real-space basis of the form in Eq. \eqref{eq: hamiltonian in Kekule basis}. This transformation is completely generic and generally does not respect any symmetry.

\subsubsection{Band Structures for Example Models}

\begin{figure}
    \centering
    \includegraphics[width=1\linewidth]{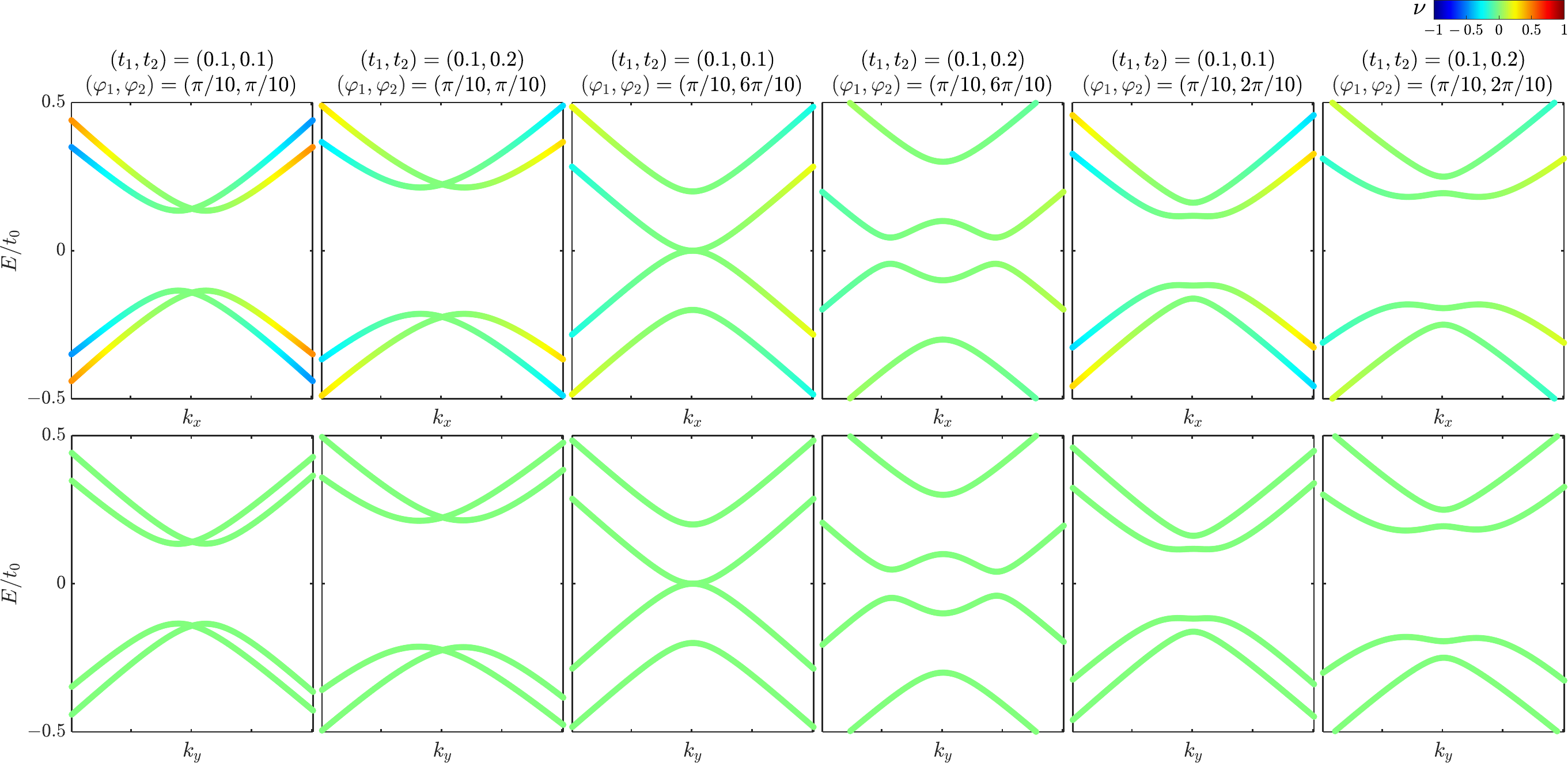}
    \caption{\textbf{Band structures for the model defined in Eq. \eqref{eq: model 1}.} The band structures on the top (bottom) panel are traced along the $k_x$ ($k_y$) direction. We notice that Bloch states are much more strongly mixed in valley character along the $k_y$ direction compared to the $k_x$ direction.}
    \label{fig:c3_models_band_structure}
\end{figure}

In this section, we study the band structures of various example models in the vicinity of $\overline{\Gamma}$ of the Kekul\'{e} Brillouin zone. Let us first study the example in Eq. \eqref{eq: tau_i sigma_j Hamiltonian} with $a = t_1e^{i\varphi_1}$and $b = -c = t_2e^{i\varphi_2}/\sqrt{3}.$ With this minor relabeling, we have
\begin{equation}
\label{eq: model 1}
    \hat{\mathcal{H}}_{\tau_i\otimes\sigma_j}(\mathbf{k}) = \hat{\mathcal{H}}_0(\mathbf{k}) +\begin{pmatrix}
        0 & t_1e^{i\varphi_1}e^{i \mathbf{k} \cdot \boldsymbol{\delta}_1} & 0 & \frac{t_2e^{i\varphi_2}}{\sqrt{3}}e^{i \mathbf{k} \cdot \boldsymbol{\delta}_2} & 0 & -\frac{t_2e^{i\varphi_2}}{\sqrt{3}}e^{i \mathbf{k} \cdot \boldsymbol{\delta}_3} \\
        t_1e^{-i\varphi_1}e^{-i \mathbf{k} \cdot \boldsymbol{\delta}_1} & 0 & -\frac{t_2e^{-i\varphi_2}}{\sqrt{3}}e^{-i \mathbf{k} \cdot \boldsymbol{\delta}_2} & 0 & \frac{t_2e^{-i\varphi_2}}{\sqrt{3}}e^{-i \mathbf{k} \cdot \boldsymbol{\delta}_3} & 0 \\
        0 & -\frac{t_2e^{i\varphi_2}}{\sqrt{3}}e^{i \mathbf{k} \cdot \boldsymbol{\delta}_2} & 0 & t_1e^{i\varphi_1}e^{i \mathbf{k} \cdot \boldsymbol{\delta}_3} & 0 & \frac{t_2e^{i\varphi_2}}{\sqrt{3}}e^{i \mathbf{k} \cdot \boldsymbol{\delta}_1} \\
        \frac{t_2e^{-i\varphi_2}}{\sqrt{3}}e^{-i \mathbf{k} \cdot \boldsymbol{\delta}_2} & 0 & t_1e^{-i\varphi_1}e^{-i \mathbf{k} \cdot \boldsymbol{\delta}_3} & 0 & -\frac{t_2e^{-i\varphi_2}}{\sqrt{3}}e^{-i \mathbf{k} \cdot \boldsymbol{\delta}_1} & 0 \\
        0 & \frac{t_2e^{i\varphi_2}}{\sqrt{3}}e^{i \mathbf{k} \cdot \boldsymbol{\delta}_3} & 0 & -\frac{t_2e^{i\varphi_2}}{\sqrt{3}}e^{i \mathbf{k} \cdot \boldsymbol{\delta}_1} & 0  & t_1e^{i\varphi_1}e^{i \mathbf{k} \cdot \boldsymbol{\delta}_2} \\
        -\frac{t_2e^{-i\varphi_2}}{\sqrt{3}}e^{-i \mathbf{k} \cdot \boldsymbol{\delta}_3} & 0 & \frac{t_2e^{-i\varphi_2}}{\sqrt{3}}e^{-i \mathbf{k} \cdot \boldsymbol{\delta}_1} & 0 & t_1e^{-i\varphi_1}e^{-i \mathbf{k} \cdot \boldsymbol{\delta}_2} & 0
    \end{pmatrix}.    
\end{equation}
The energies at $\mathbf{k}=\mathbf{0}$ for the four low-energy states are given by 
\begin{equation}
    E = \pm \sqrt{t_1^2+t_2^2\pm 2t_1t_2\sin\left(\varphi_1-\varphi_2\right)}.
\end{equation}
When $(\varphi_1-\varphi_2)\mod\pi = 0,$ the spectrum is degenerate with two states at $E = + \sqrt{t_1^2+t_2^2}$ and two states at $E = -\sqrt{t_1^2+t_2^2}.$ When $(\varphi_1-\varphi_2)\mod\pi = \pi/2,$ we generically have four non-degenerate states with $E = \pm |t_1\pm t_2|$ unless $t_1=0$ or $t_2 =0,$ in which case, we have two upper degenerate states and low lower degenerate states. If instead, $t_1 = \pm t_2,$ then we have two degenerate states at $E=0.$ Band structures for some representative values are shown in Fig. \ref{fig:c3_models_band_structure}. There, each state $\ket{\psi}$ is labeled by its valley polarization $\nu$, defined as 
\begin{equation}
    \nu(\psi) = \abs{\bra{\psi}\ket{\psi_A(\mathbf{K})}}^2+\abs{\bra{\psi}\ket{\psi_B(\mathbf{K})}}^2-\abs{\bra{\psi}\ket{\psi_A(\mathbf{K}')}}^2-\abs{\bra{\psi}\ket{\psi_B(\mathbf{K}')}}^2.
\end{equation}
This value ranges in $\left[-1,1\right],$ with $-1$ indicating polarization in the $K'$ valley and $+1$ indicating polarization in the $K$ valley. As shown in Fig. \ref{fig:c3_models_band_structure}, all the states are intervalley-hybridized states as diagnosed by $|\nu| < 1.$ The bands are highly anisotropic along the two perpendicular directions shown. It is immediately apparent from Fig. \ref{fig:c3_models_band_structure} that Bloch states along the $k_y$ direction are much more strongly mixed in valley character, i.e. $\nu \approx 0$, than states along the $k_x$ direction. This is sensible because the $k_y$ direction runs along the armchair direction while the $k_x$ direction is parallel to the zigzag direction.

\begin{figure}
    \centering
    \includegraphics[width=\linewidth]{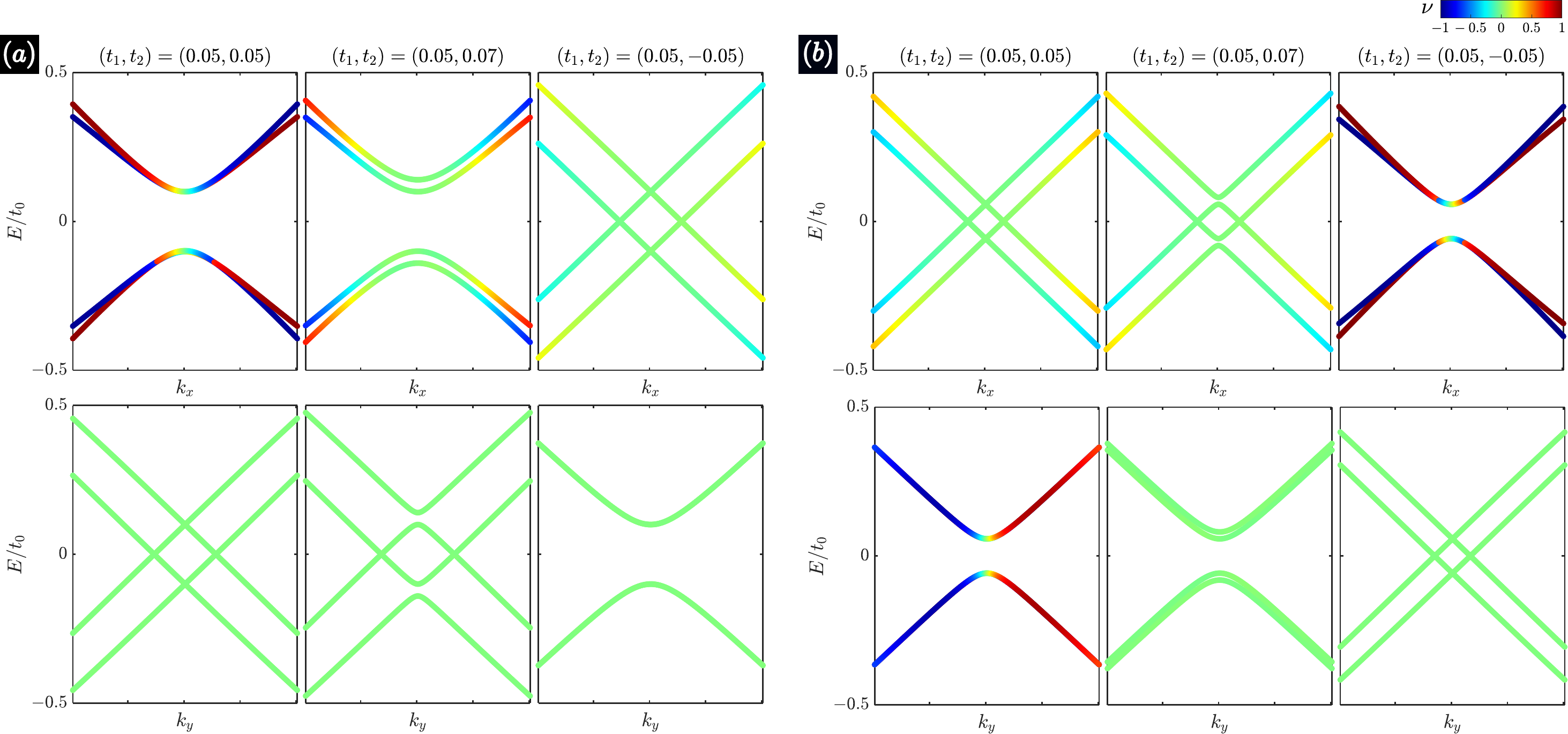}
    \caption{\textbf{Band structures for the models defined in Eqs. \eqref{eq: model 2} and \eqref{eq: model 3}.} (a) $\tau_1$ model and (b) $\tau_2$ model.}
    \label{fig:c3_breaking_models_band_structure}
\end{figure}

Next, we examine the example in Eq. \eqref{eq: tau1 Hamiltonian}, which, when the momentum dependence is restored, takes the following form
\begin{equation}
\label{eq: model 2}
    \hat{\mathcal{H}}_{\tau_1}(\mathbf{k}) = \hat{\mathcal{H}}_0(\mathbf{k})+\frac{1}{3}\begin{pmatrix}
        0 & 0 & t_1 g^\dagger(\mathbf{k}) & 0 & t_1g(\mathbf{k}) & 0 \\
        0 & 0 & 0 & t_2g(\mathbf{k}) & 0 & t_2g^\dagger(\mathbf{k}) \\
        t_1g(\mathbf{k}) & 0 & 0 & 0 & -2t_1g^\dagger(\mathbf{k}) & 0 \\
        0 & t_2g^\dagger(\mathbf{k}) & 0 & 0 & 0 & -2t_2g(\mathbf{k}) \\
        t_1g^\dagger(\mathbf{k}) & 0 & -2t_1g(\mathbf{k}) & 0 & 0 & 0 \\
        0 & t_2g(\mathbf{k}) & 0 & -2t_2g^\dagger(\mathbf{k}) & 0 & 0 \\
    \end{pmatrix}, 
\end{equation}
where $g(\mathbf{k}) = e^{i\mathbf{k} \cdot \mathbf{a}_1}+e^{i\mathbf{k} \cdot \mathbf{a}_2}+e^{-i\mathbf{k} \cdot \left(\mathbf{a}_1+\mathbf{a}_2\right)}.$ This model corresponds to $\tau_1 \otimes \sigma_0$ and $\tau_1 \otimes \sigma_3.$ In the corresponding effective Hamiltonian in the valley-orbital basis, the energies are given by 
\begin{equation} 
    E = \pm 2t_1, \pm 2t_2.
\end{equation}
When $|t_1| = |t_2|,$ there are near-degenerate states (actually, in the Kekul\'{e} basis, some of these degeneracies are broken), as shown in Fig. \ref{fig:c3_breaking_models_band_structure}(a). For $\tau_2 \otimes \sigma_0$ and $\tau_2 \otimes \sigma_3$, we study the following Hamiltonian
\begin{equation}
\label{eq: model 3}
    \hat{\mathcal{H}}_{\tau_2}(\mathbf{k}) = \hat{\mathcal{H}}_0(\mathbf{k}) +\frac{1}{3} \begin{pmatrix}
        0 & 0 & t_1g^\dagger(\mathbf{k}) & 0 & -t_1 g(\mathbf{k}) & 0 \\
        0 & 0 & 0 & t_2g(\mathbf{k}) & 0 & -t_2g^\dagger(\mathbf{k}) \\
        t_1g(\mathbf{k}) & 0 & 0 & 0 & 0 & 0 \\
        0 & t_2g^\dagger(\mathbf{k}) & 0 & 0 & 0 & 0 \\
        -t_1g^\dagger(\mathbf{k}) & 0 & 0 & 0 & 0 & 0 \\
        0 & -t_2g(\mathbf{k}) & 0 & 0 & 0 & 0 \\
    \end{pmatrix}.
\end{equation}
The band structures are shown in Fig. \ref{fig:c3_breaking_models_band_structure}(b).

\begin{figure}
    \centering
    \includegraphics[width=\linewidth]{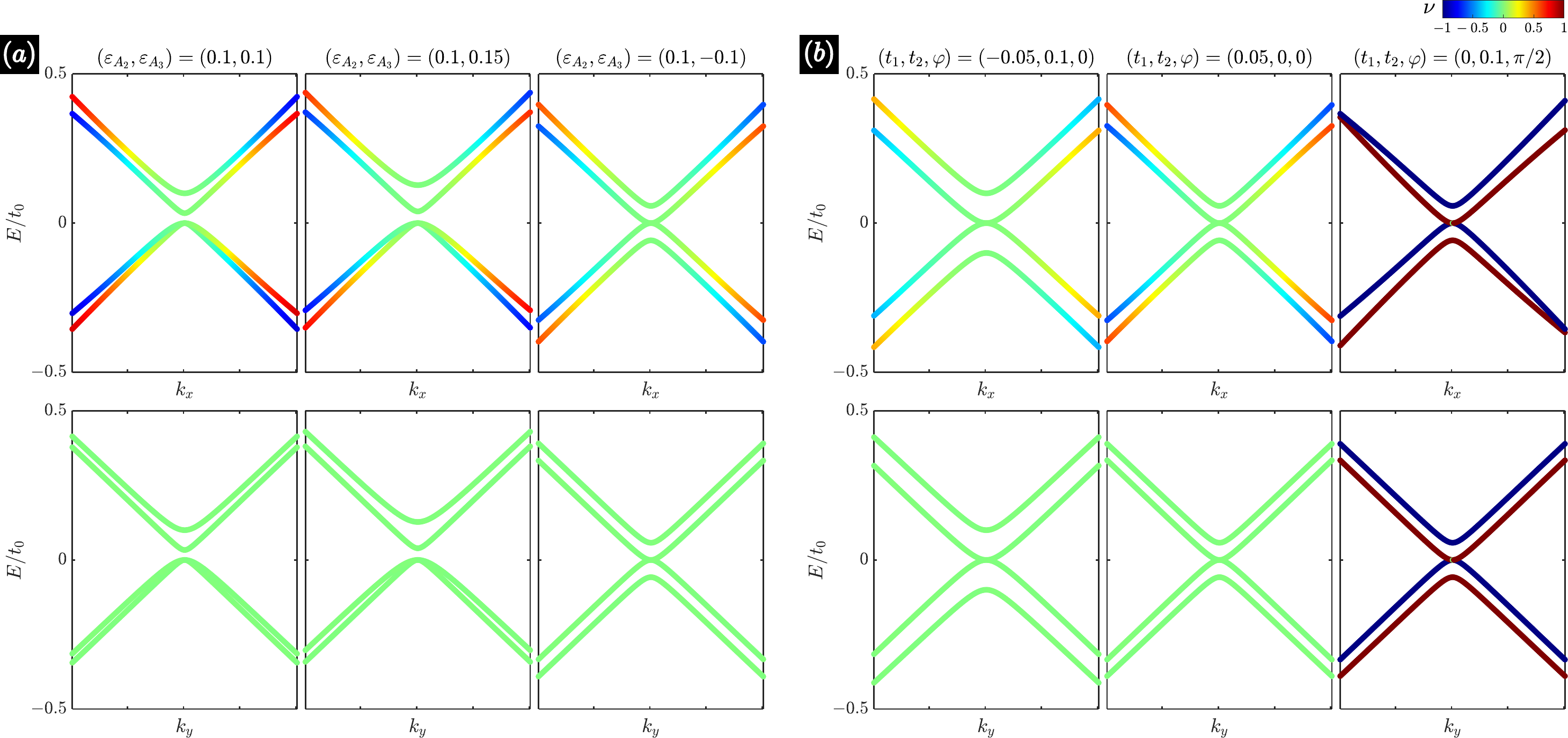}
    \caption{\textbf{Band structures for the models defined in Eqs. \eqref{eq: model 4} and \eqref{eq: model 5}.} (a) On-site model defined in Eq. \eqref{eq: model 4}. (b) Hopping model defined in Eq. \eqref{eq: model 5}.}
    \label{fig:site_center_C3_models}
\end{figure}

Finally, we consider two models that respect $C_3(\vartriangle).$ The first model is simply a charge modulation on the $A$ sublattice around the Kekul\'{e} unit cell:
\begin{equation}
\label{eq: model 4}
    \hat{\mathcal{H}}_{\tau_i\otimes(\sigma_0+\sigma_3)}(\mathbf{k}) = \hat{\mathcal{H}}_0(\mathbf{k}) + \begin{pmatrix}
        0 & 0 & 0 & 0 & 0 & 0\\
        0 & 0 & 0 & 0 & 0 & 0\\
        0 & 0 & \varepsilon_{A_2} & 0 & 0 & 0\\
        0 & 0 & 0 & 0 & 0 & 0\\
        0 & 0 & 0 & 0 & \varepsilon_{A_3} & 0\\
        0 & 0 & 0 & 0 & 0 & 0\\
    \end{pmatrix}.
\end{equation}
In this model, $\varepsilon_{A_2}+\varepsilon_{A_3}$ is proportional to $\tau_1 \otimes \left(\sigma_0+\sigma_3 \right)$ while $\varepsilon_{A_2}-\varepsilon_{A_3}$ is proportional to $\tau_2 \otimes \left(\sigma_0+\sigma_3 \right)$. Some band structures are shown in Fig. \ref{fig:site_center_C3_models}(a). The second model requires modulation of bonds between atoms of the $A$ sublattice
\begin{equation}
\label{eq: model 5}
    \hat{\mathcal{H}}_{\tau_i\otimes(\sigma_0+\sigma_3)}(\mathbf{k}) = \hat{\mathcal{H}}_0(\mathbf{k}) + \frac{1}{3}\begin{pmatrix}
        0 & 0 & \left(-t_1-t_2\cos\varphi\right)g^\dagger(\mathbf{k}) & 0 & t_1g(\mathbf{k}) & 0\\
        0 & 0 & 0 & 0 & 0 & 0\\
        \left(-t_1-t_2\cos\varphi\right)g(\mathbf{k}) & 0 & 0 & 0 & t_2e^{i\varphi}g^\dagger(\mathbf{k}) & 0\\
        0 & 0 & 0 & 0 & 0 & 0\\
        t_1 g^\dagger(\mathbf{k}) & 0 & t_2e^{-i\varphi}g(\mathbf{k}) & 0 & 0 & 0\\
        0 & 0 & 0 & 0 & 0 & 0\\
    \end{pmatrix}.
\end{equation}
This model allows us to tune continuously from $\tau_1 \otimes (\sigma_0+\sigma_3)$ to $\tau_2\otimes (\sigma_0+\sigma_3)$ to $\tau_3\otimes (\sigma_0+\sigma_3)$ by varying the three available parameters. $\tau_1 \otimes (\sigma_0+\sigma_3)$ is isolated when $\varphi = 0$ and $2t_1+t_2 = 0,$ $\tau_2 \otimes (\sigma_0+\sigma_3)$ is isolated when $t_2 = 0,$ and $\tau_3 \otimes (\sigma_0+\sigma_3)$ is isolated when $\varphi = \pi/2$ and $t_1 = 0.$ All three cases are shown in Fig. \ref{fig:site_center_C3_models}(b).

\subsection{Multilayer Models}

\subsubsection{General Symmetry Considerations}

\begin{figure}
    \centering
    \includegraphics[width=0.7\linewidth]{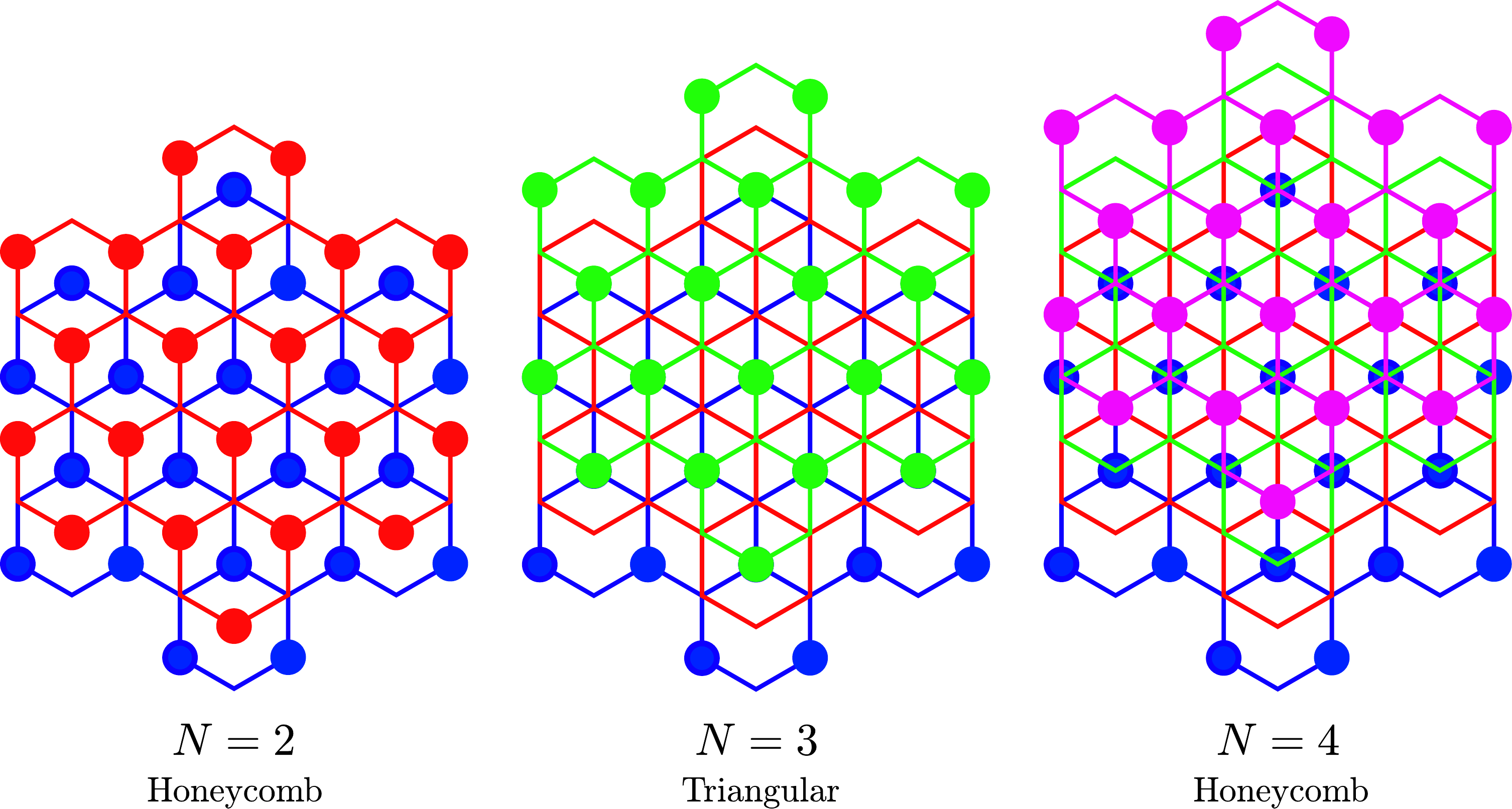}
    \caption{\textbf{Sublattice structure for rhombohedral multilayer stacks.} For $N$ even, the structures are honeycomb, while for $N$ odd, the structures are triangular.}
    \label{fig:multilayer_schematic}
\end{figure}

To generalize our results in Sec. \ref{sec: Monolayer Toy Models} to $N$-layer rhombohedral graphene, we first rewrite the tight-binding Hamiltonian in the reconstructed Kekul\'{e} zone. We use the same lattice structure for layer 1 as shown in Fig. \ref{fig:sqrt(3)sqrt(3)}(a). Layer 2 has exactly the same lattice structure as layer 1 but shifted by $\left(0,a/\sqrt{3}\right)$ laterally in the $y$-direction. In a similar way, every layer $\ell$ is shifted by $\left(0,a/\sqrt{3}\right)$ relative to layer $\ell-1,$ as shown in Fig. \ref{fig:multilayer_schematic}. The valley states for layer $\ell$ are
\begin{equation}
\label{eq: Kekula basis 2}
    \ket{\psi_{A^{(\ell)}}(\mathbf{K})} = \frac{\mathbb{1}_\ell}{\sqrt{3}} \otimes \begin{pmatrix}
        1 \\ 0 \\ \omega \\ 0 \\ \omega^\dagger \\ 0 
    \end{pmatrix}, \quad     \ket{\psi_{B^{(\ell)}}(\mathbf{K})} = \frac{\mathbb{1}_\ell}{\sqrt{3}} \otimes \begin{pmatrix}
        0 \\ 1 \\ 0 \\ \omega^\dagger \\ 0 \\ \omega
    \end{pmatrix}\ \quad \ket{\psi_{A^{(\ell)}}(\mathbf{K}')} = \frac{\mathbb{1}_\ell}{\sqrt{3}} \otimes \begin{pmatrix}
        1 \\ 0 \\ \omega^\dagger  \\ 0 \\ \omega\\ 0 
    \end{pmatrix}, \quad     \ket{\psi_{B^{(\ell)}}(\mathbf{K}')} = \frac{\mathbb{1}_\ell}{\sqrt{3}} \otimes \begin{pmatrix}
        0 \\ 1 \\ 0 \\ \omega \\ 0 \\ \omega^\dagger
    \end{pmatrix},
\end{equation}
where $\mathbb{1}_\ell$ is a vector of size $N\times1$ with one on row $\ell$ and zero elsewhere. The valley states on each layer are identical to each other because a shift in the $y$ direction does not acquire any additional phase since it is orthogonal to the valley vectors $\mathbf{K}_\tau.$ In the Kekul\'{e} basis, the only uneclipsed sites are $A_1^{(1)},$ $A_2^{(1)},$ $A_3^{(1)},$ $B_1^{(N)},$ $B_2^{(N)},$ and $B_3^{(N)}.$ Threefold rotation centered on $A_1^{(1)}$ (or equivalently, on $A_2^{(1)}$ or on $A_3^{(1)}$) affects different layers in disparate fashions because of the relative shifts of rotation centers. On layer 1, it is implemented by the same lateral operator $\hat{C}_3(\vartriangle)$ as in the monolayer case. On layer 2, the rotation center is  located at a hexagon center; consequently, this rotation on layer 2 is implemented by $\hat{C}_3(\varhexagon).$ On layer 3, the rotation center is on the $B$ sublattice; therefore, the action of rotation on this layer is $\hat{C}_3(\triangledown).$ On layers 4 and beyond, the rotation centers cycle through the same ordered sequence as described for layers 1-3. In total, the rotation operator for an $N$-layer stack about an axis going through $A_1^{(1)}$ is
\begin{equation}
    \hat{C}_3(\vartriangle_N) = \begin{pmatrix}
        \hat{C}_3(\vartriangle) & 0 & 0 & 0 & \dots \\
        0 & \hat{C}_3(\varhexagon) & 0 & 0 & \dots \\
        0 & 0 & \hat{C}_3(\triangledown) & 0 & \dots \\
        0 & 0 & 0 & \hat{C}_3(\vartriangle) & \dots \\
        \vdots & \vdots & \vdots & \vdots & \ddots.
    \end{pmatrix}
\end{equation}
As is well known, the low-energy physics is dominated by states localized primarily on $A^{(1)}$ and $B^{(N)}.$ Projecting to just this subspace, we obtain the following representations for $C_3(\vartriangle_N)$ in the valley-orbital-layer basis $\lbrace KA^{(1)},KB^{(N)},K'A^{(1)},K'B^{(N)}\rbrace.$
\begin{equation}
         \hat{C}_3(\vartriangle_N) = \left\{\begin{array}{lr}
        \begin{pmatrix}
            1 & 0 & 0 & 0 \\
            0 & \omega & 0 & 0 \\
            0 & 0 & 1 & 0 \\
            0 & 0 & 0 & \omega^\dagger
        \end{pmatrix}, & \text{if } \mod(N,3)=1,\\
        \begin{pmatrix}
            1 & 0 & 0 & 0 \\
            0 & \omega^\dagger & 0 & 0 \\
            0 & 0 & 1 & 0 \\
            0 & 0 & 0 & \omega
        \end{pmatrix}, & \text{if } \mod(N,3)=2,\\
        \begin{pmatrix}
            1 & 0 & 0 & 0 \\
            0 & 1 & 0 & 0 \\
            0 & 0 & 1 & 0 \\
            0 & 0 & 0 & 1
        \end{pmatrix}, & \text{if } \mod(N,3)=0.\\
        \end{array}\right.
\end{equation}
For $N = 4,7,10,...,$ the projected representation of $C_3(\Delta_N)$ looks exactly like $C_3(\vartriangle)$ in the monolayer case. Inspecting the $N=4$ case in Fig. \ref{fig:multilayer_schematic}, we find that the effective degrees of freedom, shown by solid circles, populate the vertices of a honeycomb lattice in the exact same orientation as in the monolayer configuration. Therefore, it must be the case that the two representations of $C_3(\Delta_N)$ agree. For $N = 2,5,8,...,$ the projected representation of $C_3(\Delta_N)$ looks exactly like $C_3(\triangledown)$ in the monolayer case (if the two sublattices were switched). Again, referring to $N=2$ in Fig. \ref{fig:multilayer_schematic}, we see that the effective degrees of freedom are located at the vertices of a honeycomb lattice but \textit{rotated by} $180^\circ$ relative to the monolayer lattice of Sec. \ref{sec: Monolayer Toy Models}. For $N = 3,6,9,...,$ $\hat{C}_3(\vartriangle_N)$ is the identity because the effective lattice is triangular, not honeycomb, as shown for $N=3$ in Fig.  \ref{fig:multilayer_schematic}. 

Spinless time-reversal symmetry is a completely local operation. So, it does not couple to the sublattice or layer degrees of freedom; it simply takes every state at $K$ to the corresponding state at $K'.$ Therefore, its representation in the multilayer generalization remains the same as before
\begin{equation}
    \hat{\mathcal{T}} = \tau_1 \mathcal{K}.
\end{equation}
With these considerations, we can classify all possible perturbations at the reconstructed $\overline{\Gamma}$ point of rhombohedral $N$-layer graphene that are invariant under $\hat{C}_3(\vartriangle_N)$ and $\hat{\mathcal{T}},$ as listed in Table \ref{tab:valley-orbital operators in N layers}. For $\mod(N,3)=1$ or $2,$ the allowed terms are the same as those found before in the monolayer case. Interestingly, for $\mod(N,3) = 0,$ all 15 nontrivial coupling terms $\tau_i \otimes \sigma_j$ are possible because of the trivial action of $C_3(\vartriangle_N)$ on this projected basis, i.e. it is the identity. In particular, in this setting, it is possible to have $\mathcal{T}$-breaking terms that mix valleys: $\tau_1\otimes \sigma_2$ and $\tau_2\otimes \sigma_2.$ These terms must necessarily have sublattice coherence, which \textit{might} be difficult to achieve because the top and bottom layers can be far apart. We postpone the consideration of these terms to later analysis. In this section, we focus on the layer-projected $\tau_1\otimes \left( \sigma_0 + \sigma_3 \right),$ $\tau_2\otimes \left( \sigma_0 + \sigma_3 \right),$ and $\tau_3\otimes \left( \sigma_0 + \sigma_3 \right),$ which are consistent with $C_3(\Delta_N)$ symmetry regardless of the number of layers.

\begin{table}[]
\begin{center}
\begin{tabular}{c |c |c} 
 \hline
  $\mod(N,3)=0$& $\mathcal{T}$ even & $\mathcal{T}$ odd\\
 \hline
 Intravalley & $\sigma_3,\sigma_1, \tau_3\otimes\sigma_2$ & $\sigma_2,\tau_3,\tau_3\otimes \sigma_3, \tau_3\otimes\sigma_1$ \\
 Intervalley & $\tau_1,\tau_2,\tau_1\otimes \sigma_1,\tau_1\otimes \sigma_3,\tau_2\otimes\sigma_1,\tau_2\otimes\sigma_3$ & $\tau_1\otimes \sigma_2,\tau_2\otimes \sigma_2$ \\
 \hline
 \hline
  $\mod(N,3)=1$ or $2$ & $\mathcal{T}$ even & $\mathcal{T}$ odd\\
 \hline
 Intravalley & $\sigma_3$ & $\tau_3,\tau_3\otimes \sigma_3$ \\
 Intervalley & $\tau_1 \otimes (\sigma_0 +\sigma_3),\tau_2 \otimes (\sigma_0 + \sigma_3)$ & none \\
 \hline

 \end{tabular}
\end{center}
    \caption{\textbf{Symmetry classification of operators in valley-orbital space for rhombohedral $N$-layer graphene at the $\overline{\Gamma}$ point in the reconstructed Kekul\'{e} lattice.} $\mathcal{T}$ is spinless time-reversal symmetry.}
    \label{tab:valley-orbital operators in N layers}
\end{table}

\subsubsection{Full $N$-Layer Models in the Kekul\'{e} Basis}
\label{sec: Full $N$-Layer Models in the Kekule Basis}

The full Hamiltonian is built from  sub-Hamiltonians in the following way
\begin{equation}
\label{eq: pristine Hamiltonian}
    \hat{\mathcal{H}}(\mathbf{k}) = \begin{pmatrix}
        \hat{\mathcal{K}}(\mathbf{k}) & \hat{\mathcal{U}}_1(\mathbf{k}) & \hat{\mathcal{U}}_2 & 0 & \dots & 0 \\
        \hat{\mathcal{U}}_1^\dagger(\mathbf{k}) & \hat{\mathcal{K}}(\mathbf{k}) & \hat{\mathcal{U}}_1(\mathbf{k}) & \hat{\mathcal{U}}_2 & \dots & 0 \\
        \hat{\mathcal{U}}_2^\dagger & \hat{\mathcal{U}}_1^\dagger(\mathbf{k}) & \hat{\mathcal{K}}(\mathbf{k}) & \hat{\mathcal{U}}_1(\mathbf{k})  & \dots & 0 \\
        0 & \hat{\mathcal{U}}_2^\dagger & \hat{\mathcal{U}_1^\dagger}(\mathbf{k}) & \hat{\mathcal{K}}(\mathbf{k}) & \dots & 0 \\
        \vdots & \vdots & \vdots& \vdots & \ddots & \vdots \\
        0 & 0 & 0 & 0 & \dots &\hat{\mathcal{K}}(\mathbf{k})
    \end{pmatrix} + \Delta \begin{pmatrix}
        \frac{N-1}{2} & 0 & 0 & 0 & \dots & 0 \\
        0 & \frac{N-3}{2} & 0 & 0 & \dots & 0 \\
        0 & 0 & \frac{N-5}{2} & 0 & \dots & 0 \\
        0 & 0 & 0 & \frac{N-7}{2} & \dots & 0 \\
        \vdots & \vdots & \vdots& \vdots & \ddots & \vdots \\
        0 & 0 & 0 & 0 & \dots &-\frac{N-1}{2}
    \end{pmatrix},
\end{equation}
where $\hat{\mathcal{K}}(\mathbf{k})$ is the in-plane kinetic energy, which is identical to $\hat{\mathcal{H}}_0$ given in Eq. \eqref{eq: unperturbed monolayer Hamiltonian} and repeated here for convenience,
\begin{equation}
    \hat{\mathcal{K}}(\mathbf{k}) = - \gamma_0\begin{pmatrix}
        0 &  e^{i \mathbf{k} \cdot \boldsymbol{\delta}_1} & 0 &   e^{i \mathbf{k} \cdot \boldsymbol{\delta}_2} & 0 &   e^{i \mathbf{k} \cdot \boldsymbol{\delta}_3} \\
          e^{-i \mathbf{k} \cdot \boldsymbol{\delta}_1} & 0 &   e^{-i \mathbf{k} \cdot \boldsymbol{\delta}_2} &0 &   e^{-i \mathbf{k} \cdot \boldsymbol{\delta}_3} & 0 \\
        0 &   e^{i \mathbf{k} \cdot \boldsymbol{\delta}_2} & 0 &   e^{i \mathbf{k} \cdot \boldsymbol{\delta}_3} & 0 &   e^{i \mathbf{k} \cdot \boldsymbol{\delta}_1} \\
          e^{-i \mathbf{k} \cdot \boldsymbol{\delta}_2} & 0 &   e^{-i \mathbf{k} \cdot \boldsymbol{\delta}_3} & 0 &   e^{-i \mathbf{k} \cdot \boldsymbol{\delta}_1} & 0 \\
        0 &   e^{i \mathbf{k} \cdot \boldsymbol{\delta}_3} & 0 &   e^{i \mathbf{k} \cdot \boldsymbol{\delta}_1} & 0 &   e^{i \mathbf{k} \cdot \boldsymbol{\delta}_2} \\
          e^{-i \mathbf{k} \cdot \boldsymbol{\delta}_3} & 0 &   e^{-i \mathbf{k} \cdot \boldsymbol{\delta}_1} & 0 &   e^{-i \mathbf{k} \cdot \boldsymbol{\delta}_2} & 0
    \end{pmatrix},
\end{equation}
$\hat{\mathcal{U}}_1$ is the interlayer hopping matrix between adjacent layers, 
\begin{equation}
        \hat{\mathcal{U}}_1(\mathbf{k}) = \begin{pmatrix}
        -\gamma_4 e^{i\mathbf{k} \cdot \boldsymbol{\delta}_1} & -\gamma_3 e^{-i\mathbf{k} \cdot \boldsymbol{\delta}_1} & -\gamma_4 e^{i\mathbf{k} \cdot \boldsymbol{\delta}_3} & -\gamma_3 e^{-i\mathbf{k} \cdot \boldsymbol{\delta}_3} & -\gamma_4 e^{i\mathbf{k} \cdot \boldsymbol{\delta}_2} & -\gamma_3 e^{-i\mathbf{k} \cdot \boldsymbol{\delta}_2} \\
        \gamma_1 & -\gamma_4 e^{i\mathbf{k} \cdot \boldsymbol{\delta}_1} & 0 & -\gamma_4 e^{i\mathbf{k} \cdot \boldsymbol{\delta}_2} & 0 & -\gamma_4 e^{i\mathbf{k} \cdot \boldsymbol{\delta}_3} \\
        -\gamma_4 e^{i \mathbf{k} \cdot \boldsymbol{\delta}_2} & -\gamma_3 e^{-i \mathbf{k} \cdot \boldsymbol{\delta}_3} & -\gamma_4 e^{i \mathbf{k} \cdot \boldsymbol{\delta}_1} & -\gamma_3 e^{-i \mathbf{k} \cdot \boldsymbol{\delta}_2} & -\gamma_4 e^{i \mathbf{k} \cdot \boldsymbol{\delta}_3} & -\gamma_3 e^{-i \mathbf{k} \cdot \boldsymbol{\delta}_1} \\
        0 & -\gamma_4 e^{i\mathbf{k}\cdot \boldsymbol{\delta}_3} & 0 & -\gamma_4 e^{i\mathbf{k}\cdot \boldsymbol{\delta}_1} & \gamma_1 & -\gamma_4 e^{i\mathbf{k}\cdot \boldsymbol{\delta}_2} \\
        -\gamma_4 e^{i\mathbf{k}\cdot \boldsymbol{\delta}_3} & -\gamma_3 e^{-i\mathbf{k}\cdot \boldsymbol{\delta}_2} & -\gamma_4 e^{i\mathbf{k}\cdot \boldsymbol{\delta}_2} & -\gamma_3 e^{-i\mathbf{k}\cdot \boldsymbol{\delta}_1} & -\gamma_4 e^{i\mathbf{k}\cdot \boldsymbol{\delta}_1} & -\gamma_3 e^{-i\mathbf{k}\cdot \boldsymbol{\delta}_3} \\
        0 & -\gamma_4 e^{i\mathbf{k}\cdot \boldsymbol{\delta}_2} & \gamma_1 & -\gamma_4 e^{i\mathbf{k}\cdot \boldsymbol{\delta}_3} & 0 & -\gamma_4 e^{i\mathbf{k}\cdot \boldsymbol{\delta}_1} 
    \end{pmatrix},
\end{equation}
and $\hat{\mathcal{U}}_2$ is the interlayer hopping matrix between layers that are separated by another in the middle,
\begin{equation}
    \hat{\mathcal{U}}_2 = \begin{pmatrix}
        0 & \gamma_2/2 & 0 & 0 & 0 & 0 \\
        0 & 0 & 0 & 0 & 0 & 0 \\
        0 & 0 & 0 & 0 & 0 & \gamma_2/2 \\
        0 & 0 & 0 & 0 & 0 & 0 \\
        0 & 0 & 0 & \gamma_2/2 & 0 & 0 \\
        0 & 0 & 0 & 0 & 0 & 0 
    \end{pmatrix}.
\end{equation}
Here, $\Delta$ is the field-induced layer potential energy and $\gamma_i$ are the hopping parameters. We ignore various on-site energies due to dimerization since they do not qualitatively matter when the displacement field is reasonably large. To demonstrate that this Hamiltonian respects $C_3(\vartriangle_N),$ we have verified the following identity
\begin{equation}
\begin{split}
    &\begin{pmatrix}
        \hat{C}_3(\vartriangle)^\dagger & 0 & 0 & 0 & \dots \\
        0 & \hat{C}_3(\varhexagon)^\dagger & 0 & 0 & \dots \\
        0 & 0 & \hat{C}_3(\triangledown)^\dagger & 0 & \dots \\
        0 & 0 & 0 & \hat{C}_3(\vartriangle)^\dagger & \dots \\
        \vdots & \vdots & \vdots & \vdots & \ddots.
    \end{pmatrix}\begin{pmatrix}
        \hat{\mathcal{K}}(\mathbf{k}) & \hat{\mathcal{U}}_1(\mathbf{k}) & \hat{\mathcal{U}}_2 & 0 & \dots  \\
        \hat{\mathcal{U}}_1^\dagger(\mathbf{k}) & \hat{\mathcal{K}}(\mathbf{k}) & \hat{\mathcal{U}}_1(\mathbf{k}) & \hat{\mathcal{U}}_2 & \dots \\
        \hat{\mathcal{U}}_2^\dagger & \hat{\mathcal{U}}_1^\dagger(\mathbf{k}) & \hat{\mathcal{K}}(\mathbf{k}) & \hat{\mathcal{U}}_1(\mathbf{k})  & \dots  \\
        0 & \hat{\mathcal{U}}_2^\dagger & \hat{\mathcal{U}_1^\dagger}(\mathbf{k}) & \hat{\mathcal{K}}(\mathbf{k}) & \dots  \\
        \vdots & \vdots & \vdots& \vdots & \ddots & 
    \end{pmatrix}\begin{pmatrix}
        \hat{C}_3(\vartriangle) & 0 & 0 & 0 & \dots \\
        0 & \hat{C}_3(\varhexagon) & 0 & 0 & \dots \\
        0 & 0 & \hat{C}_3(\triangledown) & 0 & \dots \\
        0 & 0 & 0 & \hat{C}_3(\vartriangle) & \dots \\
        \vdots & \vdots & \vdots & \vdots & \ddots.
    \end{pmatrix}    \\
    =&\begin{pmatrix}
        \hat{C}_3(\vartriangle)^\dagger\hat{\mathcal{K}}(\mathbf{k})\hat{C}_3(\vartriangle) & \hat{C}_3(\vartriangle)^\dagger\hat{\mathcal{U}}_1(\mathbf{k})\hat{C}_3(\varhexagon) & \hat{C}_3(\vartriangle)^\dagger\hat{\mathcal{U}}_2 \hat{C}_3(\triangledown)  & 0 & \dots  \\
        \hat{C}_3(\varhexagon)^\dagger\hat{\mathcal{U}}_1^\dagger(\mathbf{k})\hat{C}_3(\vartriangle) & \hat{C}_3(\varhexagon)^\dagger\hat{\mathcal{K}}(\mathbf{k})\hat{C}_3(\varhexagon) & \hat{C}_3(\varhexagon)^\dagger\hat{\mathcal{U}}_1(\mathbf{k}) \hat{C}_3(\triangledown)  & \hat{C}_3(\varhexagon)^\dagger\hat{\mathcal{U}}_2\hat{C}_3(\vartriangle) & \dots \\
        \hat{C}_3(\triangledown)^\dagger\hat{\mathcal{U}}_2^\dagger\hat{C}_3(\vartriangle) & \hat{C}_3(\triangledown)^\dagger\hat{\mathcal{U}}_1^\dagger(\mathbf{k})\hat{C}_3(\varhexagon) & \hat{C}_3(\triangledown)^\dagger\hat{\mathcal{K}}(\mathbf{k}) \hat{C}_3(\triangledown)  & \hat{C}_3(\triangledown)^\dagger\hat{\mathcal{U}}_1(\mathbf{k})\hat{C}_3(\vartriangle)  & \dots  \\
        0 & \hat{C}_3(\vartriangle)^\dagger\hat{\mathcal{U}}_2^\dagger\hat{C}_3(\varhexagon) & \hat{C}_3(\vartriangle)^\dagger\hat{\mathcal{U}_1^\dagger}(\mathbf{k}) \hat{C}_3(\triangledown)  & \hat{C}_3(\vartriangle)^\dagger\hat{\mathcal{K}}(\mathbf{k})\hat{C}_3(\vartriangle) & \dots  \\
        \vdots & \vdots & \vdots& \vdots & \ddots & 
    \end{pmatrix}\\
    =&\begin{pmatrix}
        \hat{\mathcal{K}}(R_3\mathbf{k}) & \hat{\mathcal{U}}_1(R_3\mathbf{k}) & \hat{\mathcal{U}}_2 & 0 & \dots  \\
        \hat{\mathcal{U}}_1^\dagger(R_3\mathbf{k}) & \hat{\mathcal{K}}(R_3\mathbf{k}) & \hat{\mathcal{U}}_1(R_3\mathbf{k}) & \hat{\mathcal{U}}_2 & \dots \\
        \hat{\mathcal{U}}_2^\dagger & \hat{\mathcal{U}}_1^\dagger(R_3\mathbf{k}) & \hat{\mathcal{K}}(R_3\mathbf{k}) & \hat{\mathcal{U}}_1(R_3\mathbf{k})  & \dots  \\
        0 & \hat{\mathcal{U}}_2^\dagger & \hat{\mathcal{U}_1^\dagger}(R_3\mathbf{k}) & \hat{\mathcal{K}}(R_3\mathbf{k}) & \dots  \\
        \vdots & \vdots & \vdots& \vdots & \ddots & 
    \end{pmatrix},
\end{split}
\end{equation}
which simplifies to confirming the following equalities:
\begin{equation}
    \begin{split}
        \hat{C}_3(\vartriangle)^\dagger\hat{\mathcal{K}}(\mathbf{k})\hat{C}_3(\vartriangle) &= \hat{\mathcal{K}}(R_3\mathbf{k}), \quad \hat{C}_3(\varhexagon)^\dagger\hat{\mathcal{K}}(\mathbf{k})\hat{C}_3(\varhexagon) = \hat{\mathcal{K}}(R_3\mathbf{k}), \quad \hat{C}_3(\triangledown)^\dagger\hat{\mathcal{K}}(\mathbf{k})\hat{C}_3(\triangledown) = \hat{\mathcal{K}}(R_3\mathbf{k}),\\
        \hat{C}_3(\vartriangle)^\dagger\hat{\mathcal{U}}_1(\mathbf{k})\hat{C}_3(\varhexagon) &= \hat{\mathcal{U}}_1(R_3\mathbf{k}), \quad \hat{C}_3(\varhexagon)^\dagger\hat{\mathcal{U}}_1(\mathbf{k})\hat{C}_3(\triangledown) = \hat{\mathcal{U}}_1(R_3\mathbf{k}), \quad \hat{C}_3(\triangledown)^\dagger\hat{\mathcal{U}}_1(\mathbf{k})\hat{C}_3(\vartriangle) = \hat{\mathcal{U}}_1(R_3\mathbf{k}),\\
        \hat{C}_3(\vartriangle)^\dagger\hat{\mathcal{U}}_2\hat{C}_3(\triangledown) &= \hat{\mathcal{U}}_2, \quad \hat{C}_3(\varhexagon)^\dagger\hat{\mathcal{U}}_2\hat{C}_3(\vartriangle) = \hat{\mathcal{U}}_2, \quad \hat{C}_3(\triangledown)^\dagger\hat{\mathcal{U}}_2\hat{C}_3(\varhexagon) = \hat{\mathcal{U}}_2,
    \end{split}
\end{equation}
where $R_3$ rotates wavevectors $\mathbf{k}$ by $2\pi/3:$ $R_3(k_x,k_y) = (-k_x+\sqrt{3}k_y,-\sqrt{3}k_x-k_y)/2.$ In the Kekul\'{e} basis, the three different symmetry operators are given elsewhere already but are repeated here for convenience:
\begin{equation}
    \hat{C}_3(\varhexagon) = \begin{pmatrix}
        0 & 0 & 1  & 0 & 0 & 0 \\
        0 & 0 & 0  & 1 & 0 & 0  \\
        0 & 0 & 0  & 0 & 1 & 0 \\
        0 & 0 & 0  & 0 & 0 & 1 \\
        1 & 0 & 0  & 0 & 0 & 0 \\
        0 & 1 & 0  & 0 & 0 & 0
    \end{pmatrix}, \quad \hat{C}_3(\vartriangle) = \begin{pmatrix}
        1 & 0 & 0 & 0 & 0 & 0 \\
        0 & 0 & 0 & 0 & 0 & 1 \\
        0 & 0 & 1 & 0 & 0 & 0 \\
        0 & 1 & 0 & 0 & 0 & 0 \\
        0 & 0 & 0 & 0 & 1 & 0 \\
        0 & 0 & 0 & 1 & 0 & 0 \\
    \end{pmatrix}, \quad \hat{C}_3(\triangledown) = \begin{pmatrix}
        0 & 0 & 0 & 0 & 1 & 0 \\
        0 & 1 & 0 & 0 & 0 & 0 \\
        1 & 0 & 0 & 0 & 0 & 0 \\
        0 & 0 & 0 & 1 & 0 & 0 \\
        0 & 0 & 1 & 0 & 0 & 0 \\
        0 & 0 & 0 & 0 & 0 & 1 \\
    \end{pmatrix}.
\end{equation}
Since the displacement field matrix acts uniformly and locally on each layer, it must commute with the $C_3(\Delta_N)$ symmetry operator. We have therefore checked that the full Hamiltonian is invariant under $C_3$ rotation about $A_1^{(1)}.$

\subsubsection{Band Structures for Realistic Systems}

\begin{table}[t]
    \centering
    \renewcommand{\arraystretch}{1.25}
    \begin{tabular}{c c c c c}
        \hline\hline
        $\gamma_0$ &
        $\gamma_1$ &
        $\gamma_2$ &
        $\gamma_3$ &
        $\gamma_4$ \\
        \hline
        $3100$ & $380$ & $-15$ & $-290$ & $-141$ \\
        \hline\hline
    \end{tabular}
    \caption{\textbf{Tight-binding parameters}. All values are quoted in meV.}
    \label{tab:tight_binding_params}
\end{table}

\begin{figure}
    \centering
    \includegraphics[width=1\linewidth]{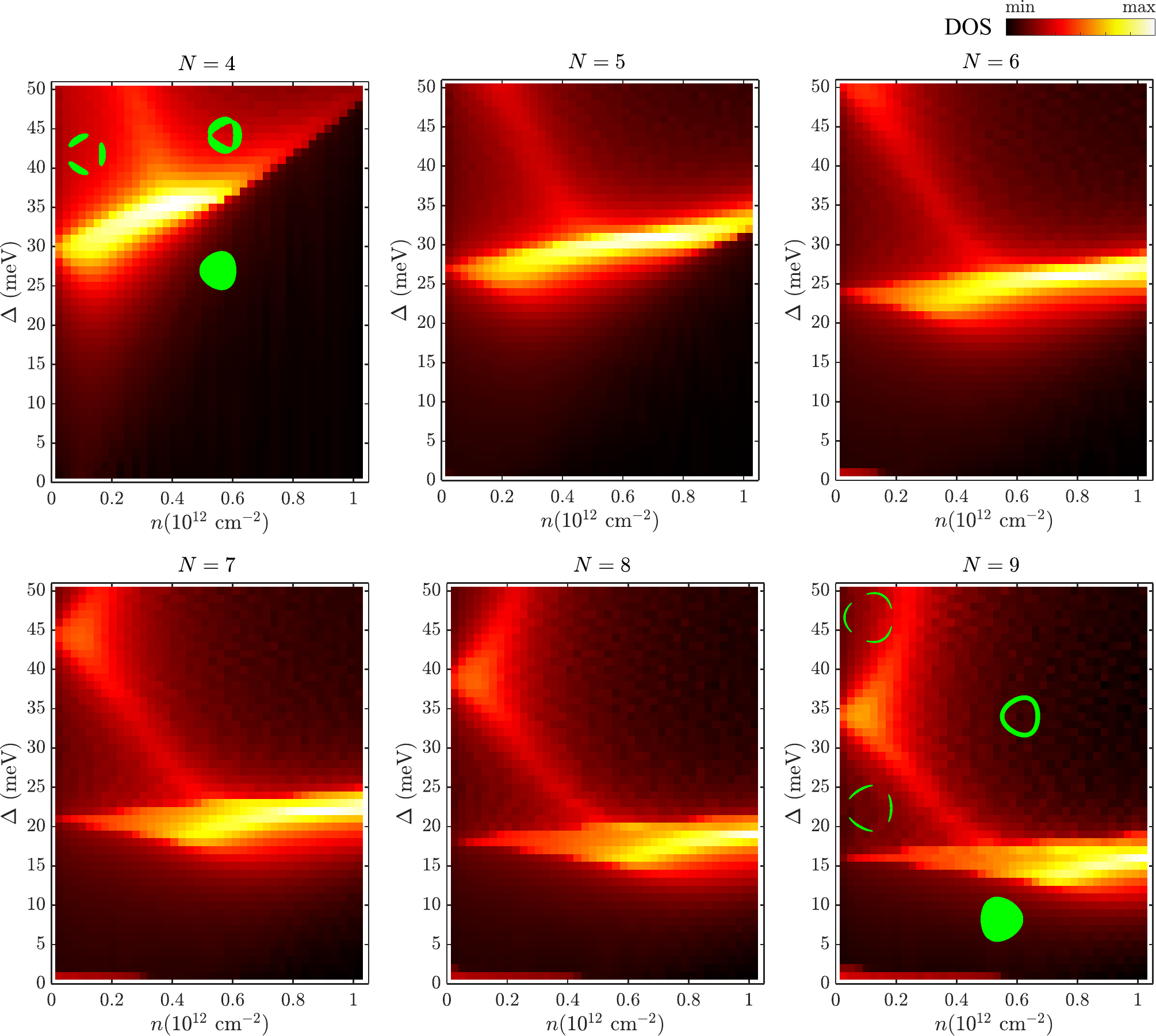}
    \caption{\textbf{Density-of-states maps as a function of electron density (in a single valley-spin) and displacement field.} High-intensity bright regions correspond to Lifshitz transitions where Fermi surface topologies change. Representative Fermi surfaces are shown for $N = 4$ and $N =9,$ with the other layer numbers displaying similar trends.}
    \label{fig:DOS_noninteracting}
\end{figure}

In this section, we study the band structures of realistic models of $N$-layer rhombohedral graphene. We first examine the single-valley Fermi topology without any perturbation. We use the  parameters listed in Table \ref{tab:tight_binding_params} for all of our simulations. Using the non-interacting band structures using Eq. \eqref{eq: pristine Hamiltonian}, we compute the density of states (DOS) as a function of electron density in a single valley-spin flavor (because we are ultimately interested only in the quarter metal phase) and displacement field using the formula
\begin{equation}
    \mathrm{DOS}(E_F) = \frac{1}{\pi} \sum_{n} \frac{\eta}{\eta^2 + (E_F-E_n)^2},
\end{equation}
where $\eta = 0.1$ meV is a numerical broadening factor, $E_F$ is the Fermi energy determined by the charge density, and $E_n$ are the energy eigenvalues. The results for $N = 4,5,6,7,8,9$ are shown in Fig. \ref{fig:DOS_noninteracting}. The DOS maps contain a few prominent peaks associated with diverging DOS due to Lifshitz transitions for all values of $N$ shown. At small values of $\Delta,$ the Fermi surface is primarily that of a simply-connected trigonally-warped geometry for the values of electron density shown across the different numbers of layers. As the displacement field increases, that simply-connected geometry evolves into one of two distinct Fermi surface topologies: a three-pocket structure at small densities or an annular surface at larger densities.  The critical fields where these Liftshitz transitions occur are inversely related to increasing layer number. In between these Lifshitz transitions, there are many more topologically distinct Fermi surface topologies, such as a four-pocket structure, but they require fine-tuning parameters that we do not consider further. For our purpose, we will only consider the three Fermi surface topologies mentioned, although the primary focus is on the simply-connected topology.

\begin{figure}
    \centering
    \includegraphics[width=1\linewidth]{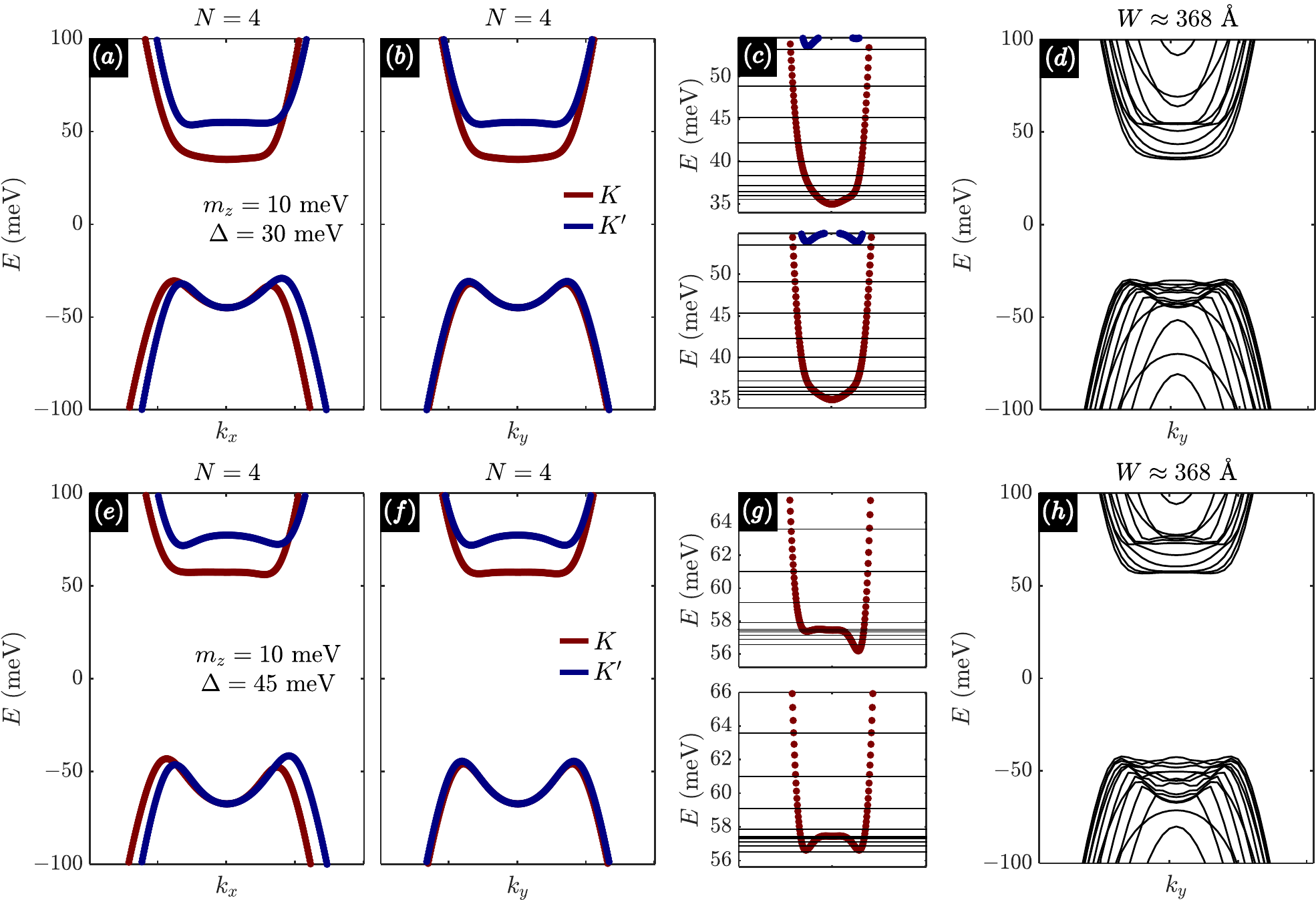}
    \caption{\textbf{Band structures for $N=4$ with intervalley imbalance.} (a,b) Band structures along $k_x$ and $k_y$ for $m_z = 10$ meV and $\Delta = 30$ meV. (c)  we zoom in on the bottoms of the conduction bands using black horizontal lines to show the chemical potentials at $n = 0.1,0.2,...,0.8,0.9,1.0 \times 10^{12}$ cm$^{-2}.$ (d) Band structure of an armchair nanoribbon with width $W = 368$ \AA. (a-d) The top panel shows band structures at $\Delta = 30$ meV, while (e-h) the bottom panel shows band structures at $\Delta = 45$ meV. }
    \label{fig:bandstructure1}
\end{figure}

\begin{figure}
    \centering
    \includegraphics[width=1\linewidth]{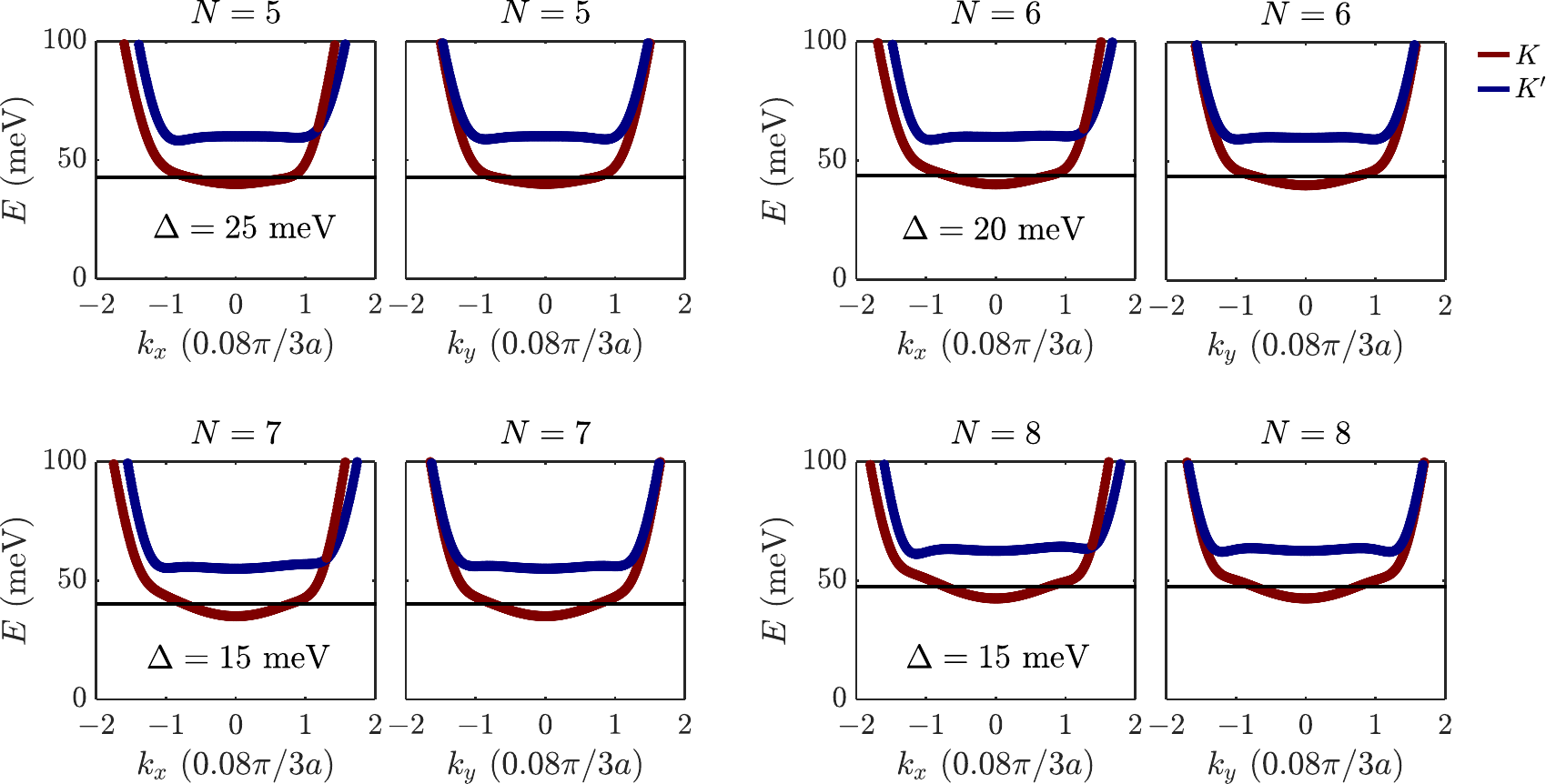}
    \caption{\textbf{Band structures for $N=5-8$ with intervalley imbalance.} For each $N,$ we show the band structure along $k_x$ and $k_y.$ The black horizontal lines indicate the chemical potentials at $n = 5\times10^{11}$ cm$^{-2}.$ In all cases, $m_z = 10$ meV. Here, there is no intervalley interaction.}
    \label{fig:band_structure_increasing_N}
\end{figure}

Next, we add intervalley imbalance and hybridization. We implement intervalley imbalance on the $A^{(1)}$ sublattice to which we assume most of the charge density has been driven by the interlayer displacement field. For intervalley imbalance, we use a version of the Haldane model, which in the Kekul\'{e} basis, takes the following form on layer 1:
\begin{equation}
    \delta\hat{\mathcal{H}}_{\tau_3}(\mathbf{k})  = \frac{m_z}{3\sqrt{3}i}\begin{pmatrix}
        0 & 0 & -g^\dagger(\mathbf{k}) & 0 & g(\mathbf{k}) & 0 \\
        0 & 0 & 0 & 0 & 0 & 0 \\
       g(\mathbf{k}) & 0 & 0 & 0 &  -g^\dagger(\mathbf{k})  & 0 \\
        0 & 0 & 0 & 0 & 0 & 0 \\
       - g^\dagger(\mathbf{k}) & 0 & g(\mathbf{k}) & 0 & 0 & 0 \\
        0 & 0 & 0 & 0 & 0 & 0 
    \end{pmatrix},
\end{equation}
where $g(\mathbf{k}) =  \sum_{i=1}^3 e^{i \mathbf{k} \cdot \mathbf{a}_i}$ as defined before. At the $\bar{\Gamma}$ point, this Hamiltonian generates an intervalley imbalance with magnitude $-m_z \tau_3 \otimes \left( \sigma_0-\sigma_3\right)/2.$ In the valley-sublattice basis, the Hamiltonian takes the following minimal form 
\begin{equation}
    \delta\hat{\mathcal{H}}_{\tau_3}(\mathbf{0})  = \begin{pmatrix}
        \Delta\frac{N-1}{2} - m_z & 0 & 0 & 0 \\
        0 & -\Delta\frac{N-1}{2} & 0 & 0 \\
        0 & 0 & \Delta\frac{N-1}{2}+m_z & 0  \\
        0 & 0 & 0 & -\Delta\frac{N-1}{2}  \\
    \end{pmatrix}.
\end{equation}
To a first approximation, the valence bands are not affected by this perturbation, while the conduction bands are valley-split by an energy amount of $2m_z$ at $\mathbf{k} = \mathbf{0}.$ The situation away from $\bar{\Gamma}$ is much more complicated because the two valleys experience different effective mass terms, which as we have shown earlier, can deform the bands in various different ways to give distinct Fermi surface topologies. To ensure that we are always in the quarter metal phase, $m_z$ needs to be large enough so that the chemical potential lies entirely in a single valley. Using Fig. \ref{fig:DOS_noninteracting} as a guide, we now inspect  some specific combinations of $\Delta$ and $m_z$ to ensure that within the range of density of interest, the Fermi surface lies entirely in a single valley. In Fig. \ref{fig:bandstructure1}, we show the band structures for $N = 4$ using $m_z = 10$ meV showing that this is enough to keep both valleys separated in the range density $n \in \left[ 0,1.0 \right] \times10^{12}$ cm$^{-2}.$ In Fig. \ref{fig:band_structure_increasing_N}, we generalize this to $N = 5-8,$ showing that $m_z = 10$ meV is enough to keep the conduction electrons in only one valley for $n = 5\times10^{11}$ cm$^{-2}$.

Finally, we implement $\tau_1$ and $\tau_2$ with the following
\begin{equation}
\begin{split}
        \delta\hat{\mathcal{H}}_{\tau_1}(\mathbf{k})  &= m \cos \theta\begin{pmatrix}
        0 & 0 & - \frac{1}{2}g^\dagger(\mathbf{k}) & 0 & - \frac{1}{2}g(\mathbf{k}) & 0 \\
        0 & 0 & 0 & 0 & 0 & 0 \\
       - \frac{1}{2}g(\mathbf{k}) & 0 & 0 & 0 &  g^\dagger(\mathbf{k})  & 0 \\
        0 & 0 & 0 & 0 & 0 & 0 \\
       - \frac{1}{2}g^\dagger(\mathbf{k}) & 0 & g(\mathbf{k}) & 0 & 0 & 0 \\
        0 & 0 & 0 & 0 & 0 & 0 
    \end{pmatrix}, \\
    \delta\hat{\mathcal{H}}_{\tau_2}(\mathbf{k})  &= m \sin \theta \begin{pmatrix}
        0 & 0 &  \frac{\sqrt{3}}{2}g^\dagger(\mathbf{k}) & 0 & - \frac{\sqrt{3}}{2}g(\mathbf{k}) & 0 \\
        0 & 0 & 0 & 0 & 0 & 0 \\
       \frac{\sqrt{3}}{2}g(\mathbf{k}) & 0 & 0 & 0 &  0  & 0 \\
        0 & 0 & 0 & 0 & 0 & 0 \\
       - \frac{\sqrt{3}}{2}g^\dagger(\mathbf{k}) & 0 & 0 & 0 & 0 & 0 \\
        0 & 0 & 0 & 0 & 0 & 0 
    \end{pmatrix},
\end{split}
\end{equation}
where $m$ is the magnitude of the perturbation and $\theta$ rotates the $\tau$ matrices.

\section{Transport Along and Across Domain Walls}

\subsection{ Reflections at an Armchair Termination}
\label{sec: Reflections at an Armchair Termination}
\subsubsection{Continuum Description}

We consider the continuum description of an armchair termination for monolayer graphene where the bulk extends to the negative $x$ direction. This calculation will naturally extend to the domain-wall scenario that we will show later. In the bulk, the continuum Hamiltonian is
\begin{equation}
    \hat{\mathcal{H}}_\mathrm{eff}(\mathbf{k}) = \begin{pmatrix}
        \hat{\mathcal{H}}_K(\mathbf{k}) & 0 \\
        0 & \hat{\mathcal{H}}_{K'}(\mathbf{k}) 
    \end{pmatrix} = \begin{pmatrix}
        \varepsilon_K & \hbar v_0 \left( k_x - ik_y \right) & 0 & 0 \\
        \hbar v_0 \left( k_x + ik_y \right) & -\varepsilon_K & 0 & 0 \\
        0 & 0 & \varepsilon_{K'} & \hbar v_0 \left( -k_x - ik_y \right) \\
        0 & 0 & \hbar v_0 \left( -k_x + ik_y \right)& -\varepsilon_{K'}
    \end{pmatrix}.
\end{equation}
In real space, we implement the replacement $k_i = -i \partial_i.$ We place the armchair edge at $\mathbf{r} = \left(0, y\right)$ and impose the boundary condition $\psi_K(0,y) = \psi_{K'}(0,y).$ By mapping $k_x \mapsto -k_x$ in the $K'$ sector, we can map the two-valley wavefunction in the half-plane to a one-valley wavefunction in the entire plane where the boundary condition connecting the two valleys in the former representation is converted to a continuity requirement of the one-valley wavefunction at the origin in the latter representation. Because of translational symmetry along the $y$ direction, $k_y$ remains a good quantum number.  For concreteness, let us assume $0 < \varepsilon_K \leq \varepsilon_{K'}.$ This non-essential assumption follows from the more general assumption that the large displacement-field-induced gap is topologically trivial. We use it here only so that we do not have to look for topological edge states. With these simplifications, eigenvalue problem in real space becomes a pair of coupled differential equations
\begin{equation}
    \begin{pmatrix}
        \varepsilon(x) & \hbar v_0 \left( -i\partial_x - ik_y \right) \\
        \hbar v_0 \left( -i\partial_x + ik_y \right) & -\varepsilon(x)
    \end{pmatrix} \begin{pmatrix}
        \psi_A(x) \\ 
        \psi_B(x)
    \end{pmatrix} = E \begin{pmatrix}
        \psi_A(x) \\ 
        \psi_B(x)
    \end{pmatrix},
\end{equation}
 where $\varepsilon(x\leq 0) = \varepsilon_K$ and $\varepsilon(x> 0) = \varepsilon_{K'}.$ We have converted the problem of an armchair boundary into a \textit{scattering} problem where an incoming wave from valley $K$ is either reflected or transmitted. Reflection in the recast problem corresponds to \textit{valley-preserving} reflection (a $K$ state is reflected to a $K$ state) in the original formalism while transmission in the recast problem corresponds to \textit{valley-exchanging} reflection (a $K$ state is reflected to a $K'$ state) in the original formalism.

To begin, let us consider the familiar problem where $\varepsilon_K = \varepsilon_{K'} = \varepsilon.$ We take the following ansatz for the wavefunction 
\begin{equation}
    \begin{pmatrix}
        \psi_A(x\leq 0) \\ 
        \psi_B(x\leq 0)
    \end{pmatrix} = e^{ik_-x}\begin{pmatrix}
        \psi_A^- \\ 
        \psi_B^-
    \end{pmatrix} \quad \text{and} \quad     \begin{pmatrix}
        \psi_A(x> 0) \\ 
        \psi_B(x> 0)
    \end{pmatrix} = e^{ik_+x}\begin{pmatrix}
        \psi_A^+ \\ 
        \psi_B^+
    \end{pmatrix},
\end{equation}
where $k_\pm$ are real assuming that $E \geq \varepsilon.$ The boundary condition is satisfied if 
\begin{equation}
    \begin{pmatrix}
        \psi_A^- \\ 
        \psi_B^-
    \end{pmatrix} = \begin{pmatrix}
        \psi_A^+ \\ 
        \psi_B^+
    \end{pmatrix}.
\end{equation}
Substituting this ansatz into the differential equations, we obtain
\begin{equation}
    \begin{split}
        \left(\varepsilon - E \right) \psi_A^- + \hbar v_0 \left(k_- -ik_y \right) \psi_B^- & = 0,\\
        \hbar v_0 \left(k_- +ik_y \right) \psi_A^- +\left(-\varepsilon - E \right) \psi_B^- & = 0, \\
        \left(\varepsilon - E \right) \psi_A^+ + \hbar v_0 \left(k_+ -ik_y \right) \psi_B^+ & = 0,\\
        \hbar v_0 \left(k_+ +ik_y \right) \psi_A^+ +\left(-\varepsilon - E \right) \psi_B^+ & = 0.
    \end{split}
\end{equation}
We notice that the first two equations are identical to the last two equations; therefore, solutions to the first two are also solutions to the second two. If either the $A$ or the $B$ sublattice wavefunction vanishes, then $E = \pm \varepsilon$ and $k_\pm = k_y = 0,$ e.g. states at the band extrema. For $E \neq 0,$ the following conditions must be true:
\begin{equation}
    \begin{split}
        -\frac{\psi_A^-}{\psi_B^-} & = \frac{\hbar v_0 \left(k_- -ik_y \right)}{\varepsilon - E} = \frac{-\varepsilon - E}{\hbar v_0 \left(k_- +ik_y \right)}, \quad \text{and} \quad
        -\frac{\psi_A^+}{\psi_B^+}  = \frac{\hbar v_0 \left(k_+ -ik_y \right)}{\varepsilon - E} = \frac{-\varepsilon - E}{\hbar v_0 \left(k_+ +ik_y \right)}. 
    \end{split}
\end{equation}
Rearranging, we recover the familiar energy dispersion $E^2 = \varepsilon^2 + (\hbar v_0 k_y)^2 + (\hbar v_0k_\pm)^2.$ The constraint from the boundary condition is $k_+ = k_-.$ This shows that an incoming $K$ wave is completely scattered into a reflected $K'$ wave and vice versa. In other words, when $\varepsilon_K = \varepsilon_{K'},$ scattering at an armchair edge completely exchanges the valley flavor.

The preceding analysis seems to pose a contradiction when $\varepsilon_{K} < \varepsilon_{K'}.$ In the energy window $ \varepsilon_K < E < \varepsilon_{K'},$ there are no propagating states in the $K'$ valley into which to an incoming $K$ wave can scatter. So to where does an incoming $K$ wave scatter? Of course, an incoming $K$ wave can \textit{only} scatter into an outgoing, reflected $K$ wave since these are the only propagating states available in the bulk. In this energy window, there are no \textit{propagating} $K'$ states but there are \textit{evanescent} $K'$ states which act as scattering centers to reflect the incoming $K$ waves into outgoing $K$ waves. For higher energies $E > \varepsilon_{K'} > \varepsilon_{K},$ there are both $K$ and $K'$ waves into which an incoming $K$ wave can scatter; and in general, this reflection process depolarizes the valley character of the outgoing wave. Only in the special case where $\varepsilon_K = \varepsilon_{K'}$ is the reflected wave also valley polarized (but in the opposite valley compared to the incoming wave). We now validate these claims with scattering calculations, which are similar in spirit to the calculation done in Ref. \cite{Takahashi2011Gapless}. 

\begin{figure}
    \centering
    \includegraphics[width=0.5\linewidth]{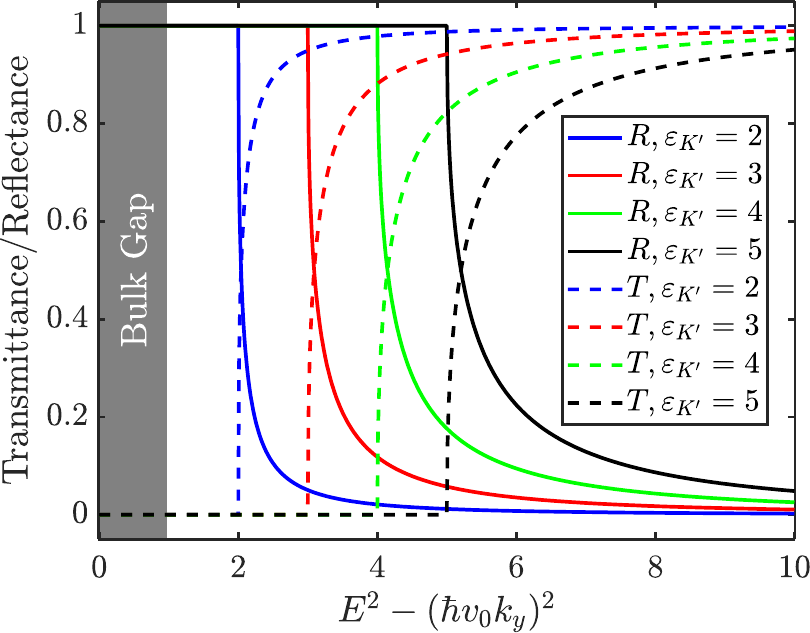}
    \caption{\textbf{Transmittance and reflectance as a function of energy, as defined in Eq. \eqref{eq: R and T of Dirac model}.} Here, $\varepsilon_K = 1$ sets the energy scale. Solid (dashed) lines show the reflectance (transmittance). For energies below $\varepsilon_{K'},$ we have unity reflectance. For large energies, we have near unity transmittance. For energies below $\varepsilon_K,$ neither transmittance nor reflectance is defined since inside the bulk gap, there are propagating states.}
    \label{fig:transmission}
\end{figure}

For $\sqrt{\varepsilon_K^2+\left(\hbar v_0 k_y \right)^2} \leq E < \sqrt{\varepsilon_{K'}^2+\left(\hbar v_0 k_y \right)^2} ,$ assuming positive $E$ throughout, the bulk states are given by 
\begin{equation}
\begin{split}
     \begin{pmatrix}
        \psi_A(x\leq 0) \\ 
        \psi_B(x\leq 0)
    \end{pmatrix} &= e^{ik_xx} \begin{pmatrix}
        \varepsilon_{K} + E \\
        \hbar v_0\left(k_x+ik_y\right)
    \end{pmatrix} +  \mathfrak{r}e^{-ik_xx} \begin{pmatrix}
        \varepsilon_{K} + E \\
        \hbar v_0\left(-k_x+ik_y\right)
    \end{pmatrix} ,  \\
    \begin{pmatrix}
        \psi_A(x> 0) \\ 
        \psi_B(x> 0)
    \end{pmatrix} &= \mathfrak{t}e^{-\kappa_x x} \begin{pmatrix}
        \varepsilon_{{K}'}+ E \\
        i\hbar v_0\left(\kappa_x+k_y\right)
    \end{pmatrix},
\end{split}
\end{equation}
where $E = \sqrt{\varepsilon_K^2 + \hbar^2 v_0^2 k_x^2 + \hbar^2 v_0^2 k_y^2 } =  \sqrt{\varepsilon_{K'}^2 - \hbar^2 v_0^2 \kappa_x^2 + \hbar^2 v_0^2 k_y^2 }.$ Matching the boundary condition, we find
\begin{equation}
    \begin{split}
        \mathfrak{r} &= -1 + \frac{2 (\varepsilon_{K'} + E) k_x}{
    E (i \kappa_x + k_x) + \varepsilon_{K'} (k_x - i k_y) + i \varepsilon_{K} (\kappa_x + k_y)}, \\
  \mathfrak{t} &= -\frac{2 i (\varepsilon_{K} + E) k_x}{
    E (\kappa_x - i k_x) - \varepsilon_{K'}(i  k_x + k_y) + \varepsilon_{K} (\kappa_x + k_y)}.
    \end{split}
\end{equation}
We have verified that $|\mathfrak{r}| = 1,$ demonstrating that the entire incoming wave is reflected to the \textit{same} valley. It is worth pointing out that the evanescent wave carries current in the direction parallel to the armchair edge. The current operator along the $y$-direction is $\hat{J}_y = -ev_0 \sigma_2.$ Even for $k_y = 0,$ this current is nonzero:
\begin{equation}
    \langle \hat{J}_y \rangle \propto -ev_0 \mathfrak{t}^\dagger \mathfrak{t}(\varepsilon_{K'}+E) \neq 0 .
\end{equation}
Now, for $E  \geq \sqrt{\varepsilon_{K'}^2+ \left(\hbar v_0 k_y \right)^2} > \sqrt{\varepsilon_K^2 +\left(\hbar v_0 k_y \right)^2},$ we study first the scattering of an incoming $K$ wave
\begin{equation}
\label{eq: scattering wavefunction 1}
\begin{split}
     \begin{pmatrix}
        \psi_A(x\leq 0) \\ 
        \psi_B(x\leq 0)
    \end{pmatrix} &= e^{ik_xx} \begin{pmatrix}
        \varepsilon_{K} + E \\
        \hbar v_0\left(k_x+ik_y\right)
    \end{pmatrix} +  \mathfrak{r}e^{-ik_xx} \begin{pmatrix}
        \varepsilon_{K} + E \\
        \hbar v_0\left(-k_x+ik_y\right)
    \end{pmatrix} ,  \\
    \begin{pmatrix}
        \psi_A(x> 0) \\ 
        \psi_B(x> 0)
    \end{pmatrix} &= \mathfrak{t}e^{ip_x x} \begin{pmatrix}
        \varepsilon_{{K}'}+ E \\
        \hbar v_0\left(p_x+ik_y\right)
    \end{pmatrix},
\end{split}
\end{equation}
where $E = \sqrt{\varepsilon_K^2 + \hbar^2 v_0^2 k_x^2 + \hbar^2 v_0^2 k_y^2 } =  \sqrt{\varepsilon_{K'}^2 + \hbar^2 v_0^2 p_x^2 + \hbar^2 v_0^2 k_y^2 }.$ Matching the boundary condition, we find
\begin{equation}
\begin{split}
  \mathfrak{r} &= -1 + \frac{2 (\varepsilon_{K'} + E) k_x}{\varepsilon_{K} p_x + E p_x + \varepsilon_{K'} k_x + E k_x + i (\varepsilon_{K} - \varepsilon_{K'}) k_y},  \\
  \mathfrak{t} &= \frac{2 (\varepsilon_{K} + E) k_x}{\varepsilon_{K} p_x + E p_x + \varepsilon_{K'} k_x + E k_x + i (\varepsilon_{K} - \varepsilon_{K'}) k_y}.
\end{split}
\end{equation}
As a check of consistency, if $\varepsilon_K = \varepsilon_{K'},$ then $k_x = p_x,$ $\mathfrak{r} = 0,$ and $\mathfrak{t} = 1,$ showing once again that the entire incoming wave is reflected to the \textit{opposite} valley. In general, the reflectance and transmittance are given by 
\begin{equation}
\label{eq: R and T of Dirac model}
    \begin{split}
        \mathfrak{R}(E,k_y) &= \frac{E^2 -\varepsilon_{K} \varepsilon_{K'} - \left(\hbar v_0 k_y \right)^2 - 
 \sqrt{E^2 -\varepsilon_{K}^2 - \left(\hbar v_0 k_y \right)^2} \sqrt{E^2-\varepsilon_{K'}^2  - \left(\hbar v_0 k_y \right)^2}}{
E^2 -\varepsilon_{K} \varepsilon_{K'} - \left(\hbar v_0 k_y \right)^2 + 
 \sqrt{E^2 -\varepsilon_{K}^2 - \left(\hbar v_0 k_y \right)^2} \sqrt{E^2-\varepsilon_{K'}^2  - \left(\hbar v_0 k_y \right)^2}}, \\
  \mathfrak{T}(E,k_y) &= \frac{2 \sqrt{E^2 -\varepsilon_{K}^2 - \left(\hbar v_0 k_y \right)^2} \sqrt{E^2-\varepsilon_{K'}^2  - \left(\hbar v_0 k_y \right)^2}}{E^2 -\varepsilon_{K} \varepsilon_{K'} - \left(\hbar v_0 k_y \right)^2 + 
 \sqrt{E^2 -\varepsilon_{K}^2 - \left(\hbar v_0 k_y \right)^2} \sqrt{E^2-\varepsilon_{K'}^2  - \left(\hbar v_0 k_y \right)^2}} .\\
    \end{split}
\end{equation}
Again, these formulas are valid when $E^2 - \left(\hbar v_0 k_y \right)^2 \geq \varepsilon_{K'}^2 > \varepsilon_K^2,$ where we clearly have $\mathfrak{R}(E,k_y) +\mathfrak{T}(E,k_y) = 1.$ From these formulas, it is clear that if $E^2 - \left(\hbar v_0 k_y \right)^2 = \varepsilon_{K'}^2,$ we have perfect reflection to the \textit{same} valley. Furthermore, we have
\begin{equation}
    \lim_{E^2-(\hbar v_0k_y)^2 \rightarrow \infty} \mathfrak{T}(E,k_y) = 1 \quad \text{and} \quad \lim_{E^2-(\hbar v_0k_y)^2 \rightarrow \infty} \mathfrak{R}(E,k_y) = 0.
\end{equation}
In other words, when $E^2-(\hbar v_0k_y)^2$ is much larger than both $\varepsilon_K$ and $\varepsilon_{K'},$ we revert back to the situation where an incoming $K$ wave is completely scattered into an outgoing $K'$ wave. A plot of $\mathfrak{R}$ and $\mathfrak{T}$ for general energies is shown in Fig. \ref{fig:transmission}. All of the limits noted above can be observed in this plot as well. For completeness, we now consider the situation where the incoming wave comes from the $K'$ valley. In this case, we 
\begin{equation}
\label{eq: scattering wavefunction 2}
\begin{split}
     \begin{pmatrix}
        \psi_A(x\leq 0) \\ 
        \psi_B(x\leq 0)
    \end{pmatrix} &=   \mathfrak{t}e^{-ik_xx} \begin{pmatrix}
        \varepsilon_{K} + E \\
        \hbar v_0\left(-k_x+ik_y\right)
    \end{pmatrix} ,  \\
    \begin{pmatrix}
        \psi_A(x> 0) \\ 
        \psi_B(x> 0)
    \end{pmatrix} &= e^{-ip_x x} \begin{pmatrix}
        \varepsilon_{{K}'}+ E \\
        \hbar v_0\left(-p_x+ik_y\right)
    \end{pmatrix} + \mathfrak{r}e^{ip_x x} \begin{pmatrix}
        \varepsilon_{{K}'}+ E \\
        \hbar v_0\left(p_x+ik_y\right)
    \end{pmatrix},
\end{split}
\end{equation}
where, again, $E = \sqrt{\varepsilon_K^2 + \hbar^2 v_0^2 k_x^2 + \hbar^2 v_0^2 k_y^2 } =  \sqrt{\varepsilon_{K'}^2 + \hbar^2 v_0^2 p_x^2 + \hbar^2 v_0^2 k_y^2 }.$ Eq. \eqref{eq: scattering wavefunction 2} is formally equivalent to Eq. \eqref{eq: scattering wavefunction 1} with $\varepsilon_{K} \leftrightarrow \varepsilon_{K'}$ and $k_x \leftrightarrow -p_x.$ Therefore, all of the preceding analysis applies; in particular, the transmittance and reflectance functions remain the same.

The above results are easily generalized to a slightly different situation where the sublattice gap is only rigidly shifted, instead of being modulated, depending on valleys. In this case, the Hamiltonian takes the following form
\begin{equation}
    \hat{\mathcal{H}}_\mathrm{eff}(\mathbf{k}) = \begin{pmatrix}
        \hat{\mathcal{H}}_K(\mathbf{k}) & 0 \\
        0 & \hat{\mathcal{H}}_{K'}(\mathbf{k}) 
    \end{pmatrix} = \begin{pmatrix}
        \varepsilon & \hbar v_0 \left( k_x - ik_y \right) & 0 & 0 \\
        \hbar v_0 \left( k_x + ik_y \right) & -\varepsilon & 0 & 0 \\
        0 & 0 & \varepsilon + \delta\varepsilon & \hbar v_0 \left( -k_x - ik_y \right) \\
        0 & 0 & \hbar v_0 \left( -k_x + ik_y \right)& -\varepsilon +\delta\varepsilon
    \end{pmatrix},
\end{equation}
where we assume $0< \delta\varepsilon < \varepsilon.$ When $\sqrt{\varepsilon^2+(\hbar v_0 k_y)^2} \leq E < \delta \varepsilon + \sqrt{\varepsilon^2+(\hbar v_0 k_y)^2},$ we must have perfect reflection to the same valley. When $E > \delta \varepsilon + \sqrt{\varepsilon^2+(\hbar v_0 k_y)^2},$ the scattering states with an incoming $K$ wave are
\begin{equation}
\label{eq: scattering wavefunction 3}
\begin{split}
     \begin{pmatrix}
        \psi_A(x\leq 0) \\ 
        \psi_B(x\leq 0)
    \end{pmatrix} &= e^{ik_xx} \begin{pmatrix}
        \varepsilon + E \\
        \hbar v_0\left(k_x+ik_y\right)
    \end{pmatrix} +  \mathfrak{r}e^{-ik_xx} \begin{pmatrix}
        \varepsilon + E \\
        \hbar v_0\left(-k_x+ik_y\right)
    \end{pmatrix} ,  \\
    \begin{pmatrix}
        \psi_A(x> 0) \\ 
        \psi_B(x> 0)
    \end{pmatrix} &= \mathfrak{t}e^{ip_x x} \begin{pmatrix}
        \varepsilon+ E - \delta\varepsilon \\
        \hbar v_0\left(p_x+ik_y\right)
    \end{pmatrix},
\end{split}
\end{equation}
where $E = \sqrt{\varepsilon^2 + \hbar^2 v_0^2 k_x^2 + \hbar^2 v_0^2 k_y^2 } =  \delta \varepsilon +\sqrt{\varepsilon^2 + \hbar^2 v_0^2 p_x^2 + \hbar^2 v_0^2 k_y^2 }.$ Matching boundary condition, we obtain
\begin{equation}
    \begin{split}
        \mathfrak{R}(E,k_y) &= \frac{(E-\delta\varepsilon ) E-\varepsilon^2 - \left(\hbar v_0 k_y\right)^2 - 
 \sqrt{(E-\delta\varepsilon)^2-\varepsilon^2  - \left(\hbar v_0 k_y\right)^2} \sqrt{E^2-\varepsilon^2 - \left(\hbar v_0 k_y\right)^2}}{(E-\delta\varepsilon ) E-\varepsilon^2 - \left(\hbar v_0 k_y\right)^2 + 
 \sqrt{(E-\delta\varepsilon)^2-\varepsilon^2  - \left(\hbar v_0 k_y\right)^2} \sqrt{E^2-\varepsilon^2 - \left(\hbar v_0 k_y\right)^2}}, \\
 \mathfrak{T}(E,k_y) &= \frac{2\sqrt{(E-\delta\varepsilon)^2-\varepsilon^2  - \left(\hbar v_0 k_y\right)^2} \sqrt{E^2-\varepsilon^2 - \left(\hbar v_0 k_y\right)^2}}{(E-\delta\varepsilon ) E-\varepsilon^2 - \left(\hbar v_0 k_y\right)^2 + 
 \sqrt{(E-\delta\varepsilon)^2-\varepsilon^2  - \left(\hbar v_0 k_y\right)^2} \sqrt{E^2-\varepsilon^2 - \left(\hbar v_0 k_y\right)^2}}, \\
    \end{split}
\end{equation}
which is nearly identical to the previous result. Therefore, the general conclusions regarding wave transmission and reflection in the presence of a valley imbalance are robust. In particular, without valley imbalance, an incoming $K$ wave is reflected perfectly into an outgoing $K'$ wave. In the presence of a valley imbalance, say where there are propagating $K$ states but only localized $K'$ states, then an incoming $K$ wave is reflected entirely into an outgoing $K$ wave, as it must. For higher energies, where there are both propagating $K$ and $K'$ states but at different wavevectors due to valley imbalance, then an incoming $K$ wave is generically reflected into both an outing $K$ and $K'$ wave.

We end this section by considering the $N$-layer generalization with Hamiltonian given by 
\begin{equation}
    \hat{\mathcal{H}}_\mathrm{eff}(\mathbf{k}) = \begin{pmatrix}
        \hat{\mathcal{H}}_K(\mathbf{k}) & 0 \\
        0 & \hat{\mathcal{H}}_{K'}(\mathbf{k}) 
    \end{pmatrix} = \begin{pmatrix}
        \varepsilon_K & \left( k_x - ik_y \right)^N & 0 & 0 \\
         \left( k_x + ik_y \right)^N & -\varepsilon_K & 0 & 0 \\
        0 & 0 & \varepsilon_{K'} &  \left( -k_x - ik_y \right)^N \\
        0 & 0 & \left( -k_x + ik_y \right)^N& -\varepsilon_{K'}
    \end{pmatrix}.
\end{equation}
Here, we work in units where $k^N$ has the same dimension as energy. It is instructive to look at the particular case $N = 2$ first. In this case, the (positive) energies are given by $E = \sqrt{\varepsilon_{K,K'}^2 + \left(k_x^2+k_y^2 \right)^2}.$ Solving for $k_x,$ we obtain
\begin{equation}
    k_x = \pm \sqrt{\pm \sqrt{E^2-\varepsilon_{K,K'}^2}-k_y^2}.
\end{equation}
If $\sqrt{\varepsilon_{K,K'}^2+k_y^4} < E$, we have two real roots and two purely imaginary roots (only one of which is normalizable). If $E <\varepsilon_{K,K'} < \sqrt{\varepsilon_{K,K'}^2+k_y^4}, $ we have four complex roots (only two of which are normalizable). If $\varepsilon_{K,K'} < E < \sqrt{\varepsilon_{K,K'}^2+k_y^4}, $ we have four purely imaginary roots (only two of which are normalizable). The presence of many \textit{additional} evanescent modes contrasts the higher-order theory with the linear-momentum theory previously considered. The number of additional modes is proportional to $N$ since these complex wavevectors originate from taking the $N^\mathrm{th}$ roots of the energy dispersion. In general, we write the wavevectors
\begin{equation}
    k_{\pm n} = \pm \sqrt{|E^2-\varepsilon_{K,K'}^2|^{1/N} \exp \left( \frac{i\pi\left[1-\mathrm{sign}(E^2-\varepsilon_{K,K'}^2) \right]}{2N}+ \frac{2\pi i n}{N} \right)-k_y^2}.
\end{equation}
For complex roots, we only keep the ones with negative (positive) imaginary part for states on the left (right) side of the boundary that extends to negative (positive) $x$ direction to enforce normalizability. For real roots, we keep both the positive and negative solutions. If $\sqrt{\varepsilon_{{K},K'}^2+k_y^{2N}} < E,$ which we always assume for the left side, then we have two real roots and $N-1$ admissible complex roots. If $E < \sqrt{\varepsilon_{{K},K'}^2+k_y^{2N}} ,$ then there are $N$ complex roots. The boundary condition is derived by enforcing vanishing current at the boundary $\bra{\psi}\hat{J}_x \ket{\psi} = 0.$ One such choice is obtained by requiring the wavefunction and all of its derivatives up to the $(N-1)^\mathrm{th}$ one to be continuous at the interface. The most general scattering wavefunction with an incoming $K$ wave in this case where there are no propagating states on the right side can be written as 
\begin{equation}
\label{eq: scattering wavefunction 4}
\begin{split}
     \begin{pmatrix}
        \psi_A(x\leq 0) \\ 
        \psi_B(x\leq 0)
    \end{pmatrix} &= e^{ik_{+0}x} \begin{pmatrix}
        \varepsilon_K + E \\
        \left(k_{+0}+ik_y\right)^N
    \end{pmatrix} +  \mathfrak{r}e^{-ik_{+0}x} \begin{pmatrix}
        \varepsilon_K + E \\
        \left(-k_{+0}+ik_y\right)^N
    \end{pmatrix} + \sum_{n=1}^{N-1} \mathfrak{a}_n e^{i \kappa_{-n} x}  \begin{pmatrix}
        \varepsilon_K + E \\
        \left(\kappa_{-n}+ik_y\right)^N
    \end{pmatrix},  \\
    \begin{pmatrix}
        \psi_A(x> 0) \\ 
        \psi_B(x> 0)
    \end{pmatrix} &= \sum_{n=0}^{N-1} \mathfrak{b}_n e^{i \rho_{+n} x}  \begin{pmatrix}
        \varepsilon_{K'} + E \\
        \left(\rho_{+n}+ik_y\right)^N
    \end{pmatrix},
\end{split}
\end{equation}
where $\kappa_{-n}$ are complex numbers with negative imaginary parts (so that the exponents contain $e^{|\mathrm{Im}\kappa_{-n}|x}$ which decay as $x \rightarrow -\infty$) while $\rho_{+n}$ are complex numbers with positive imaginary parts (so that the exponents contain $e^{-\mathrm{Im}\rho_{+n}x}$ which decay as $x \rightarrow +\infty$). Matching boundary condition, we find
\begin{equation}
    \begin{split}
         \left[   +  \mathfrak{r} \left(-ik_{+0} \right)^m + \sum_{n=1}^{N-1} \mathfrak{a}_n \left(i \kappa_{-n} \right)^m\right] \left[ \varepsilon_K+E \right] - \left[ \sum_{n=0}^{N-1}  \mathfrak{b}_n \left(i \rho_{+n} \right)^m \right] \left[ \varepsilon_{K'}+E \right] &= - \left(ik_{+0}\right)^m\left[ \varepsilon_K+E \right], \\
         \mathfrak{r} \left(-i k_{+0} \right)^m \left(-k_{+0}+ik_y\right)^N + \sum_{n=1}^{N-1} \mathfrak{a}_n \left( i \kappa_{-n}\right)^m\left(\kappa_{-n}+ik_y\right)^N - \sum_{n=0}^{N-1} \mathfrak{b}_n \left(i \rho_{+n} \right)^m \left(\rho_{+n}+ik_y\right)^N &=-\left(i k_{+0} \right)^m \left(k_{+0}+ik_y\right)^N,
    \end{split}
\end{equation}
for $m = 0,...,N-1.$ We note that there are $2N$ unknowns and $2N$ equations. Therefore, this system of equations is solvable. The reflectance is $|\mathfrak{r}|^2,$ which we have checked numerically is always unity. Now, moving onto the case where are propagating modes on both sides of the boundary, the most general scattering wavefunction with an incoming $K$ wave is modified slightly to 
\begin{equation}
\label{eq: scattering wavefunction 5}
\begin{split}
     \begin{pmatrix}
        \psi_A(x\leq 0) \\ 
        \psi_B(x\leq 0)
    \end{pmatrix} &= e^{ik_{+0}x} \begin{pmatrix}
        \varepsilon_K + E \\
        \left(k_{+0}+ik_y\right)^N
    \end{pmatrix} +  \mathfrak{r}e^{-ik_{+0}x} \begin{pmatrix}
        \varepsilon_K + E \\
        \left(-k_{+0}+ik_y\right)^N
    \end{pmatrix} + \sum_{n=1}^{N-1} \mathfrak{a}_n e^{i \kappa_{-n} x}  \begin{pmatrix}
        \varepsilon_K + E \\
        \left(\kappa_{-n}+ik_y\right)^N
    \end{pmatrix},  \\
    \begin{pmatrix}
        \psi_A(x> 0) \\ 
        \psi_B(x> 0)
    \end{pmatrix} &=  \mathfrak{t} e^{i p_{+0} x}  \begin{pmatrix}
        \varepsilon_{K'} + E \\
        \left(p_{+0}+ik_y\right)^N
    \end{pmatrix} + \sum_{n=1}^{N-1} \mathfrak{b}_n e^{i \rho_{+n} x}  \begin{pmatrix}
        \varepsilon_{K'} + E \\
        \left(\rho_{+n}+ik_y\right)^N
    \end{pmatrix}.
\end{split}
\end{equation}
Matching the boundary condition is done exactly as before by demanding continuity of the wavefunction and its derivatives at $x = 0.$ The transmittance and reflectance are defined in the usual way
\begin{equation}
\mathfrak{R}(E,k_y) = \abs{\mathfrak{r}(E,k_y)}^2 \quad \text{and} \quad \mathfrak{T} = \abs{\mathfrak{t}(E,k_y)}^2 \frac{\left(\varepsilon_{K'} + E\right) \left( p_{+0}^2+ k_y^2 \right)^{N-1}p_{+0}}{\left(\varepsilon_{K} + E\right) \left( k_{+0}^2+ k_y^2 \right)^{N-1}k_{+0}}.
\end{equation}
We have checked numerically that $\mathfrak{R}(E,k_y) + \mathfrak{T}(E,k_y) = 1$ in the appropriate range of energy and parallel momentum. All of the \textit{qualitative} conclusions for the monolayer model apply to the $N$-layer models without modification. Namely:
\begin{enumerate}
    \item For $\sqrt{\varepsilon_K^2+k_y^{2N}}<E <\sqrt{\varepsilon_{K'}^2+k_y^{2N}},$ we have total intravalley reflection: an incoming $K$ wave is reflected entirely into an outgoing $K$ wave.
    \item The evanescent modes can carry current in the direction parallel to the armchair edge even for $k_y = 0.$
    \item For $\sqrt{\varepsilon_K^2+k_y^{2N}} <\sqrt{\varepsilon_{K'}^2+k_y^{2N}} < E,$ we have partial intravalley reflection and partial intervalley reflection: an incoming $K$ wave is reflected into both an outgoing $K$ wave and an outgoing $K'$ wave. 
    \item For $E \rightarrow \infty,$ we have total intervalley reflection: an incoming $K$ wave is reflected entirely into an outgoing $K'$ wave.
\end{enumerate}

\subsubsection{Semi-Infinite Plane}
\label{sec: Semi-Infinite Plane}

We employ the standard iterative Green's function method to calculate the surface Green's function \cite{Guinea1983Effective, sancho1985highly}. We partition the Hamiltonian in ``layers" (this is \textit{not} the same as the number of layers in a rhombohedral stack) that are coupled to each other in the following way
\begin{equation}
    \hat{\mathcal{H}}  = \begin{pmatrix}
        \mathcal{H}_{0,0} & \mathcal{V}_0  & 0 & 0 & \dots \\
        \mathcal{V}^\dagger_0 & \mathcal{H}_{1,1} & \mathcal{V} & 0 & \dots \\
        0 & \mathcal{V}^\dagger & \mathcal{H}_{2,2} & \mathcal{V} & \dots \\
        0 & 0 & \mathcal{V}^\dagger & \mathcal{H}_{3,3} & \dots \\
        \vdots & \vdots & \vdots & \vdots & \ddots
    \end{pmatrix}.
\end{equation}
$\mathcal{H}_{0,0}$ is the surface Hamiltonian. For a \textit{bulk} homogeneous system, $\mathcal{H}_{i\neq0,i\neq0} = \mathcal{H}_{1,1},$ which we now assume. The  Green's function is defined as the resolvent of the Hamiltonian $\hat{\mathcal{G}}(\omega) = \left(\omega-\hat{\mathcal{H}}\right)^{-1}.$ Internal indices, including parallel momentum, are left implicit. Using the identity $\left(\omega - \hat{\mathcal{H}}\right) \hat{\mathcal{G}}(\omega) = \mathbb{1},$ we obtain the following
\begin{equation}
    \sum_{a}\left(\omega \delta_{i, a} - \mathcal{H}_{i,a}\right) \mathcal{G}_{a,j}(\omega) = \delta_{i,j}.
\end{equation}
Because the Hamiltonian is tridiagonal, this equation simplifies significantly
\begin{equation}
\begin{split}
\left(\omega  - \mathcal{H}_{0,0}\right) \mathcal{G}_{0,0}(\omega) &= 1 + \mathcal{V}_0 \mathcal{G}_{1,0}(\omega), \\
\left(\omega  - \mathcal{H}_{1,1}\right) \mathcal{G}_{1,0}(\omega) &= \mathcal{V}^\dagger_0 \mathcal{G}_{0,0}(\omega) + \mathcal{V} \mathcal{G}_{2,0}(\omega) \\
&\vdots\\
\left(\omega  - \mathcal{H}_{1,1}\right) \mathcal{G}_{i,0}(\omega) &= \mathcal{V}^\dagger \mathcal{G}_{i-1,0}(\omega) + \mathcal{V} \mathcal{G}_{i+1,0}(\omega).
\end{split}
\end{equation}
We solve for $\mathcal{G}_{0,0}(\omega)$ iteratively as follows. In the first step, we substitute in $\mathcal{G}_{1,0}(\omega)$ from the second equation into the first equation to obtain
\begin{equation}
    \left( \omega - \mathcal{H}_{0,0} - \mathcal{V}_0 \left[\omega-\mathcal{H}_{1,1} \right]^{-1} \mathcal{V}_0^\dagger \right) \mathcal{G}_{0,0}(\omega) = 1 + \mathcal{V}_0\left[\omega-\mathcal{H}_{1,1} \right]^{-1} \mathcal{V} \mathcal{G}_{2,0}(\omega).
\end{equation}
Because we need $\mathcal{G}_{2,0}(\omega)$ to solve for $\mathcal{G}_{0,0}(\omega)$ now, we use the following equation to relate Green's functions that differ by two (inner) layers
\begin{equation}
\begin{split}
\left( \omega - \mathcal{H}_{1,1} - \mathcal{V}^\dagger \left[ \omega - \mathcal{H}_{1,1} \right]^{-1}\mathcal{V} - \mathcal{V} \left[\omega - \mathcal{H}_{1,1} \right]^{-1} \mathcal{V}^\dagger \right) \mathcal{G}_{2,0}(\omega) &=  \mathcal{V}^\dagger \left[\omega - \mathcal{H}_{1,1}\right]^{-1} \mathcal{V}^\dagger_0 \mathcal{G}_{0,0}(\omega)    + \mathcal{V} \left[\omega - \mathcal{H}_{1,1}\right]^{-1} \mathcal{V} \mathcal{G}_{4,0}(\omega),\\
\left( \omega - \mathcal{H}_{1,1} - \mathcal{V}^\dagger \left[ \omega - \mathcal{H}_{1,1} \right]^{-1}\mathcal{V} - \mathcal{V} \left[\omega - \mathcal{H}_{1,1} \right]^{-1} \mathcal{V}^\dagger \right) \mathcal{G}_{4,0}(\omega) &=  \mathcal{V}^\dagger \left[\omega - \mathcal{H}_{1,1}\right]^{-1} \mathcal{V}^\dagger \mathcal{G}_{2,0}(\omega)    + \mathcal{V} \left[\omega - \mathcal{H}_{1,1}\right]^{-1} \mathcal{V} \mathcal{G}_{6,0}(\omega),\\
&\vdots
\end{split}
\end{equation}
To simplify, we define (the superscripts denote iteration step)
\begin{equation}
\begin{split}
    \mathcal{E}^{(1)}_0 &= \mathcal{H}_{0,0} + \mathcal{V}_0 \left[\omega-\mathcal{H}_{1,1}\right]^{-1} \mathcal{V}_0^\dagger, \\
    \mathcal{E}^{(1)} &= \mathcal{H}_{1,1} + \mathcal{V}^\dagger \left[ \omega - \mathcal{H}_{1,1} \right]^{-1}\mathcal{V} + \mathcal{V} \left[\omega - \mathcal{H}_{1,1} \right]^{-1} \mathcal{V}^\dagger,\\
    \alpha_0^{(1)} &= \mathcal{V}_0 \left[\omega - \mathcal{H}_{1,1}\right]^{-1} \mathcal{V}, \\
    \beta_0^{(1)} &= \mathcal{V}^\dagger \left[\omega - \mathcal{H}_{1,1}\right]^{-1} \mathcal{V}^\dagger_0, \\
    \alpha^{(1)} &= \mathcal{V} \left[\omega - \mathcal{H}_{1,1}\right]^{-1} \mathcal{V}, \\
    \beta^{(1)} &= \mathcal{V}^\dagger \left[\omega - \mathcal{H}_{1,1}\right]^{-1} \mathcal{V}^\dagger, \\
\end{split}
\end{equation}
to obtain 
\begin{equation}
\begin{split}
\left(\omega  - \mathcal{E}_0^{(1)}\right) \mathcal{G}_{0,0}(\omega) &= 1 + \alpha^{(1)}_0 \mathcal{G}_{2,0}(\omega), \\
\left(\omega  - \mathcal{E}^{(1)}\right) \mathcal{G}_{2,0}(\omega) &= \beta^{(1)}_0 \mathcal{G}_{0,0}(\omega) + \alpha^{(1)} \mathcal{G}_{4,0}(\omega),\\
\left(\omega  - \mathcal{E}^{(1)}\right) \mathcal{G}_{4,0}(\omega) &= \beta^{(1)} \mathcal{G}_{2,0}(\omega) + \alpha^{(1)} \mathcal{G}_{6,0}(\omega),\\
&\vdots
\end{split}
\end{equation}
In the second iteration, we obtain
\begin{equation}
    \begin{split}
        &\left( \omega - \mathcal{E}_0^{(1)} -  \alpha_0^{(1)} \left[\omega - \mathcal{E}^{(1)}  \right]^{-1}  \beta_0^{(1)} \right) \mathcal{G}_{0,0}(\omega) = 1 + \alpha_0^{(1)} \left[\omega - \mathcal{E}^{(1)} \right]^{-1} \alpha^{(1)} \mathcal{G}_{4,0}(\omega), \\
        &\left( \omega - \mathcal{E}^{(1)}- \beta^{(1)}\left[\omega-\mathcal{E}^{(1)} \right]^{-1}\alpha^{(1)}- \alpha^{(1)}\left[\omega-\mathcal{E}^{(1)} \right]^{-1}\beta^{(1)}\right) \mathcal{G}_{4,0}(\omega) \\
        &=  \beta^{(1)} \left[\omega-\mathcal{E}^{(1)} \right]^{-1} \beta^{(1)}_0 \mathcal{G}_{0,0}(\omega) +\alpha^{(1)} \left[\omega-\mathcal{E}^{(1)} \right]^{-1} \alpha^{(1)} \mathcal{G}_{8,0}(\omega),\\
        &\left( \omega - \mathcal{E}^{(1)}- \beta^{(1)}\left[\omega-\mathcal{E}^{(1)} \right]^{-1}\alpha^{(1)}- \alpha^{(1)}\left[\omega-\mathcal{E}^{(1)} \right]^{-1}\beta^{(1)}\right) \mathcal{G}_{8,0}(\omega) \\
        &=  \beta^{(1)} \left[\omega-\mathcal{E}^{(1)} \right]^{-1} \beta^{(1)} \mathcal{G}_{4,0}(\omega) +\alpha^{(1)} \left[\omega-\mathcal{E}^{(1)} \right]^{-1} \alpha^{(1)} \mathcal{G}_{12,0}(\omega),\\
        &\vdots
    \end{split}
\end{equation}
from which it is evident that we can define the iterative procedure at the $n^\mathrm{th}$ step as
\begin{equation}
    \begin{split}
        \mathcal{E}^{(n)}_0 & = \mathcal{E}_0^{(n-1)} +  \alpha^{(n-1)}_0 \left[\omega - \mathcal{E}^{(n-1)}  \right]^{-1}  \beta^{(n-1)}_0, \\
        \mathcal{E}^{(n)} &= \mathcal{E}^{(n-1)}+ \beta^{(n-1)}\left[\omega-\mathcal{E}^{(n-1)} \right]^{-1}\alpha^{(n-1)}+ \alpha^{(n-1)}\left[\omega-\mathcal{E}^{(n-1)} \right]^{-1}\beta^{(n-1)}, \\
        \alpha_0^{(n)} &= \alpha_0^{(n-1)} \left[\omega-\mathcal{E}^{(n-1)} \right]^{-1} \alpha^{(n-1)}, \\
        \beta_0^{(n)} &= \beta^{(n-1)} \left[\omega-\mathcal{E}^{(n-1)} \right]^{-1} \beta_0^{(n-1)},\\
        \alpha^{(n)} &= \alpha^{(n-1)} \left[\omega-\mathcal{E}^{(n-1)} \right]^{-1} \alpha^{(n-1)}, \\
        \beta^{(n)} &= \beta^{(n-1)} \left[\omega-\mathcal{E}^{(n-1)} \right]^{-1} \beta^{(n-1)}.
    \end{split}
\end{equation}
At this step, the Green's function satisfies
\begin{equation}
\begin{split}
\left(\omega  - \mathcal{E}_0^{(n)}\right) \mathcal{G}_{0,0}(\omega) &= 1 + \alpha^{(n)}_0 \mathcal{G}_{2^n,0}(\omega).
\end{split}    
\end{equation}
We terminate the iterative loop at some critical $n^*$ when $\norm{\alpha^{(n^*)}_0} < \mathrm{tol},$ where $\mathrm{tol}$ is some small convergence parameter. At this critical step, the surface Green's function is appropriately given by 
\begin{equation}
    \mathcal{G}_{0,0}(\omega) \approx \left( \omega - \mathcal{E}_0^{(n^*)} \right)^{-1}.
\end{equation}
The spectral weight at energy $\omega$ is computed using this approximated surface Green's function by 
\begin{equation}
    A(\omega) = - \lim_{\eta \rightarrow 0^+} \frac{1}{\pi} \mathrm{Im } \mathrm{Tr } \mathcal{G}_{0,0} ( \omega= E + i \eta).
\end{equation}


\subsection{Transmission and Reflection at an  Armchair Domain Wall}

\subsubsection{Continuum Description of a Step-Function Domain Wall}

Here, we extend the results of Sec. \ref{sec: Reflections at an Armchair Termination} to study transmission and reflection due to an abrupt domain wall that switches the sense of valley polarization. The domain wall is located at $x = 0$ and runs along the armchair direction. To model this system, we first consider the Hamiltonian of a monolayer graphene sheet with a valley-dependent mass gap that varies along the $x$-direction
\begin{equation}
    \hat{\mathcal{H}}_\mathrm{eff}(\mathbf{k}) =  \begin{pmatrix}
        \varepsilon_K(x) & \hbar v_0 \left( k_x - ik_y \right) & 0 & 0 \\
        \hbar v_0 \left( k_x + ik_y \right) & -\varepsilon_K(x) & 0 & 0 \\
        0 & 0 & \varepsilon_{K'}(x) & \hbar v_0 \left( -k_x - ik_y \right) \\
        0 & 0 & \hbar v_0 \left( -k_x + ik_y \right)& -\varepsilon_{K'}(x)
    \end{pmatrix}.
\end{equation}
We take the gap function to be 
\begin{equation}
\label{eq: valley dependent mass gap}
    \varepsilon_K(x) = \left\{
\begin{array}{ll}
      \varepsilon_- & x\leq 0 \\
      \varepsilon_+ & x > 0
\end{array} 
\right. \quad \text{and} \quad \varepsilon_{K'}(x) = \left\{
\begin{array}{ll}
      \varepsilon_+ & x\leq 0 \\
      \varepsilon_- & x > 0
\end{array} 
\right. ,
\end{equation}
where we assume that $0< \varepsilon_-< \varepsilon_+.$ That is, for $x< 0,$ the gap at $K$ is smaller than the gap at $K'$ and the opposite is true for $x > 0.$ Now, let us consider the case where  $ \sqrt{\varepsilon_-^2+(\hbar v_0k_y)^2} \leq E < \sqrt{\varepsilon_+^2+(\hbar v_0k_y)^2}$ where the wavefunction can be written as 
\begin{equation}
\begin{split}
    \begin{pmatrix}
        \psi_{A,K}(x \leq 0) \\
        \psi_{B,K}(x \leq 0) \\
        \psi_{A,K'}(x \leq 0) \\
        \psi_{B,K'}(x \leq 0) \\
    \end{pmatrix} &=  e^{ik_xx} \begin{pmatrix}
         \varepsilon_- + E \\
         \hbar v_0 (k_x+ik_y) \\
         0 \\ 
         0
    \end{pmatrix} + \mathfrak{r}_K e^{-ik_xx} \begin{pmatrix}
         \varepsilon_- + E \\
         \hbar v_0 (-k_x+ik_y) \\
         0 \\ 
         0
    \end{pmatrix}  + \mathfrak{r}_{K'}e^{+\kappa_xx} \begin{pmatrix}
        0 \\
        0 \\
         \varepsilon_+ + E  \\
         i\hbar v_0 (\kappa_x+k_y) 
    \end{pmatrix}  ,   \\
    \begin{pmatrix}
        \psi_{A,K}(x > 0) \\
        \psi_{B,K}(x > 0) \\
        \psi_{A,K'}(x > 0) \\
        \psi_{B,K'}(x > 0) \\
    \end{pmatrix} &= \mathfrak{t}_K e^{-\kappa_xx} \begin{pmatrix}
         \varepsilon_+ + E \\
         i\hbar v_0 (\kappa_x+k_y) \\
         0 \\ 
         0
    \end{pmatrix}  + \mathfrak{t}_{K'}e^{+ik_xx} \begin{pmatrix}
        0 \\
        0 \\
         \varepsilon_- + E  \\
         \hbar v_0 (-k_x+ik_y) 
    \end{pmatrix}  .
\end{split}
\end{equation}
The energy dispersion is given by $E = \sqrt{\varepsilon_-^2+(\hbar v_0 k_x)^2+(\hbar v_0 k_y)^2}= \sqrt{\varepsilon_+^2-(\hbar v_0 \kappa_x)^2+(\hbar v_0 k_y)^2}.$ We enforce the boundary condition by demanding continuity of the current along the $x$-direction. This can be achieved by making the wavefunction continuous at $x = 0.$ Matching the boundary condition yields $\mathfrak{r}_{K'} =  \mathfrak{t}_{K'} = 0$ and
\begin{equation}
    \begin{split}
        \mathfrak{r}_{K} &= -1 + \frac{2 (\varepsilon_+ + E) k_x}{
    E (i \kappa_x + k_x) + \varepsilon_+ (k_x - i k_y) + i \varepsilon_- (\kappa_x + k_y)}, \\
  \mathfrak{t}_{K} &= -\frac{2 i (\varepsilon_- + E) k_x}{
    E (\kappa_x - i k_x) -  \varepsilon_+ (ik_x +  k_y) + \varepsilon_- (\kappa_x + k_y)}.
    \end{split}
\end{equation}
We find $|\mathfrak{r}_K| = 1.$ This demonstrates perfect reflection. An incoming $K$ wave is reflected  completely into an outgoing $K$ wave. There is no transmission to the other side. The calculation here bears resemblance to the calculation done for the armchair boundary above. However, it is important to note that the evanescent state here resides in the $K$ valley in contrast to the evanescent wave that resides in the $K'$ valley for the armchair boundary. Similar to before, the evanescent mode, even for $k_y = 0,$ carries current in the direction parallel to the domain wall.

We now generalize the above result to the $N$-layer situation. Again, this is very similar to the armchair calculation, but we need to keep all four degrees of freedom on both sides of the boundary. The Hamiltonian is 
\begin{equation}
    \hat{\mathcal{H}}_\mathrm{eff}(\mathbf{k}) = \begin{pmatrix}
        \varepsilon_K(x) & \left( k_x - ik_y \right)^N & 0 & 0 \\
         \left( k_x + ik_y \right)^N & -\varepsilon_K(x) & 0 & 0 \\
        0 & 0 & \varepsilon_{K'}(x) &  \left( -k_x - ik_y \right)^N \\
        0 & 0 & \left( -k_x + ik_y \right)^N& -\varepsilon_{K'}(x)
    \end{pmatrix}.
\end{equation}
The valley-dependent mass gap is taken to be of the same form as in Eq. \eqref{eq: valley dependent mass gap}. For a fixed $k_y$ and $E,$ we have $2N$ \textit{complex} wavevectors for each valley and each side of the domain wall  given by
\begin{equation}
    k_{\pm n} = \pm \sqrt{|E^2-\varepsilon_{+/-}^2|^{1/N} \exp \left( \frac{i\pi \left[1-\mathrm{sign} \left(E^2-\varepsilon_{+/-}^2 \right) \right]}{2N}  + \frac{2\pi i n}{N}\right) -k_y^2},
\end{equation}
where $ n \in \lbrace 0,1,...,N-1\rbrace $. We always work in the energy regime where there are two \textit{real} wavevectors on the left in the $K$ valley, $E > \sqrt{\varepsilon_-^2+k_y^{2N}}.$ Of the $2(N-1)$ complex roots remaining, we only take half of them which are normalizable. On the left-hand side, we only take complex roots which have negative imaginary parts, while on the right-hand side, we only take complex roots which have positive imaginary parts. If $E < \sqrt{\varepsilon_+^2+k_y^{2N}},$ then there are only complex roots, and again, we choose only half of them which are normalizable. It is worth pointing out that there are evanescent modes on both sides of the domain wall for the $N$-layer situation, which contrasts with the monolayer case where the evanescent modes reside on the opposite sides to the propagating modes. The wavefunction can be written as 
\begin{equation}
\begin{split}
    \begin{pmatrix}
        \psi_{A,K}(x \leq 0) \\
        \psi_{B,K}(x \leq 0) \\
        \psi_{A,K'}(x \leq 0) \\
        \psi_{B,K'}(x \leq 0) \\
    \end{pmatrix} &=  e^{ik_{+0}x} \begin{pmatrix}
         \varepsilon_- + E \\
         (k_{+0}+ik_y)^N\\
         0 \\ 
         0
    \end{pmatrix} + \mathfrak{r}_K e^{-ik_{+0}x} \begin{pmatrix}
         \varepsilon_- + E \\
         (-k_{+0}+ik_y)^N\\
         0 \\ 
         0
    \end{pmatrix}  \\
    &+ \sum_{n=1}^{N-1} \mathfrak{a}_n e^{i \kappa_{-n}x}\begin{pmatrix}
         \varepsilon_- + E \\
         (\kappa_{-n}+ik_y)^N\\
         0 \\ 
         0
    \end{pmatrix}
    + \sum_{n=0}^{N-1}\mathfrak{b}_n e^{i\rho_{-n}x} \begin{pmatrix}
        0 \\
        0 \\
         \varepsilon_+ + E  \\
         (-\rho_{-n}+ik_y)^N
    \end{pmatrix}  ,   \\
    \begin{pmatrix}
        \psi_{A,K}(x > 0) \\
        \psi_{B,K}(x > 0) \\
        \psi_{A,K'}(x > 0) \\
        \psi_{B,K'}(x > 0) \\
    \end{pmatrix} &= \sum_{n=0}^{N-1} \mathfrak{c}_n e^{i \rho_{+n}x}\begin{pmatrix}
         \varepsilon_+ + E \\
         (\rho_{+n}+ik_y)^N\\
         0 \\ 
         0
    \end{pmatrix}
    + \sum_{n=1}^{N-1}\mathfrak{d}_n e^{i\kappa_{+n}x} \begin{pmatrix}
        0 \\
        0 \\
         \varepsilon_- + E  \\
         (-\kappa_{+n}+ik_y)^N
    \end{pmatrix}  \\
    &+ \mathfrak{t}_{K'}e^{+ik_{+0}x} \begin{pmatrix}
        0 \\
        0 \\
         \varepsilon_- + E  \\
         (-k_{+0}+ik_y)^N
    \end{pmatrix}  .
\end{split}
\end{equation}
Here, we use a slightly different notation for the wavevectors to emphasize their signs and dependence on the energies.  $k$ is used for purely real values while $\kappa_\pm$ and $\rho_\pm$ are used to indicate complex values with $\pm$ imaginary parts. $\kappa_\pm$ uses $\varepsilon_-$ in its calculation while $\rho_\pm$ uses $\varepsilon_+$ in its calculation. Now, enforcing continuity at the domain wall, we obtain
\begin{equation}
    \begin{split}
        \left[\mathfrak{r}_K (-ik_{+0})^m + \sum_{n=1}^{N-1} \mathfrak{a}_n(i\kappa_{-n})^m \right] \left[\varepsilon_-+E\right] -  \sum_{n=0}^{N-1} \mathfrak{c}_n(i\rho_{+n})^m \left[\varepsilon_++E\right] &= - (ik_{+0})^m \left[\varepsilon_-+E\right], \\
        \mathfrak{r}_K (-ik_{+0})^m\left[-k_{+0}+ik_y\right]^N  + \sum_{n=1}^{N-1} \mathfrak{a}_n(i\kappa_{-n})^m\left[\kappa_{-n}+ik_y\right]^N   -  \sum_{n=0}^{N-1} \mathfrak{c}_n(i\rho_{+n})^m  \left[\rho_{+n}+ik_y\right]^N  &= - (ik_{+0})^m \left[k_{+0}+ik_y\right]^N, \\
        \sum_{n=0}^{N-1}\mathfrak{b}_n (i\rho_{-n})^m \left[\varepsilon_++E \right] - \left[\sum_{n=1}^{N-1} \mathfrak{d}_n (i\kappa_{+n})^m + \mathfrak{t}_{K'}(ik_{+0})^m \right] \left[ \varepsilon_-+E \right]   &= 0, \\
        \sum_{n=0}^{N-1}\mathfrak{b}_n (i\rho_{-n})^m \left[-\rho_{-n}+ik_y\right]^N -\sum_{n=1}^{N-1} \mathfrak{d}_n (i\kappa_{+n})^m\left[-\kappa_{+n}+ik_y\right]^N - \mathfrak{t}_{K'}(ik_{+0})^m \left[-k_{+0}+ik_y\right]^N    &= 0. \\
    \end{split}
\end{equation}
From here, we note that the $\mathfrak{a}_n$ and $\mathfrak{c}_n$ coefficients are completely decoupled from the $\mathfrak{b}_n$ and $\mathfrak{d}_n$ coefficients. In particular,  the $\mathfrak{b}_n$ and $\mathfrak{d}_n$ coefficients and $\mathfrak{t}_{K'}$ can always be set to zero and still satisfy the domain-wall continuity condition. We have checked numerically that $|\mathfrak{r}_K| = 1$ for any number of layer as long as the chemical potential lies in the quarter metal phase. This shows that for any number of layers, an abrupt valley domain wall is opaque. Electron waves coming in from one valley must reflect entirely back to that same valley. There can be no transmission!

\subsubsection{Green's Function Calculation of a Finite-Width Domain Wall}

To calculate transmission through a domain connected to semi-infinite right and left leads, we use the equilibrium Green's function method \cite{Datta1995,Ryndyk2016}. The domain Hamiltonian is written as $\mathcal{H}_D,$ while the left and right leads have block form, $\mathcal{H}_{L}$ and $\mathcal{H}_{R}$ connected by $\mathcal{V}_L$ and $\mathcal{V}_R$. The hoppings from the leads to the domain are denoted $\mathcal{V}_{LD}$ and $\mathcal{V}_{DR}.$ The Hamiltonian has the following form
\begin{equation}
    \hat{\mathcal{H}}  = \begin{pmatrix}
    \ddots & \vdots & \vdots & \vdots &\vdots & \vdots & \vdots & \vdots & \iddots\\
        \dots & \mathcal{H}_L & \mathcal{V}_L & 0 & 0 & 0 & 0   & 0 & \dots \\
        \dots & \mathcal{V}_L^\dagger & \mathcal{H}_L & \mathcal{V}_L & 0 & 0 & 0   & 0 & \dots \\
        \dots & 0 & \mathcal{V}_L^\dagger & \mathcal{H}_L & \mathcal{V}_{LD} & 0 & 0   & 0 & \dots \\
        \dots & 0 & 0 & \mathcal{V}_{LD}^\dagger & \mathcal{H}_D & \mathcal{V}_{DR}  & 0 & 0 & \dots \\
        \dots & 0 & 0 & 0 &\mathcal{V}^\dagger_{DR} & \mathcal{H}_R & \mathcal{V}_R & 0 & \dots \\
        \dots & 0 & 0 & 0 &0 & \mathcal{V}^\dagger_R & \mathcal{H}_R & \mathcal{V}_R & \dots \\
        \dots & 0 & 0 & 0 &0 & 0 & \mathcal{V}^\dagger_R & \mathcal{H}_R & \dots \\
        \iddots & \vdots & \vdots & \vdots &\vdots & \vdots & \vdots & \vdots & \ddots
    \end{pmatrix}.
\end{equation}
This Hamiltonian can be partitioned as follows
\begin{equation}
    \begin{split}
        \hat{\mathcal{H}} &= \begin{pmatrix}
            \tilde{\mathcal{H}}_L & \tilde{\mathcal{V}}_{LD} & 0 \\
            \tilde{\mathcal{V}}_{LD}^\dagger & \mathcal{H}_D & \tilde{\mathcal{V}}_{DR} \\
            0 & \tilde{\mathcal{V}}_{DR}^\dagger & \tilde{\mathcal{H}}_R
        \end{pmatrix}, \\
        \tilde{\mathcal{H}}_{L} &= \begin{pmatrix}
    \ddots & \vdots & \vdots & \vdots  \\
        \dots & \mathcal{H}_L & \mathcal{V}_L & 0 \\
        \dots & \mathcal{V}_L^\dagger & \mathcal{H}_L & \mathcal{V}_L \\
        \dots & 0 & \mathcal{V}_L^\dagger & \mathcal{H}_L & 
    \end{pmatrix}, \quad 
    \tilde{\mathcal{V}}_{LD} = \begin{pmatrix}
    \vdots \\
         0  \\
         0  \\
        \mathcal{V}_{LD} \\
    \end{pmatrix}, \quad
    \tilde{\mathcal{V}}_{DR} = \begin{pmatrix}
    \mathcal{V}_{DR} & 0 & 0 & \dots
    \end{pmatrix}, \quad 
    \tilde{\mathcal{H}}_{R} = \begin{pmatrix}
         \mathcal{H}_R & \mathcal{V}_R & 0 & \dots \\
        \mathcal{V}^\dagger_R & \mathcal{H}_R & \mathcal{V}_R & \dots \\
         0 & \mathcal{V}^\dagger_R & \mathcal{H}_R & \dots \\
         \vdots & \vdots & \vdots & \ddots
    \end{pmatrix}.
    \end{split} 
\end{equation}
Writing the full Green's function as 
\begin{equation}
    \begin{pmatrix}
            \omega-\tilde{\mathcal{H}}_L & -\tilde{\mathcal{V}}_{LD} & 0 \\
            -\tilde{\mathcal{V}}_{LD}^\dagger & \omega-\mathcal{H}_D & -\tilde{\mathcal{V}}_{DR} \\
            0 & -\tilde{\mathcal{V}}_{DR}^\dagger & \omega-\tilde{\mathcal{H}}_R
        \end{pmatrix}\begin{pmatrix}
            * & \tilde{\mathcal{G}}_{LD}(\omega) & * \\
            * & \mathcal{G}_{D}(\omega) & * \\
            * & \tilde{\mathcal{G}}_{DR}(\omega) & * \\
        \end{pmatrix} = \begin{pmatrix}
            1 & 0 & 0 \\
            0 & 1 & 0 \\
            0 & 0 & 1
        \end{pmatrix},
\end{equation}
we find the following
\begin{equation}
    \begin{split}
        \left( \omega-\tilde{\mathcal{H}}_L\right) \tilde{\mathcal{G}}_{LD}(\omega) &= \tilde{\mathcal{V}}_{LD} \mathcal{G}_D(\omega), \\ 
        \left( \omega-{\mathcal{H}}_D\right) \mathcal{G}_{D}(\omega) &= 1 + \tilde{\mathcal{V}}_{LD}^\dagger \tilde{\mathcal{G}}_{LD}(\omega) + \tilde{\mathcal{V}}_{DR} \tilde{\mathcal{G}}_{DR}(\omega), \\
        \left( \omega-\tilde{\mathcal{H}}_R\right) \tilde{\mathcal{G}}_{DR}(\omega) &= \tilde{\mathcal{V}}_{DR}^\dagger \mathcal{G}_{D}(\omega).
    \end{split}
\end{equation}
Solving for $\mathcal{G}_D,$ we find
\begin{equation}
    \mathcal{G}_D(\omega) = \left[ \omega - \mathcal{H}_D - \tilde{\mathcal{V}}_{LD}^\dagger \left[\omega- \tilde{\mathcal{H}}_L \right]^{-1} \tilde{\mathcal{V}}_{LD}- \tilde{\mathcal{V}}_{DR} \left[\omega- \tilde{\mathcal{H}}_R \right]^{-1} \tilde{\mathcal{V}}_{DR}^\dagger \right]^{-1}.
\end{equation}
$\left[\omega- \tilde{\mathcal{H}}_L \right]^{-1}$ and $\left[\omega- \tilde{\mathcal{H}}_R \right]^{-1}$ are the full Green's functions of the left and right leads without any effect from the domain wall respectively. However, since the coupling matrices are everywhere zero except at the domain wall, we can replace these with surface Green's functions computed using the iterative method outlined in Sec. \ref{sec: Semi-Infinite Plane}. Therefore, we obtain
\begin{equation}
    \mathcal{G}_D(\omega) = \left[ \omega - \mathcal{H}_D - \mathcal{V}_{LD}^\dagger \mathcal{G}_{0,L}(\omega) \mathcal{V}_{LD}- \mathcal{V}_{DR} \mathcal{G}_{0,R}(\omega) \mathcal{V}_{DR}^\dagger \right]^{-1}.
\end{equation}
To simplify, we define the lead self energies
\begin{equation}
    \Sigma_L(\omega) = \mathcal{V}_{LD}^\dagger \mathcal{G}_{0,L}(\omega) \mathcal{V}_{LD} \quad \text{and} \quad \Sigma_R(\omega) = \mathcal{V}_{RD}^\dagger \mathcal{G}_{0,R}(\omega) \mathcal{V}_{RD}.
\end{equation}
We also define the level-width functions
\begin{equation}
    \Gamma_L(\omega) = i \left(  \Sigma_L(\omega)-\Sigma_L^\dagger(\omega) \right) \quad \text{and} \quad \Gamma_R(\omega) = i \left(  \Sigma_R(\omega)-\Sigma_R^\dagger(\omega) \right).
\end{equation}
Using these various functions, the transmission from left to right is given by the Caroli-Fisher-Lee formula \cite{caroli1971direct,Fisher1981Relation}
\begin{equation}
    T(E) = \tr \left[ \Gamma_L(E) \mathcal{G}_D(E) \Gamma_R(E) \mathcal{G}_D^\dagger(E) \right],
\end{equation}
where $\omega = E +i0^+$ is used to define the delayed Green's functions.


\section{Models of Superconductivity}

\subsection{Monolayer Toy Model}

Because superconductivity emerges only in the large displacement field limit, it is likely that the superconducting pair function is highly layer polarized. Therefore, let us begin with a one-orbital monolayer model written in a momentum basis centered at $\mathbf{K}$ and not at $\Gamma$ in the primitive \textit{triangular} Brillouin zone (not yet in the Kekul\'{e} Brillouin zone). We have
\begin{equation}
    \hat{\mathcal{H}}_\mathrm{BdG} = \frac{1}{2} \sum_\mathbf{k}\begin{pmatrix}
        \hat{c}_\mathbf{K+\mathbf{k}}^\dagger & \hat{c}_{\mathbf{K}-\mathbf{k}}
    \end{pmatrix} \begin{pmatrix}
        \varepsilon_\mathbf{K}(\mathbf{k})-\mu & \Delta_\mathbf{K}(\mathbf{k}) \\
        \Delta_\mathbf{K}^\dagger(\mathbf{k}) & -\varepsilon^*_\mathbf{K}(-\mathbf{k})+\mu
        \end{pmatrix} \begin{pmatrix}
        \hat{c}_\mathbf{K+\mathbf{k}}\\ \hat{c}_{\mathbf{K}-\mathbf{k}}^\dagger
    \end{pmatrix}.
\end{equation}
We write the superconducting gap centered at the zone corner as 
\begin{equation}
    \hat{\Delta} = \sum_\mathbf{k} \hat{c}^\dagger_{\mathbf{K}+\mathbf{k}}\Delta_\mathbf{K}(\mathbf{k}) \hat{c}^\dagger_{\mathbf{K}-\mathbf{k}} = \sum_\mathbf{k} \hat{c}^\dagger_{\mathbf{K}+\mathbf{k}}\Delta_\mathrm{SC} \left[ \sin(\mathbf{k}\cdot \mathbf{a}_1)+  \omega\sin(\mathbf{k}\cdot \mathbf{a}_2) +  \omega^\dagger\sin(\mathbf{k}\cdot \mathbf{a}_3)\right]\hat{c}^\dagger_{\mathbf{K}-\mathbf{k}}.
\end{equation}
Here, $\omega = e^{2\pi i /3}$ and $\mathbf{a}_1 = (a,0), \mathbf{a}_2 = (-a/2,\sqrt{3}a/2),$ $ \mathbf{a}_3 = (-a/2,-\sqrt{3}a/2).$ This gap function is manifestly antisymmetric under $\mathbf{k} \mapsto-\mathbf{k}$ and has a unity phase winding, which can be seen from expanding around $\mathbf{k} = \mathbf{0}: \sin(\mathbf{k}\cdot \mathbf{a}_1)+  \omega\sin(\mathbf{k}\cdot \mathbf{a}_2) +  \omega^\dagger\sin(\mathbf{k}\cdot \mathbf{a}_3) \approx \frac{3}{2}a (k_x+ik_y)$. We rewrite this function in real space
\begin{equation}
\begin{split}
    \hat{\Delta} &= \frac{1}{N}\frac{\Delta_\mathrm{SC}}{2i}\sum_\mathbf{k} \sum_{\mathbf{r},\mathbf{r'}}e^{i\mathbf{K}\cdot( \mathbf{r}+\mathbf{r}')} e^{i\mathbf{k} \cdot (\mathbf{r}-\mathbf{r}')} \hat{c}_\mathbf{r}^\dagger  \left( e^{i\mathbf{k}\cdot \mathbf{a}_1}-e^{-i\mathbf{k}\cdot \mathbf{a}_1}+  \omega e^{i\mathbf{k}\cdot \mathbf{a}_2}-\omega e^{-i\mathbf{k}\cdot \mathbf{a}_2} +  \omega^\dagger e^{i\mathbf{k}\cdot \mathbf{a}_3}-\omega^\dagger e^{-i\mathbf{k}\cdot \mathbf{a}_3}\right) \hat{c}^\dagger_{\mathbf{r}'} \\
        &= \frac{\Delta_\mathrm{SC}}{2i}\sum_{\mathbf{r}}e^{2i\mathbf{K} \cdot \mathbf{r}} \left[  \omega^\dagger \left(\hat{c}_\mathbf{r}^\dagger\hat{c}^\dagger_{\mathbf{r}+\mathbf{a}_1} + \omega \hat{c}_\mathbf{r}^\dagger\hat{c}^\dagger_{\mathbf{r}+\mathbf{a}_2} + \omega^\dagger \hat{c}_\mathbf{r}^\dagger\hat{c}^\dagger_{\mathbf{r}+\mathbf{a}_3} \right) - \omega \left(\hat{c}_\mathbf{r}^\dagger\hat{c}^\dagger_{\mathbf{r}-\mathbf{a}_1}  + \omega  \hat{c}_\mathbf{r}^\dagger\hat{c}^\dagger_{\mathbf{r}-\mathbf{a}_2}  + \omega^\dagger  \hat{c}_\mathbf{r}^\dagger\hat{c}^\dagger_{\mathbf{r}-\mathbf{a}_3} \right)\right],    \\
\end{split}
\end{equation}
where $\mathbf{r} $ is a lattice translation vector. Because $\mathbf{K}$ is \textit{not} a reciprocal lattice vector in the triangular Brillouin zone, the overall phase $e^{2i\mathbf{K} \cdot \mathbf{r}}$ is not periodic in $\mathbf{r}$. This is a signature of \textit{finite-momentum} pairing: there must be a modulation of the pairing wavefunction on a scale \textit{incommensurate} with the lattice structure. Now, to make the phase  $e^{2i\mathbf{K} \cdot \mathbf{r}}$ periodic, we can enlarge the unit cell to a $\sqrt{3}\times \sqrt{3}$ Kekul\'{e} lattice where the sublattices are located at $\boldsymbol{\tau}_{A_1} = (0,0),$ $\boldsymbol{\tau}_{A_2} = -\mathbf{a}_3,$ and $\boldsymbol{\tau}_{A_3} = \mathbf{a}_1$ and the new primitive translation vectors are $\overline{\mathbf{a}}_1= \left(0,\sqrt{3}a \right),$ $\overline{\mathbf{a}}_2= \sqrt{3}a\left(-\sqrt{3}/2,-1/2 \right),$ and $\overline{\mathbf{a}}_3= \sqrt{3}a\left(+\sqrt{3}/2,-1/2 \right).$ In this reconstructed lattice, $\mathbf{K}$ is indeed a reciprocal lattice vector. The pairing function now takes the form
\begin{equation}
\begin{split}
    \hat{\Delta} &=  \frac{\Delta_\mathrm{SC}}{2i}\sum_{\mathbf{r}}\left[  \omega^\dagger \left(\hat{c}_{A_1,\mathbf{r}}^\dagger\hat{c}^\dagger_{A_3,\mathbf{r}} + \omega \hat{c}_{A_1,\mathbf{r}}^\dagger\hat{c}^\dagger_{A_3,\mathbf{r}+\overline{\mathbf{a}}_1+\overline{\mathbf{a}}_2} + \omega^\dagger \hat{c}_{A_1,\mathbf{r}}^\dagger\hat{c}^\dagger_{A_3,\mathbf{r}+\overline{\mathbf{a}}_2} \right) - \omega \left(\hat{c}_{A_1,\mathbf{r}}^\dagger\hat{c}^\dagger_{A_2,\mathbf{r}+\overline{\mathbf{a}}_2}  + \omega  \hat{c}_{A_1,\mathbf{r}}^\dagger\hat{c}^\dagger_{A_2,\mathbf{r}-\overline{\mathbf{a}}_1}  + \omega^\dagger  \hat{c}_{A_1,\mathbf{r}}^\dagger\hat{c}^\dagger_{A_2,\mathbf{r}} \right)\right]   \\
    &+  \frac{\Delta_\mathrm{SC}}{2i}\sum_{\mathbf{r}} \omega^\dagger \left[  \omega^\dagger \left(\hat{c}_{A_2,\mathbf{r}}^\dagger\hat{c}^\dagger_{A_1,\mathbf{r}+\overline{\mathbf{a}}_1+\overline{\mathbf{a}}_3} + \omega \hat{c}_{A_2,\mathbf{r}}^\dagger\hat{c}^\dagger_{A_1,\mathbf{r}+\overline{\mathbf{a}}_1} + \omega^\dagger \hat{c}_{A_2,\mathbf{r}}^\dagger\hat{c}^\dagger_{A_1,\mathbf{r}} \right) - \omega \left(\hat{c}_{A_2,\mathbf{r}}^\dagger\hat{c}^\dagger_{A_3,\mathbf{r}+\overline{\mathbf{a}}_1+\overline{\mathbf{a}}_2}  + \omega  \hat{c}_{A_2,\mathbf{r}}^\dagger\hat{c}^\dagger_{A_3,\mathbf{r}}  + \omega^\dagger  \hat{c}_{A_2,\mathbf{r}}^\dagger\hat{c}^\dagger_{A_3,\mathbf{r}+\overline{\mathbf{a}}_1} \right)\right]   \\
    &+  \frac{\Delta_\mathrm{SC}}{2i}\sum_{\mathbf{r}}\omega \left[  \omega^\dagger \left(\hat{c}_{A_3,\mathbf{r}}^\dagger\hat{c}^\dagger_{A_2,\mathbf{r}+\overline{\mathbf{a}}_3} + \omega \hat{c}_{A_3,\mathbf{r}}^\dagger\hat{c}^\dagger_{A_2,\mathbf{r}} + \omega^\dagger \hat{c}_{A_3,\mathbf{r}}^\dagger\hat{c}^\dagger_{A_2,\mathbf{r}-\overline{\mathbf{a}}_1} \right) - \omega \left(\hat{c}_{A_3,\mathbf{r}}^\dagger\hat{c}^\dagger_{A_1,\mathbf{r}}  + \omega  \hat{c}_{A_3,\mathbf{r}}^\dagger\hat{c}^\dagger_{A_1,\mathbf{r}+\overline{\mathbf{a}}_3}  + \omega^\dagger  \hat{c}_{A_3,\mathbf{r}}^\dagger\hat{c}^\dagger_{A_1,\mathbf{r}+\overline{\mathbf{a}}_1+\overline{\mathbf{a}}_3} \right)\right].   \\
\end{split}
\end{equation}
Here, the notation is that $\hat{c}_{\sigma,\mathbf{r}}^\dagger = \hat{c}_{\mathbf{r}+\boldsymbol{\tau}_\sigma}^\dagger$ and the $\mathbf{r}$ vectors are the translation vectors of the Kekul\'{e} lattice. In this expanded basis, we can write the pairing function with center-of-mass momentum at $\overline{\Gamma}.$ In momentum space, the Hamiltonian now takes the form
\begin{equation}
    \hat{\Delta} =  \sum_\mathbf{k} \begin{pmatrix}
        \hat{c}_{A_1,\mathbf{k}}^\dagger & \hat{c}_{A_2,\mathbf{k}}^\dagger & \hat{c}_{A_3,\mathbf{k}}^\dagger 
    \end{pmatrix}\begin{pmatrix}
        0 & -\omega f(-\mathbf{k}) & \omega^\dagger f(\mathbf{k}) \\
        \omega f(\mathbf{k}) & 0 & - f(-\mathbf{k}) \\
        - \omega^\dagger f(\mathbf{-k}) & f(\mathbf{k}) & 0 
    \end{pmatrix} \begin{pmatrix}
        \hat{c}^\dagger_{A_1,-\mathbf{k}} \\ \hat{c}^\dagger_{A_2,-\mathbf{k}} \\ \hat{c}^\dagger_{A_3,-\mathbf{k}}
    \end{pmatrix} = \sum_{\mathbf{k},\sigma,\sigma'} \hat{c}^\dagger_{\sigma,\mathbf{k}} \Delta_{\sigma \sigma'}(\mathbf{k)} \hat{c}^\dagger_{\sigma',-\mathbf{k}},
\end{equation}
where $f(\mathbf{k}) = \left(\Delta_\mathrm{SC}/2i\right) \left[\exp(i\mathbf{k} \cdot \mathbf{a}_1) + \omega \exp(i \mathbf{k} \cdot \mathbf{a}_2) + \omega^\dagger \exp(i\mathbf{k} \cdot \mathbf{a}_3) \right].$ As a check of consistency, we note that antisymmetry is satisfied explicitly because $\Delta_{\sigma,\sigma'}(\mathbf{k}) = - \Delta_{\sigma',\sigma}(-\mathbf{k})$ and that the projection to the $K$ valley holds
\begin{equation}
    \frac{1}{3} \begin{pmatrix}
        1 & \omega^\dagger & \omega
    \end{pmatrix} \begin{pmatrix}
        0 & -\omega f(-\mathbf{k}) & \omega^\dagger f(\mathbf{k}) \\
        \omega f(\mathbf{k}) & 0 & - f(-\mathbf{k}) \\
        - \omega^\dagger f(\mathbf{-k}) & f(\mathbf{k}) & 0 
    \end{pmatrix}\begin{pmatrix}
        1 \\
        \omega^\dagger \\
        \omega
    \end{pmatrix} = \Delta_\mathrm{SC} \left[ \sin(\mathbf{k}\cdot \mathbf{a}_1)+  \omega\sin(\mathbf{k}\cdot \mathbf{a}_2) +  \omega^\dagger\sin(\mathbf{k}\cdot \mathbf{a}_3)\right].
\end{equation}
To switch the chirality of the pairing function, we replace $\omega$ \textit{inside} the $f(\mathbf{k})$ functions with $\omega^\dagger.$ To switch the center-of-mass momentum, we replace $\omega$ \textit{outside} the $f(\mathbf{k})$ functions with $\omega^\dagger.$ Therefore, if we want to describe a $k_x-ik_y$ pairing function at $K',$ we simply complex conjugate all factors of $\omega$ in $\Delta_{\sigma,\sigma'}(\mathbf{k}).$ However, prior theoretical studies have suggested that the chirality of superconductivity is locked to the valley polarization.

\begin{figure}
    \centering
    \includegraphics[width=0.8\linewidth]{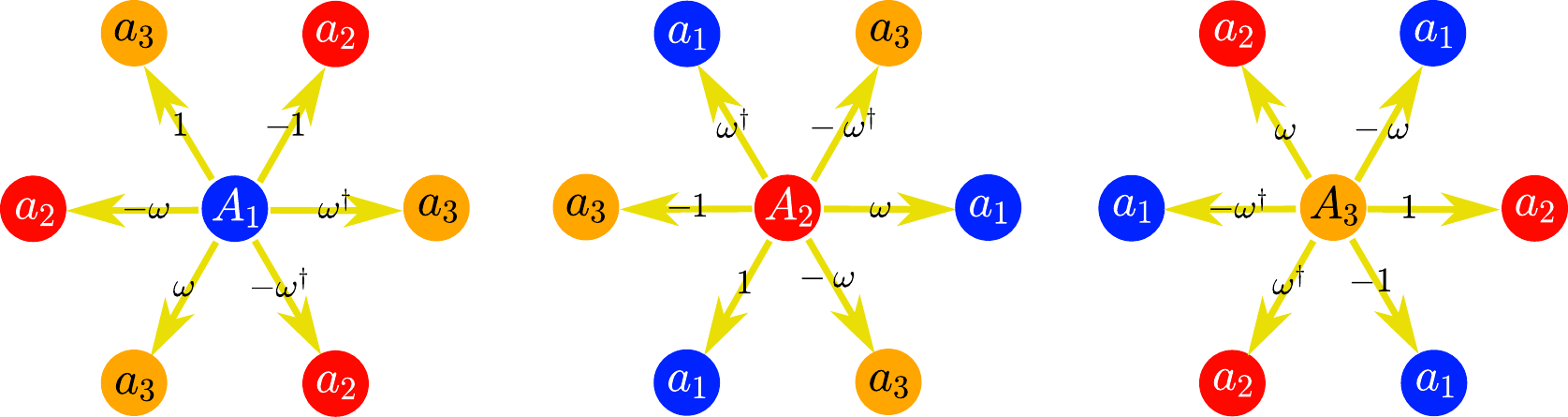}
    \caption{\textbf{Real-space representation of $k_x+ik_y$ superconducting pair function with center-of-mass momentum $2\mathbf{K}$.} Capitalized labels indicate electron sites, while lower-case labels indicate hole sites. All amplitudes should be multiplied by $\Delta_\mathrm{SC}/2i.$   To switch the chirality of the pairing function, the hopping amplitudes are reflected across the $x$-axis. To switch the center-of-mass momentum, the hopping amplitudes are complex-conjugated and reflected across the $x$-axis.}
    \label{fig:sc_gap_hoppings}
\end{figure}

The foregoing preliminary considerations allow us to construct a tight-binding model where the electron and hole sectors are independently represented by fermionic operators, $\hat{d}_{\sigma,\mathbf{k}} = \hat{c}^\dagger_{\sigma,-\mathbf{k}}.$ The gap function is represented in real space as hoppings between the electron and hole sectors, as shown in Fig. \ref{fig:sc_gap_hoppings}.


\begin{figure}
    \centering
    \includegraphics[width=1\linewidth]{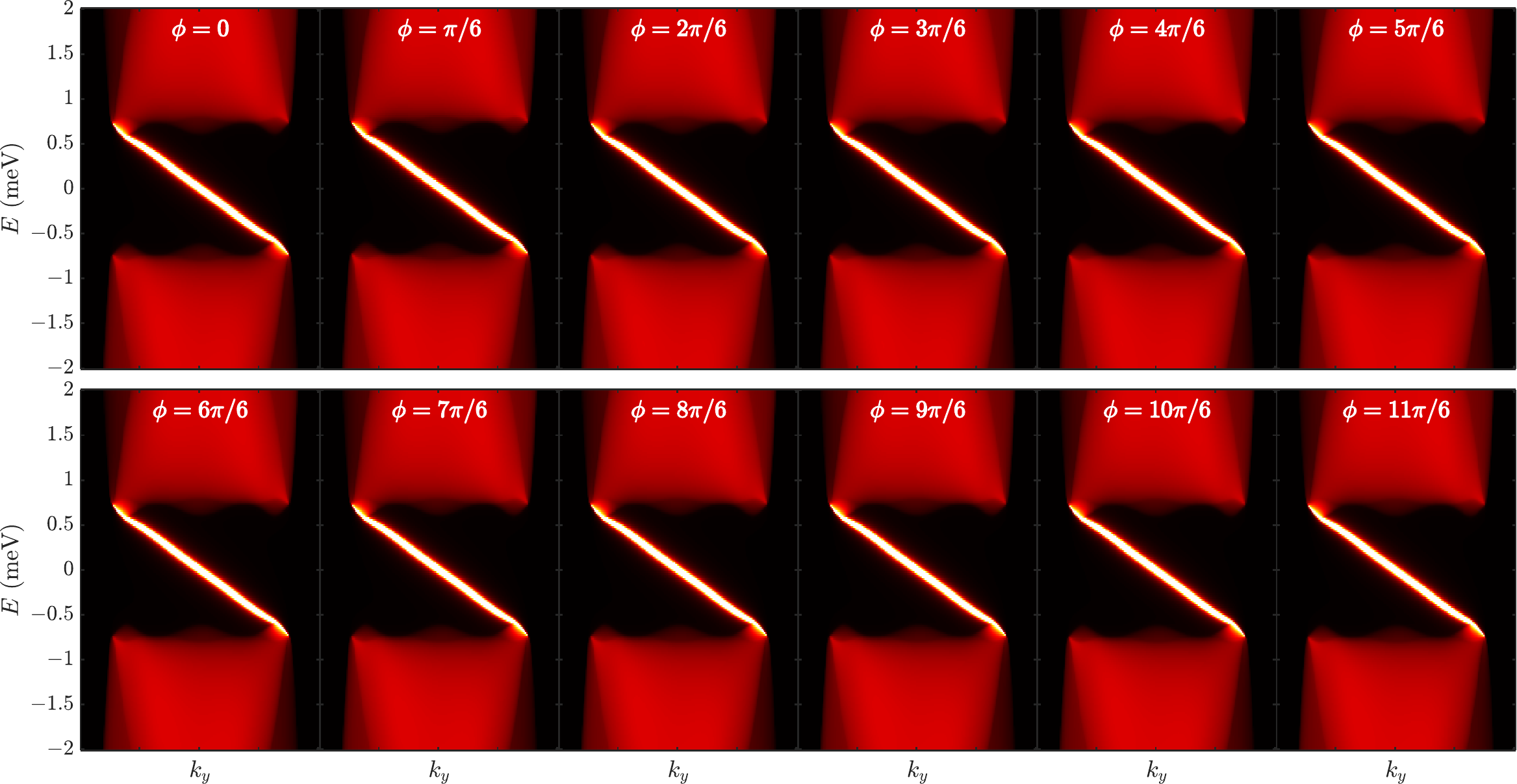}
    \caption{\textbf{BdG band structure of an SNS Josephson junction as a function of the pairing phase $\phi$}. In this calculation, the domain wall has no intervalley hybridization. Here, $N = 4,$ $m_z = 10$ meV, $n=5\times10^{11}$ cm$^{-2},$ $w = 2$ nm, $\Delta = 30$ meV, and $\Delta_\mathrm{SC} = 10$ meV.}
    \label{fig:intervalley_phase_winding_n=4_m=0}
\end{figure}

\begin{figure}
    \centering
    \includegraphics[width=1\linewidth]{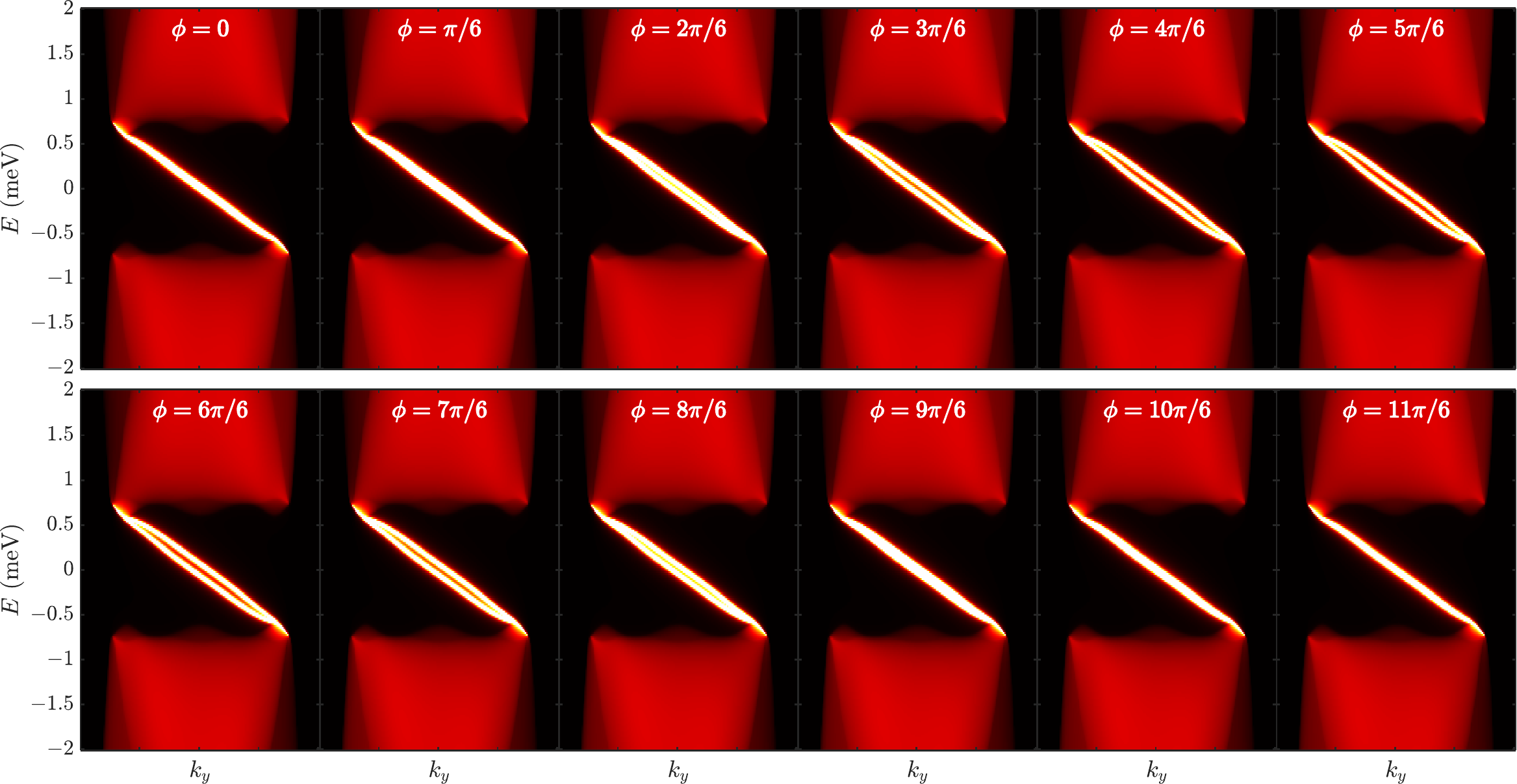}
    \caption{\textbf{BdG band structure of an SNS Josephson junction as a function of the pairing phase $\phi$}. In this calculation, the domain wall has an intervalley interaction $\tau_1$ with magnitude $m = 1$ meV. Here, $N = 4,$ $m_z = 10$ meV, $n=5\times10^{11}$ cm$^{-2},$ $w = 2$ nm, $\Delta = 30$ meV, and $\Delta_\mathrm{SC} = 10$ meV.}
    \label{fig:intervalley_phase_winding_n=4_m=1}
\end{figure}

\begin{figure}
    \centering
    \includegraphics[width=1\linewidth]{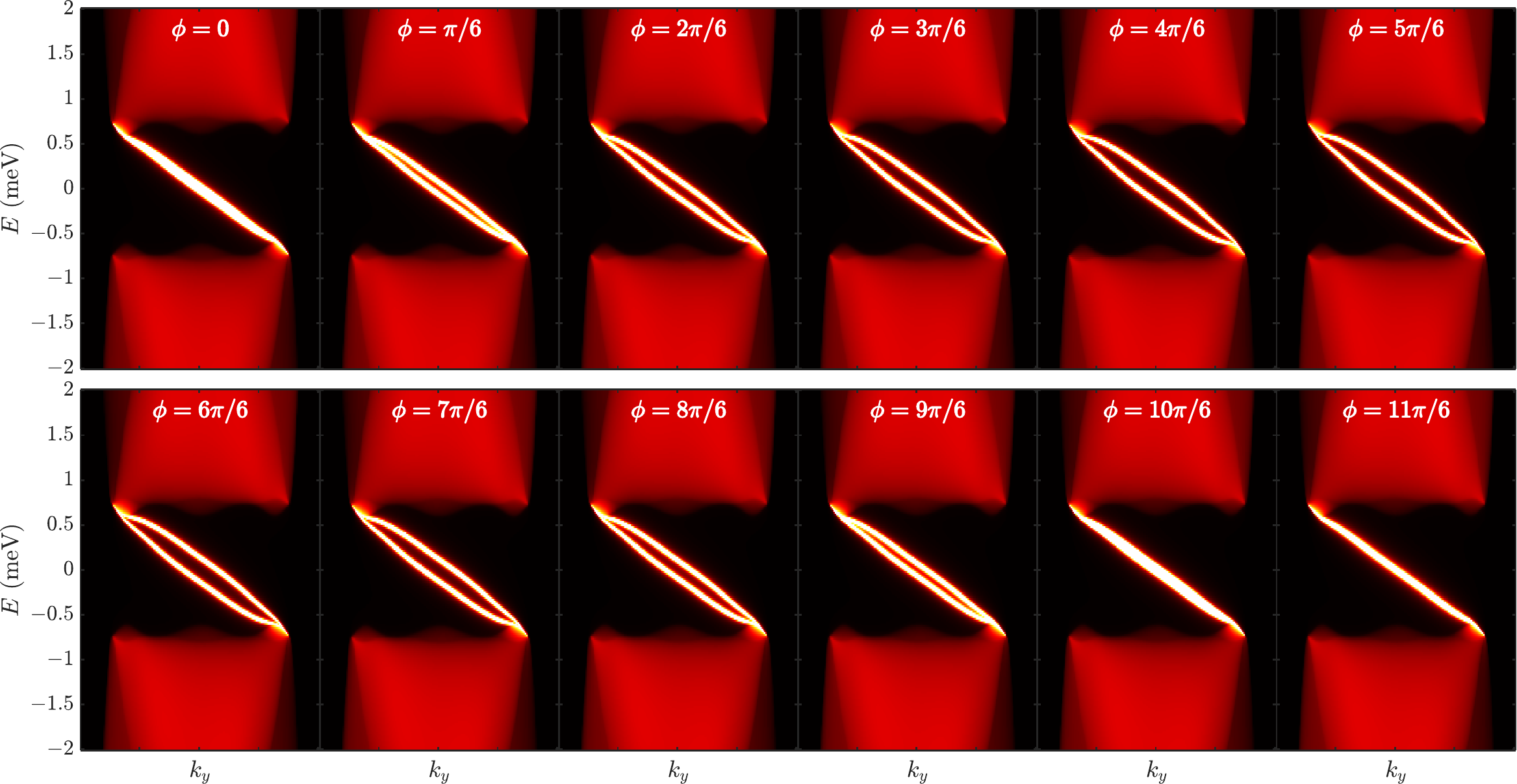}
    \caption{\textbf{BdG band structure of an SNS Josephson junction as a function of the pairing phase $\phi$}. In this calculation, the domain wall has an intervalley interaction $\tau_1$ with magnitude $m = 2$ meV. Here, $N = 4,$ $m_z = 10$ meV, $n=5\times10^{11}$ cm$^{-2},$ $w = 2$ nm, $\Delta = 30$ meV, and $\Delta_\mathrm{SC} = 10$ meV.}
    \label{fig:intervalley_phase_winding_n=4_m=2}
\end{figure}

\begin{figure}
    \centering
    \includegraphics[width=1\linewidth]{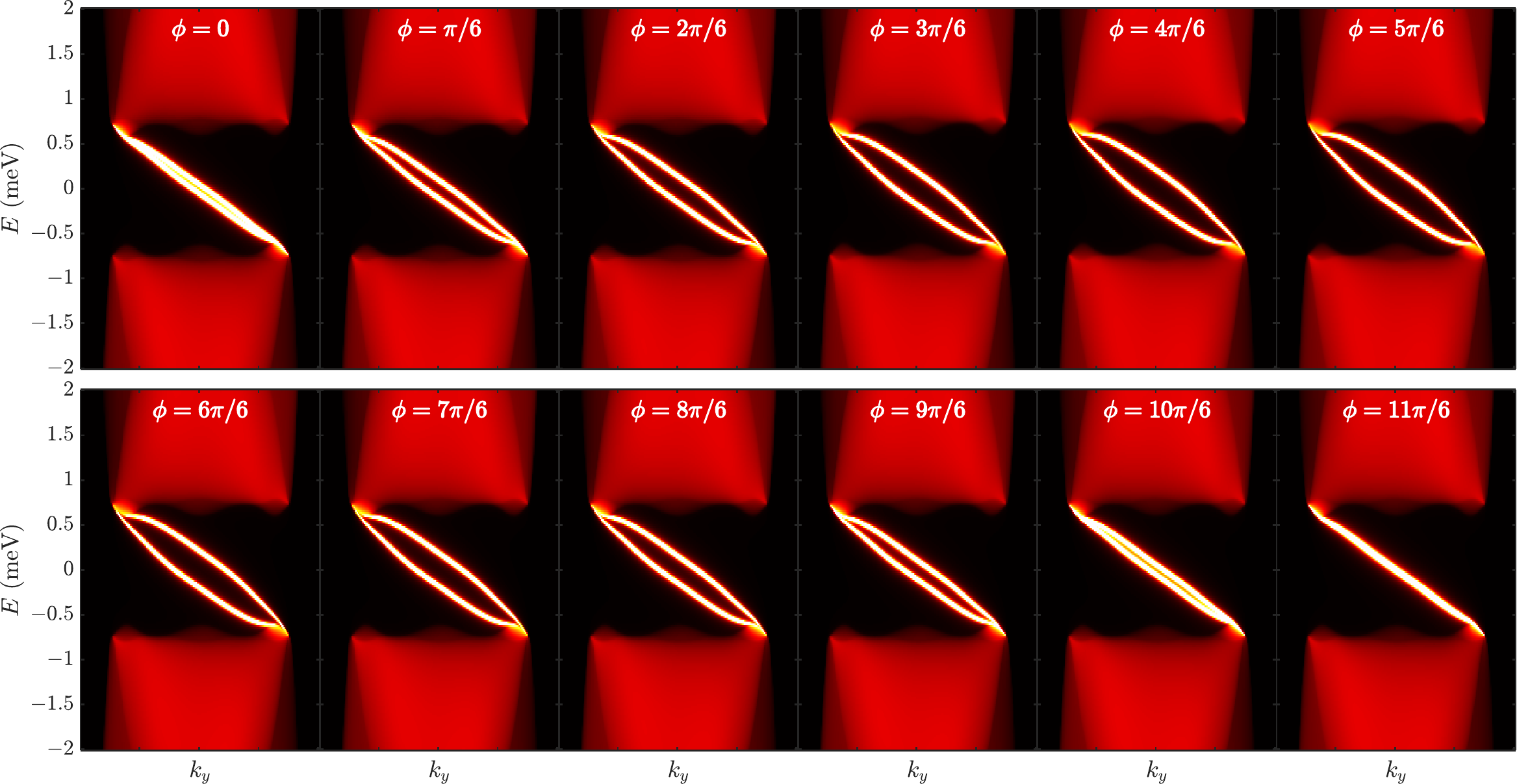}
    \caption{\textbf{BdG band structure of an SNS Josephson junction as a function of the pairing phase $\phi$}. In this calculation, the domain wall has an intervalley interaction $\tau_1$ with magnitude $m = 3$ meV. Here, $N = 4,$ $m_z = 10$ meV, $n=5\times10^{11}$ cm$^{-2},$ $w = 2$ nm, $\Delta = 30$ meV, and $\Delta_\mathrm{SC} = 10$ meV.}
    \label{fig:intervalley_phase_winding_n=4_m=3}
\end{figure}

\begin{figure}
    \centering
    \includegraphics[width=1\linewidth]{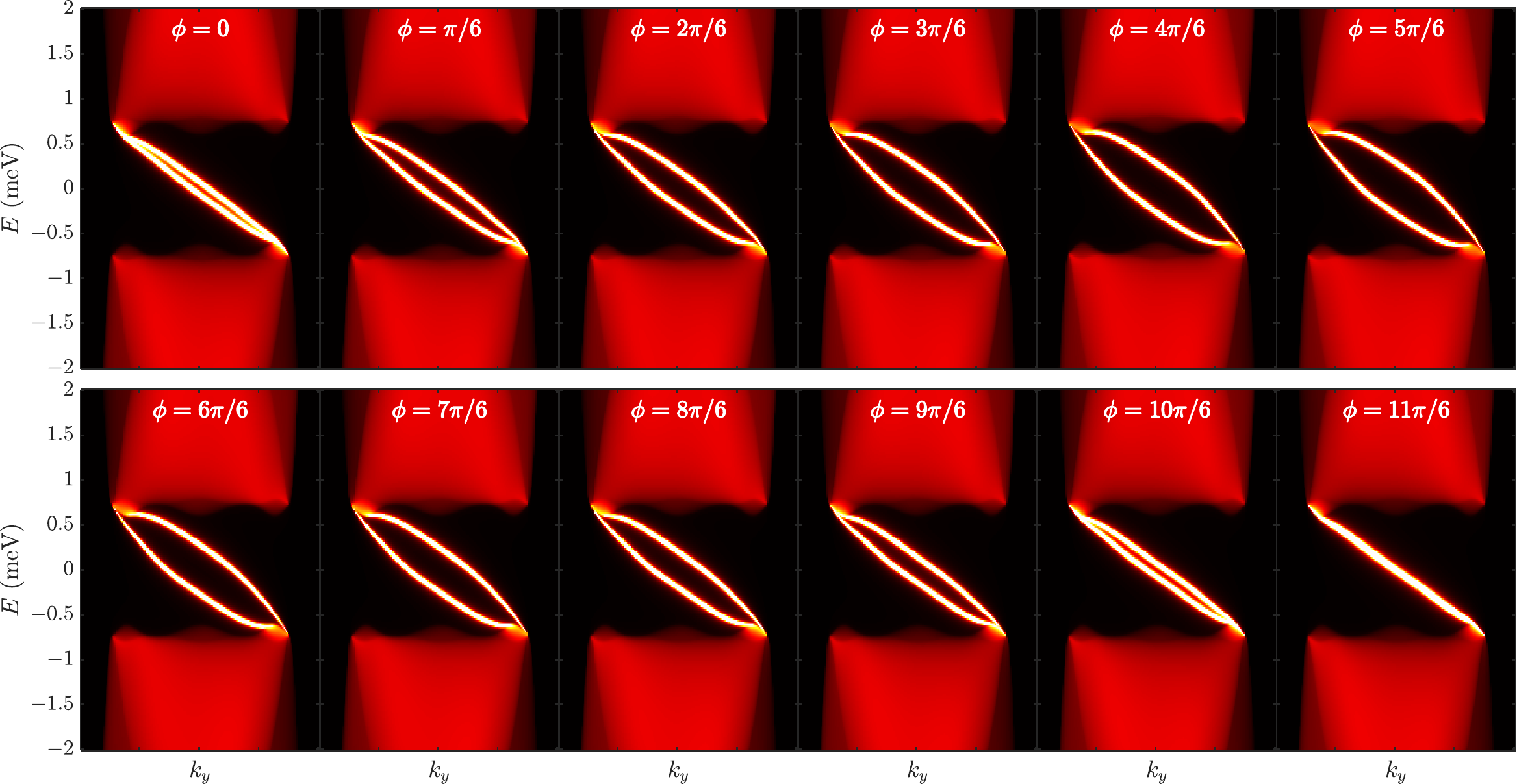}
    \caption{\textbf{BdG band structure of an SNS Josephson junction as a function of the pairing phase $\phi$}. In this calculation, the domain wall has an intervalley interaction $\tau_1$ with magnitude $m = 4$ meV. Here, $N = 4,$ $m_z = 10$ meV, $n=5\times10^{11}$ cm$^{-2},$ $w = 2$ nm, $\Delta = 30$ meV, and $\Delta_\mathrm{SC} = 10$ meV.}
    \label{fig:intervalley_phase_winding_n=4_m=4}
\end{figure}

\begin{figure}
    \centering
    \includegraphics[width=1\linewidth]{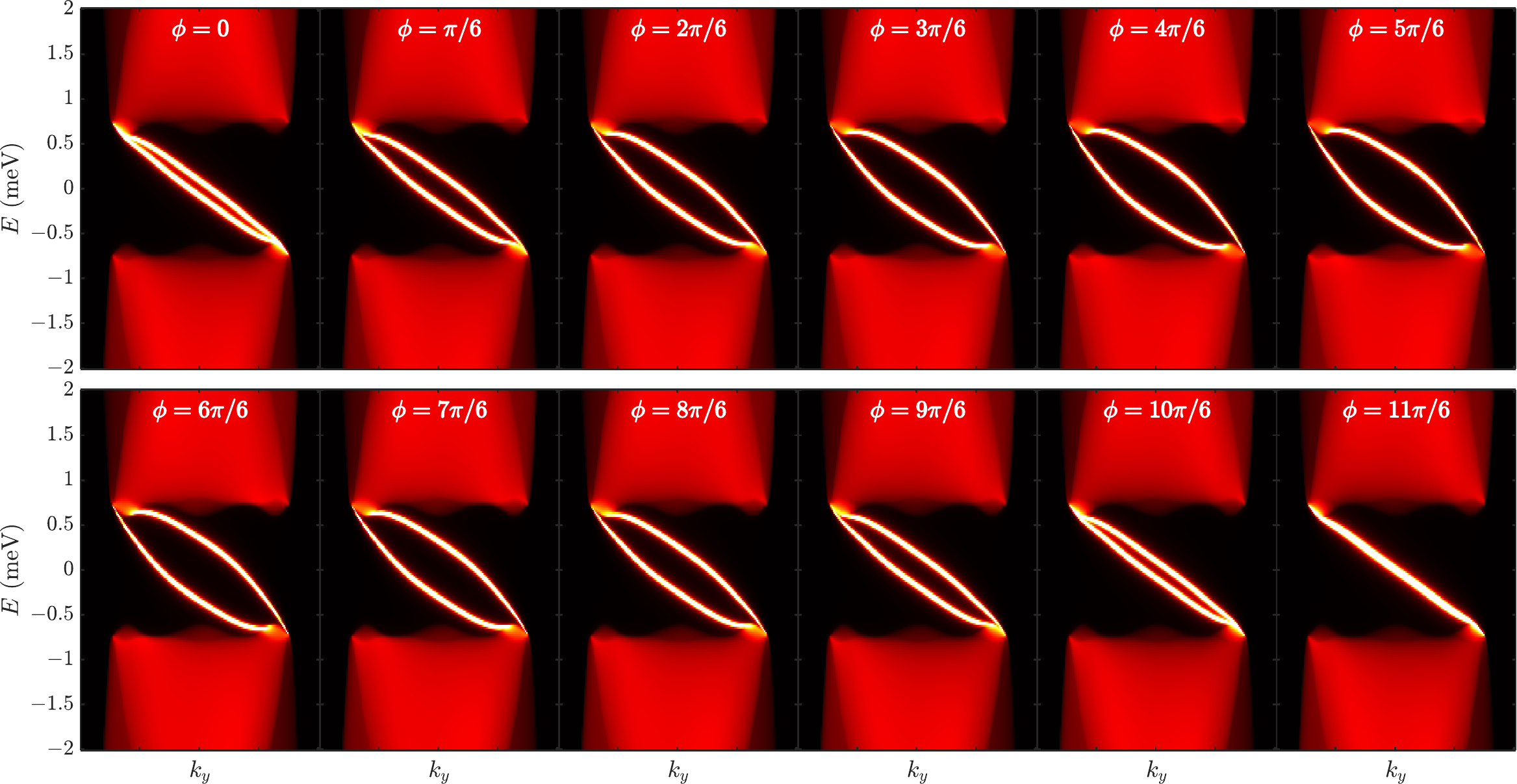}
    \caption{\textbf{BdG band structure of an SNS Josephson junction as a function of the pairing phase $\phi$}. In this calculation, the domain wall has an intervalley interaction $\tau_1$ with magnitude $m = 5$ meV. Here, $N = 4,$ $m_z = 10$ meV, $n=5\times10^{11}$ cm$^{-2},$ $w = 2$ nm, $\Delta = 30$ meV, and $\Delta_\mathrm{SC} = 10$ meV.}
    \label{fig:intervalley_phase_winding_n=4_m=5}
\end{figure}

\subsection{Supercurrent calculation}

The supercurrent is defined as the phase derivative of the free energy $\mathcal{F}(\phi)$
\begin{equation}
    I(\phi) = -\frac{2e}{\hbar} \frac{\partial \mathcal{F}}{\partial\phi}.
\end{equation}
The free energy at finite temperature for a single spin species (due to polarization) is given by
\begin{equation}
    \mathcal{F}(\phi) = - \frac{k_BT}{2} \sum_{n=-\infty}^\infty \ln \det \left[- \mathcal{G}^{-1}(i\omega_n)  \right]+...,
\end{equation}
where $\mathcal{G}(i\omega_n) = (i\omega_n-\mathcal{H}_\mathrm{BdG})^{-1}$ is the BdG Green's function and $\omega_n = (2n+1) \pi k_B T$. The factor of $1/2$ is there to avoid double counting the fermion degrees of freedom in the BdG basis \footnote{In the limit $\Delta_\mathrm{SC} \rightarrow 0,$ this definition gives exactly the free energy of the Fermi sea of a normal metal.}. Using the rule $\partial_\phi \ln \det [-\mathcal{G}^{-1}] =  \tr \mathcal{G} \partial_\phi \mathcal{G}^{-1},$ the supercurrent is given by
\begin{equation}
    I(\phi) = \frac{e}{\hbar}k_BT \sum_{n=-\infty}^\infty  \Tr \left[\mathcal{G}(i\omega_n) \frac{\partial}{\partial\phi} \mathcal{G}^{-1}(i\omega_n) \right] = -\frac{e}{\hbar} \oint_\mathcal{C} \frac{dz}{2\pi i} f(z) \Tr \left[\mathcal{G}(z) \frac{\partial}{\partial\phi} \mathcal{G}^{-1}(z) \right],
\end{equation}
where $f(z)$ is the fermion occupation function and $\mathcal{C}$ is a contour that encloses the poles of $f(z).$ For now, we leave the sum over momentum implicit for brevity. Now, assuming that $\mathcal{G}(\omega +i\eta)$ is analytic in the upper half plane and $\mathcal{G}(\omega -i\eta)$ is analytic in the lower half plane and both decay as $|z| \rightarrow \infty$, the contour integral can be turned into an integral along the real line
\begin{equation}
    I(\phi) = -\frac{e}{\hbar} \int_{-\infty}^\infty \frac{d\omega}{2\pi i} n_F(\omega) \Tr \left[\mathcal{G}(\omega + i \eta) \frac{\partial}{\partial\phi} \mathcal{G}^{-1}(\omega + i \eta) -\mathcal{G}(\omega - i \eta) \frac{\partial}{\partial\phi} \mathcal{G}^{-1}(\omega - i \eta)\right].
\end{equation}
Now, noting that $\mathcal{G}(\omega + i \eta) = \mathcal{G}^\dagger(\omega - i \eta),$ we can simplify the integrand
\begin{equation}
    I(\phi) = -\frac{e}{\hbar} \int_{-\infty}^\infty \frac{d\omega}{\pi } n_F(\omega) \mathrm{Im}\Tr \left[\mathcal{G}(\omega + i \eta) \frac{\partial}{\partial\phi} \mathcal{G}^{-1}(\omega + i \eta)\right].
\end{equation}
To simplify, we assume that only the self-energy on the right domain contains the phase $\phi,$
\begin{equation}
    I(\phi) = \frac{e}{\hbar} \int_{-\infty}^\infty \frac{d\omega}{\pi } n_F(\omega) \mathrm{Im}\Tr \left[\mathcal{G}(\omega + i \eta) \frac{\partial}{\partial\phi} \Sigma_R(\omega + i \eta)\right].
\end{equation}
Using the identity $\partial_\phi\Sigma_R = i[\gamma_z,\Sigma_R]/2,$ where $\gamma_z$ is a Pauli matrix that acts on Nambu space, we find
\begin{equation}
    I(\phi) = \frac{e}{\hbar} \mathrm{Re} \left[\int_{-\infty}^\infty \frac{d\omega}{2\pi } n_F(\omega)\Tr \left(\mathcal{G}(\omega + i \eta)[\gamma_z,\Sigma_R(\omega+i\eta)]\right)\right].
\end{equation}
At zero temperature, the integral only goes up to $\omega = 0.$ In this case, we use a change of variables to map the interval $\left(-\infty,0\right)$ to the interval $\left(-1,1 \right)$ and then use contour integration to evaluate the integral numerically. We apply the adaptive Gauss-Kronrod quadrature for fast convergence. The error on the supercurrent calculated at each $k_y$ is approximately $10^{-5}$ nA.

\bibliography{references}

\end{document}